\documentclass{ieeeaccess}

\usepackage{amsmath}
\usepackage{graphicx}
\usepackage{url}
\usepackage{hyperref}
\usepackage{cite}
\usepackage{amsfonts}
\usepackage{mathtools}
\usepackage{hhline}
\usepackage[caption=false, font=footnotesize]{subfig}

\usepackage{commath,multirow}

\usepackage{adjustbox,cleveref}

\newcommand{\etal}{\textit{et al}. }

\newcommand{\eg}{{e}.{g}. }

\makeatletter
\def\ps@IEEEtitlepagestyle{%
  \def\@oddfoot{\mycopyrightnotice}%
  \def\@evenfoot{}%
}

\makeatletter
\long\def\@makecaption#1#2{\ifx\@captype\@IEEEtablestring%
\footnotesize\begin{center}{\normalfont\footnotesize #1}\\
{\normalfont\footnotesize\scshape #2}\end{center}%
\@IEEEtablecaptionsepspace
\else
\@IEEEfigurecaptionsepspace
\setbox\@tempboxa\hbox{\normalfont\footnotesize {#1.}~~ #2}%
\ifdim \wd\@tempboxa >\hsize%
\setbox\@tempboxa\hbox{\normalfont\footnotesize {#1.}~~ }%
\parbox[t]{\hsize}{\normalfont\footnotesize \noindent\unhbox\@tempboxa#2}%
\else
\hbox to\hsize{\normalfont\footnotesize\hfil\box\@tempboxa\hfil}\fi\fi}
\makeatother

\begin{document}
\history{ }
\doi{ }

\title{Subjective and Objective Quality Assessment of High Frame Rate Videos}
\author{\uppercase{Pavan C Madhusudana}\authorrefmark{1}, \uppercase{Xiangxu Yu \authorrefmark{1}}, \uppercase{Neil Birkbeck \authorrefmark{2}}, \uppercase{Yilin Wang \authorrefmark{2}}, \uppercase{Balu Adsumilli \authorrefmark{2}and Alan C Bovik}.\authorrefmark{1}, \IEEEmembership{Fellow, IEEE}}
\address[1]{Department of Electrical and Computer Engineering, The University of Texas at Austin, Austin, TX 78705, USA.}
\address[2]{Google, Mountain View, CA 94043, USA}

\tfootnote{This work was supported by Google.}

\begin{abstract}
	High frame rate (HFR) videos are becoming increasingly common with the tremendous popularity of live, high-action streaming content such as sports. Although HFR contents are generally of very high quality, high bandwidth requirements make them challenging to deliver efficiently, while simultaneously maintaining their quality. To optimize trade-offs between bandwidth requirements and video quality, in terms of frame rate adaptation, it is imperative to understand the intricate relationship between frame rate and perceptual video quality. Towards advancing progression in this direction we designed a new subjective resource, called the LIVE-YouTube-HFR (LIVE-YT-HFR) dataset, which is comprised of 480 videos having 6 different frame rates, obtained from 16 diverse contents. In order to understand the combined effects of compression and frame rate adjustment, we also processed videos at 5 compression levels at each frame rate. To obtain subjective labels on the videos, we conducted a human study yielding 19,000 human quality ratings obtained from a pool of 85 human subjects. We also conducted a holistic evaluation of existing state-of-the-art Full and No-Reference video quality algorithms, and statistically benchmarked their performance on the new database. The LIVE-YT-HFR database has been made available online for public use and evaluation purposes, with hopes that it will help advance research in this exciting video technology direction. It may be obtained at \url{https://live.ece.utexas.edu/research/LIVE_YT_HFR/LIVE_YT_HFR/index.html}
\end{abstract}



\maketitle

\section{Introduction}
\label{sec:introduction}

\PARstart{R}{ecent} advancements in hardware technology have resulted in a dramatic visual information explosion on the Internet. Visual data such as images and videos constitute as much as 80\% of total Internet traffic \cite{cisco_vni}. Contemporaneously, increasing demands for better consumer viewing experiences has compelled streaming and social video service providers to pursue the delivery of higher quality videos. The requirements of higher video quality can involve better immersive experiences, higher spatial resolutions, larger display sizes, high dynamic ranges (HDR), and high frame rates (HFR). Indeed, the rapid development of streaming video technology has made the production and reception of superior quality videos affordable to the general public. Popular mobile capture devices have made the creation of high quality video content quite pervasive. Improved hardware supports the display of higher quality videos. Powerful GPUs are now able to display live, real-time 4K, HDR, and HFR videos on consumer displays, and virtual reality videos on head-mounted displays. Video service providers like YouTube, Netflix and Amazon Prime Video continue to offer videos having higher spatial resolutions and/or increased frame rates. 

In the past, considerable research effort has been expended on improving spatial resolution (4K/8K) \cite{ge2017toward}, HDR \cite{mai2010optimizing,kundu2017no} and multiview formats \cite{smolic2007coding,de2013toward}.  However, there has been less progress on increasing the frame rates of consumer videos, and the vast majority of streamed or shared videos are still provided at 60 frames per second (fps) or less. 

Various factors have limited the mainstream deployment of HFR videos. In the past, the necessity of sophisticated capture equipment and expensive display technologies placed HFR out of reach of the general populace. However, because of modern advanced consumer grade digital cameras such as the GoPro \cite{GoPro} and Sony RX series \cite{Sony} more casual users can capture HFR videos at a reasonable cost. While the current dearth of HFR content is a factor hindering the growth of its popularity, this is likely to change, given high interest in live action, high-speed sporting events and outdoor activities. Yet, HFR contents require higher bandwidths, making them more challenging for mass distribution by the streaming entertainment industry.

As technology evolves, HFR videos are likely to occupy a larger proportion of online videos, so it is important to understand the perceptual benefits associated with them. It is also interesting to consider the benefits conveyed to viewers' experiences when shifting from the low to high frame rate regime. While there is a general notion that HFR videos provide better perceptual quality, by reducing temporal artifacts such as flicker and motion blur, there has been little work done to validate these notions. Video Quality Assessment (VQA) has mostly addressed developments like HDR and high spatial resolution. One reason for this is the lack of subjective datasets addressing HFR videos, especially beyond 60 fps. 

Recently, there has been a renewed interest in HFR research, along with newer datasets like Waterloo HFR \cite{nasiri2015perceptual} and BVI-HFR \cite{mackin2018study}, which primarily address HFR content quality. These databases either contain only a few frame rates, and/or do not consider the joint effects of other distortions such as compression artifacts. To address these limitations and further advance progress on understanding HFR video quality, we have created a new HFR video resource, which we will refer to as the LIVE-YouTube High Frame Rate (LIVE-YT-HFR) Database. An important distinction of the new HFR database is the presence of six different frame rates with multiple spatial resolutions spread across a wide variety of contents. The new HFR database also encompasses a unique combination of compression and frame rate variations, evaluated and labeled by a large pool of volunteer subjects. Overall, the database comprises of 480 videos, making it one of the largest existing HFR video quality datasets. We also performed a holistic evaluation and benchmark study of current state-of-the-art VQA models. To help facilitate further development on HFR video quality research, we are making the new LIVE-YT-HFR dataset freely and publicly available in its entirety at \url{https://live.ece.utexas.edu/research/LIVE_YT_HFR/LIVE_YT_HFR/index.html}.

The rest of the paper is organized as follows: In Section \ref{sec:prior_work} we discuss prior work on the HFR quality problem. 
Section \ref{sec:LIVE_HFR} provides a detailed description of the new database and its construction. Section \ref{sec:subjective_quality} describes the subjective study. Section \ref{sec:objective_QA} compares and contrasts the performance of relevant VQA models on the new database. 
Finally we provide some concluding remarks in Section \ref{sec:conclusion}.

\section{Prior Work}
\label{sec:prior_work}

\subsection{Subjective VQA}
Research pertaining to video quality has made significant strides over the last decade. Several widely-used VQA databases have been proposed, including LIVE VQA \cite{seshadrinathan2010study}, LIVE Mobile \cite{moorthy2012video}, CSIQ-VQA \cite{vu2014vis3}, CDVL \cite{pinson2013consumer} etc. These generally begin with a set of less than 20 pristine video contents, on which various distortions are applied, primarily compression artifacts arising from past and present codecs, on both Standard Definition (SD) and High Definition (HD) resolutions. In all these databases, the reference and distorted sequences have the same frame rates, and therefore do not contain artifacts arising from frame rate changes. Moreover, the distortions present in these legacy databases were synthetically applied. More recent novel databases have emerged containing authentic distortions obtained from user-generated-content (UGC) videos. These include LIVE VQC \cite{sinno2018large}, KoNViD-1k \cite{hosu2017konstanz}, and YouTube UGC \cite{wang2019youtube}. Since the videos in these databases were captured by casual users, there are no pristine versions of any of the videos, hence they are primarily suited for blind video quality assessment research. Since only a single version of each content is available, these databases are not suitable for studying the perceptual impacts of frame rate changes.


Currently available datasets addressing HFR content are very limited. One of the first HFR databases was proposed by Nasiri \etal  \cite{nasiri2015perceptual}, containing SD and HD videos with frame rates no greater than 60 fps, and distorted by various compression levels. However, this database has not been made publicly available. Mackin \etal introduced the BVI-HFR \cite{mackin2018study} database, which contains videos of 4 different frame rates varying from 15 fps to 120 fps. The dataset includes 22 120 fps source sequences, where the lower fps videos were obtained by subsampling the source videos via frame averaging. Possible shortcomings of this database are that it only includes frame rate artifacts, it does not consider the effects of compression on frame rate, and it uses simple frame averaging to subsample in time. The latter strategy imposes a strong assumption on the changed videos, creates specific motion blur artifacts, and may not match practical systems.

\subsection{Objective VQA}
Generally, VQA models are broadly classified into three categories: Full-Reference (FR), Reduced-Reference (RR) and No-Reference (NR). FR VQA models require entire pristine undistorted videos against which degraded versions are perceptually compared, while RR models operate with limited reference information. NR-models predict quality with no reference knowledge.

\begin{figure*}[!t]
	\centering
	\subfloat[Runner]{\includegraphics[width=0.24\textwidth]{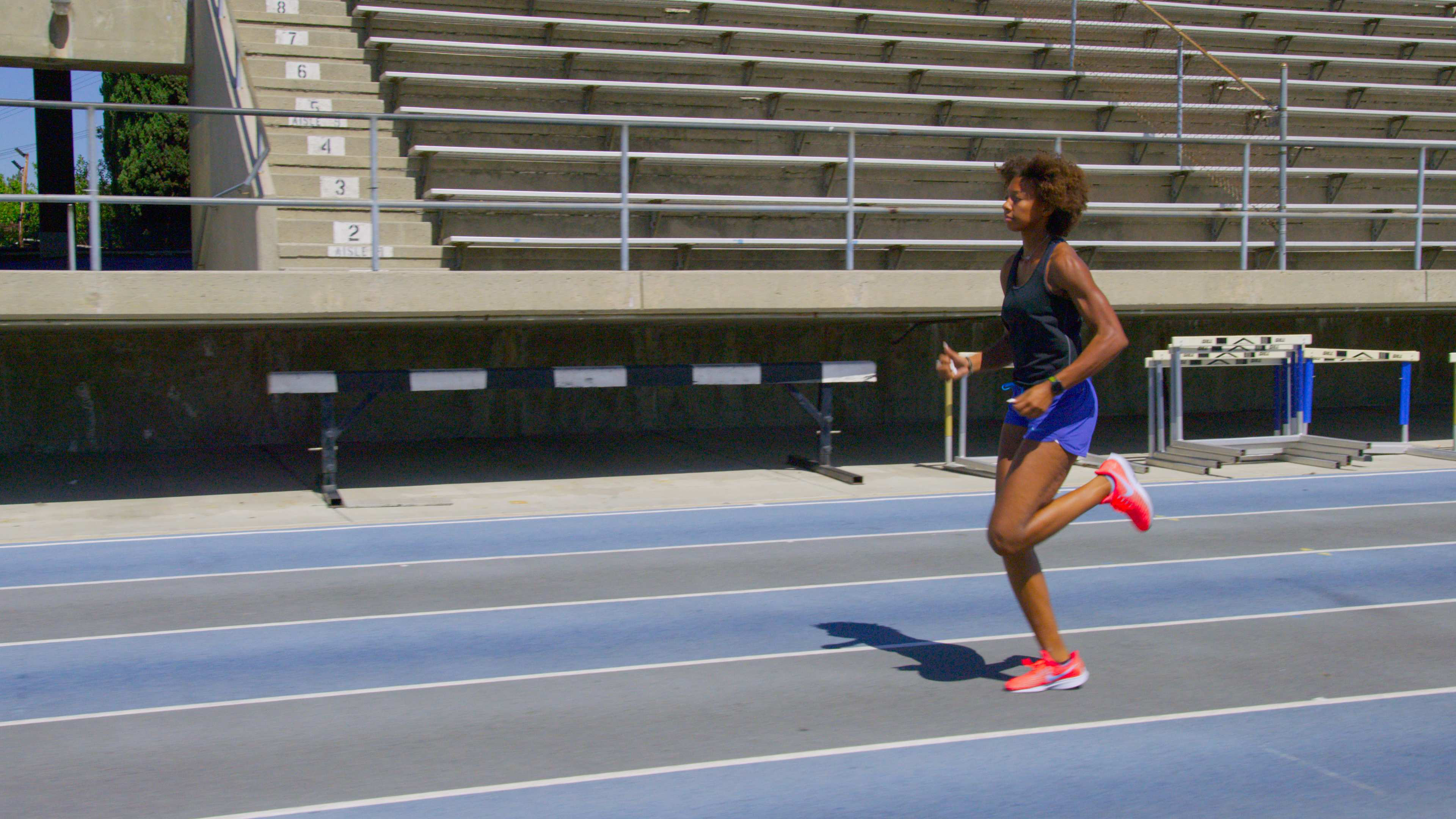}}\hfill
	\subfloat[3 Runners]{\includegraphics[width=0.24\textwidth]{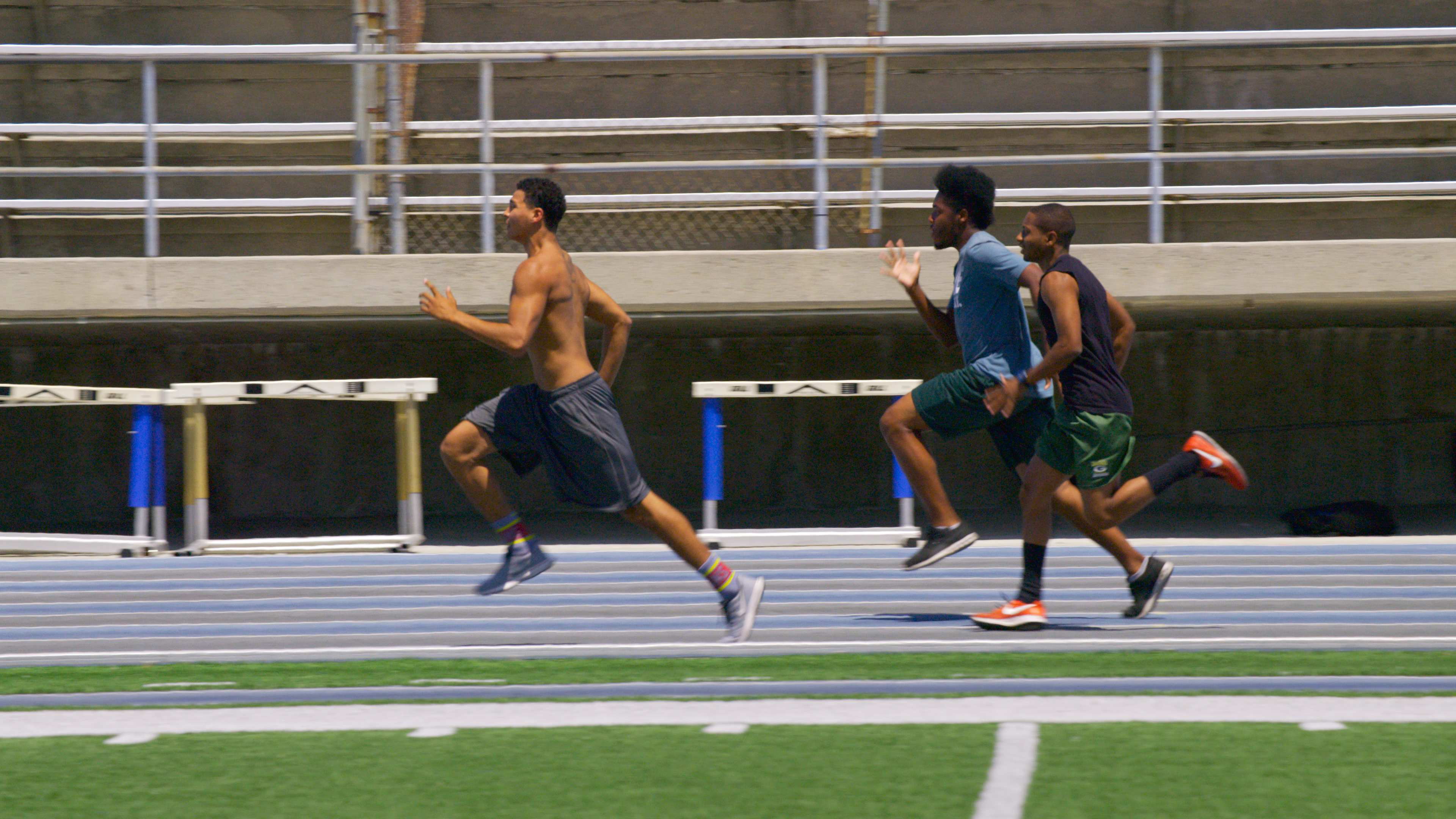}}\hfill
	\subfloat[Flips]{\includegraphics[width=0.24\textwidth]{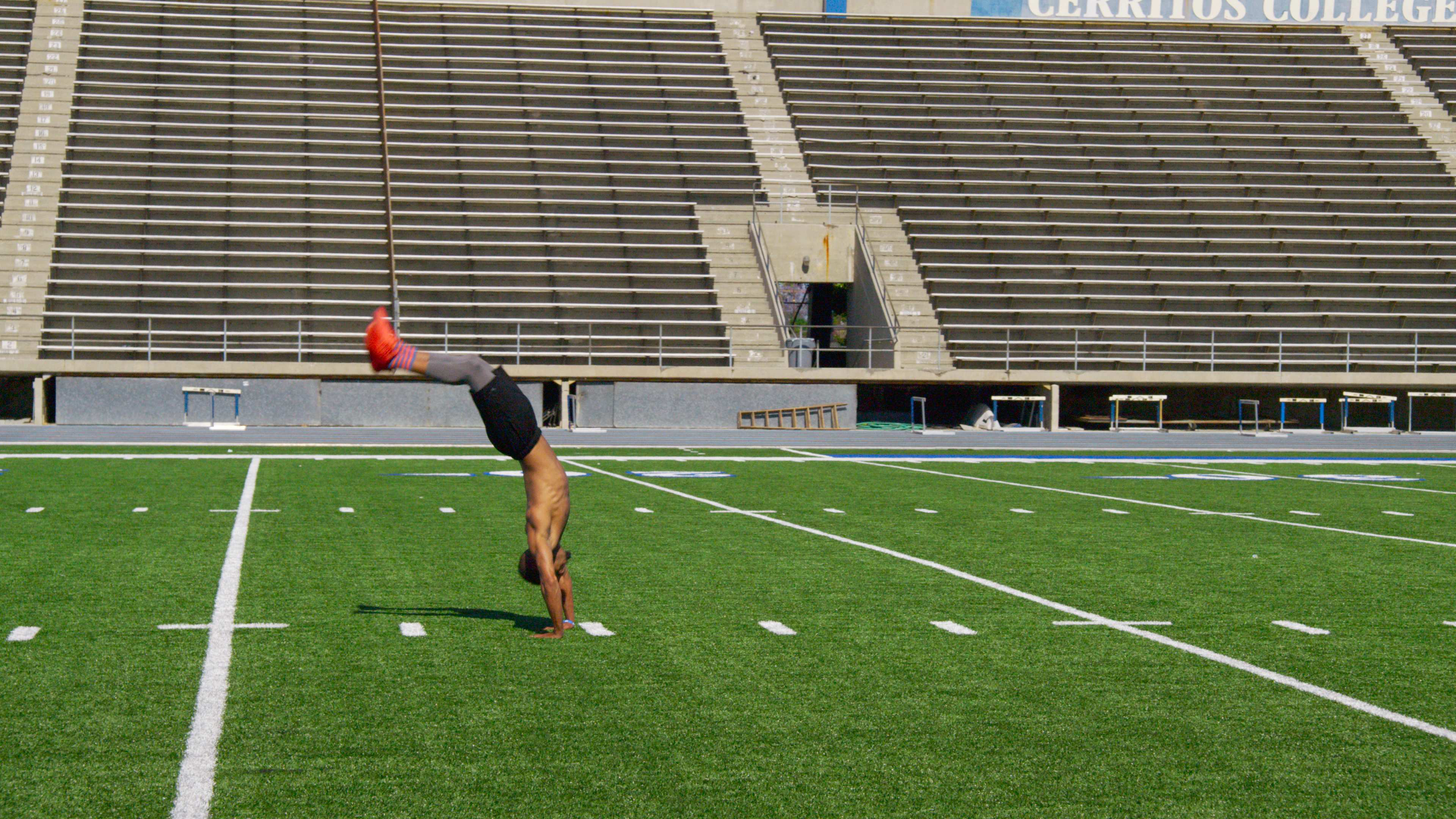}} \hfill
	\subfloat[Hurdles]{\includegraphics[width=0.24\textwidth]{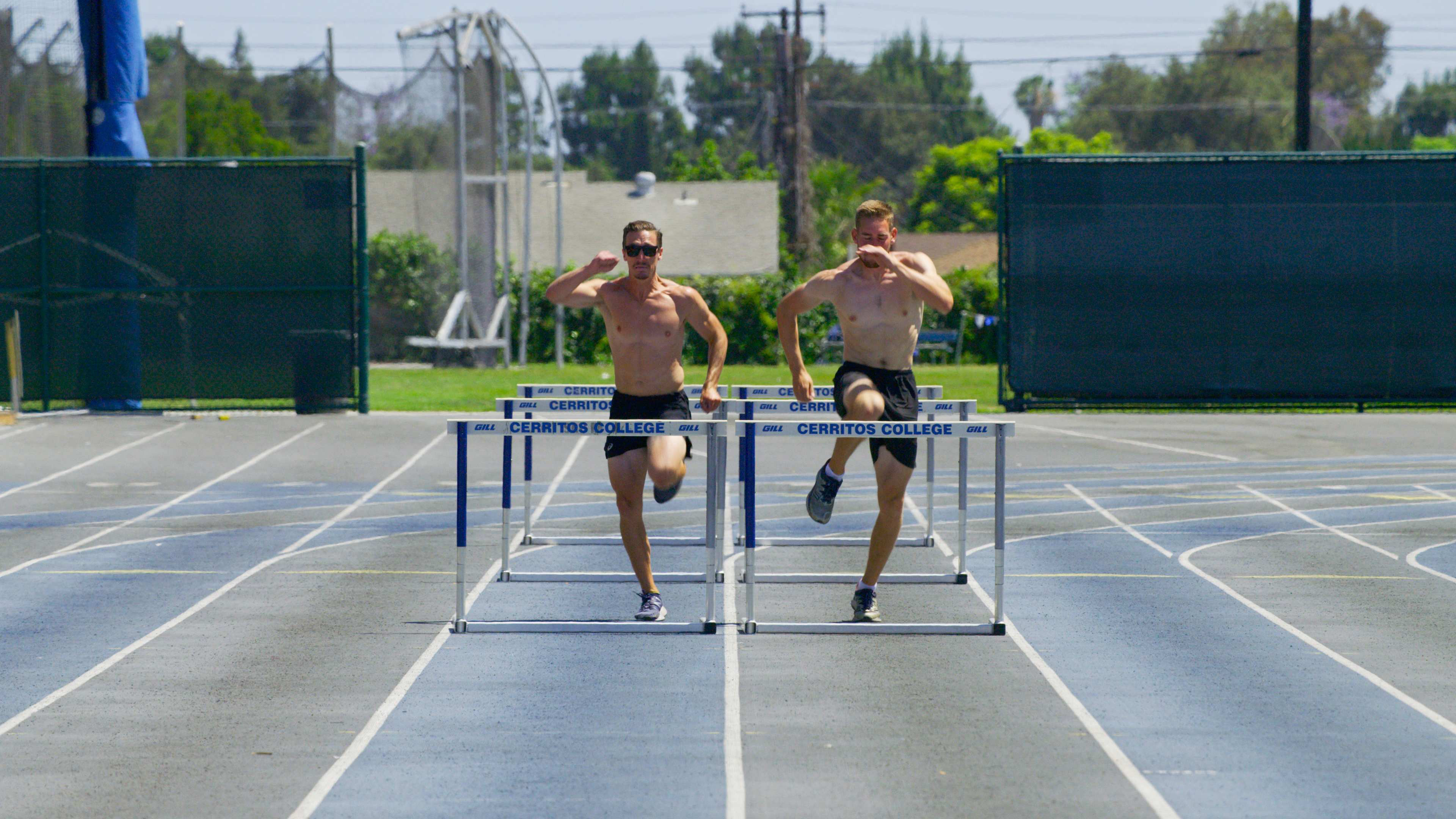}} \\
	
	\subfloat[Longjump]{\includegraphics[width=0.24\textwidth]{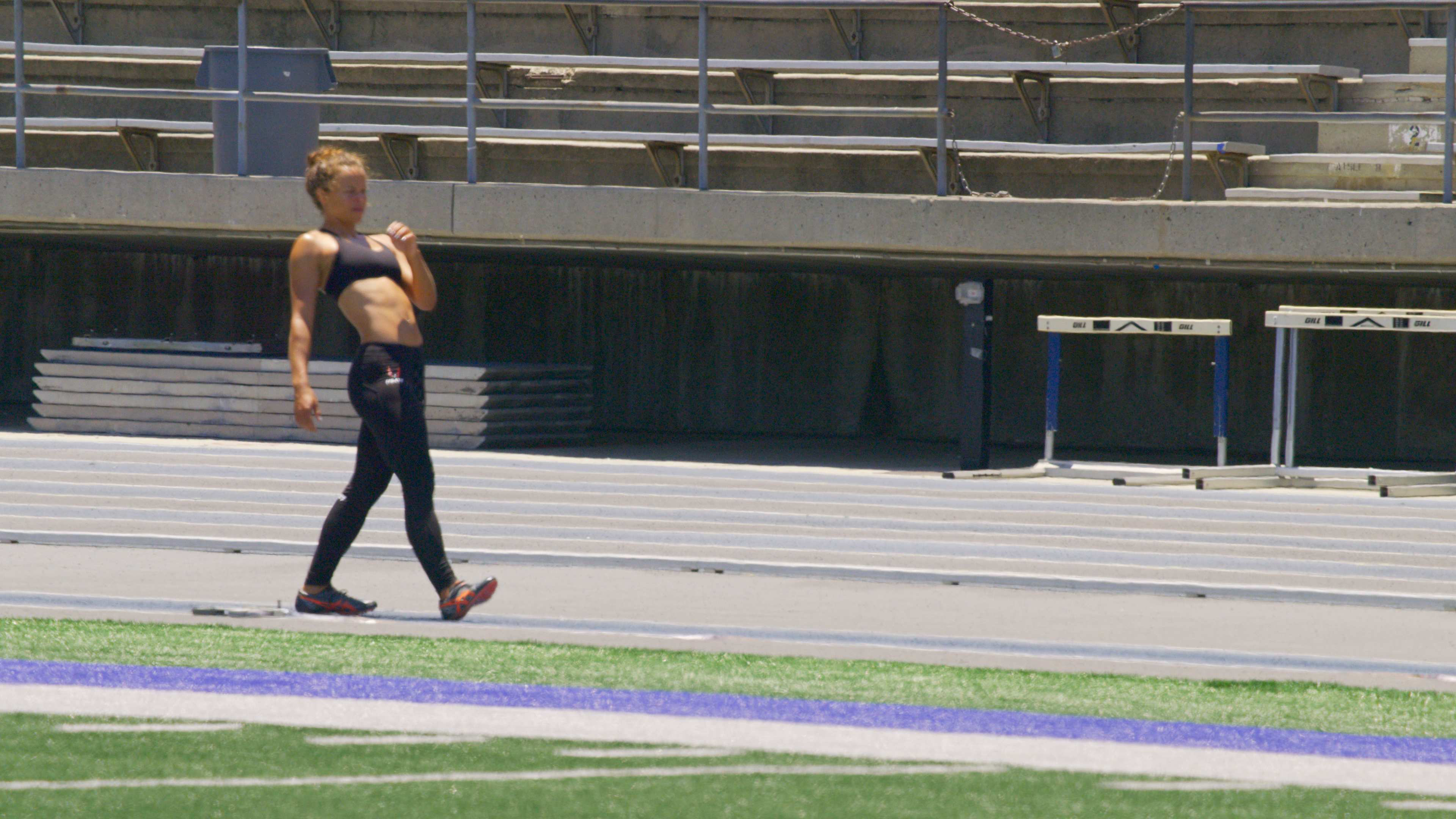}}\hfill
	\subfloat[bobblehead]{\includegraphics[width=0.24\textwidth]{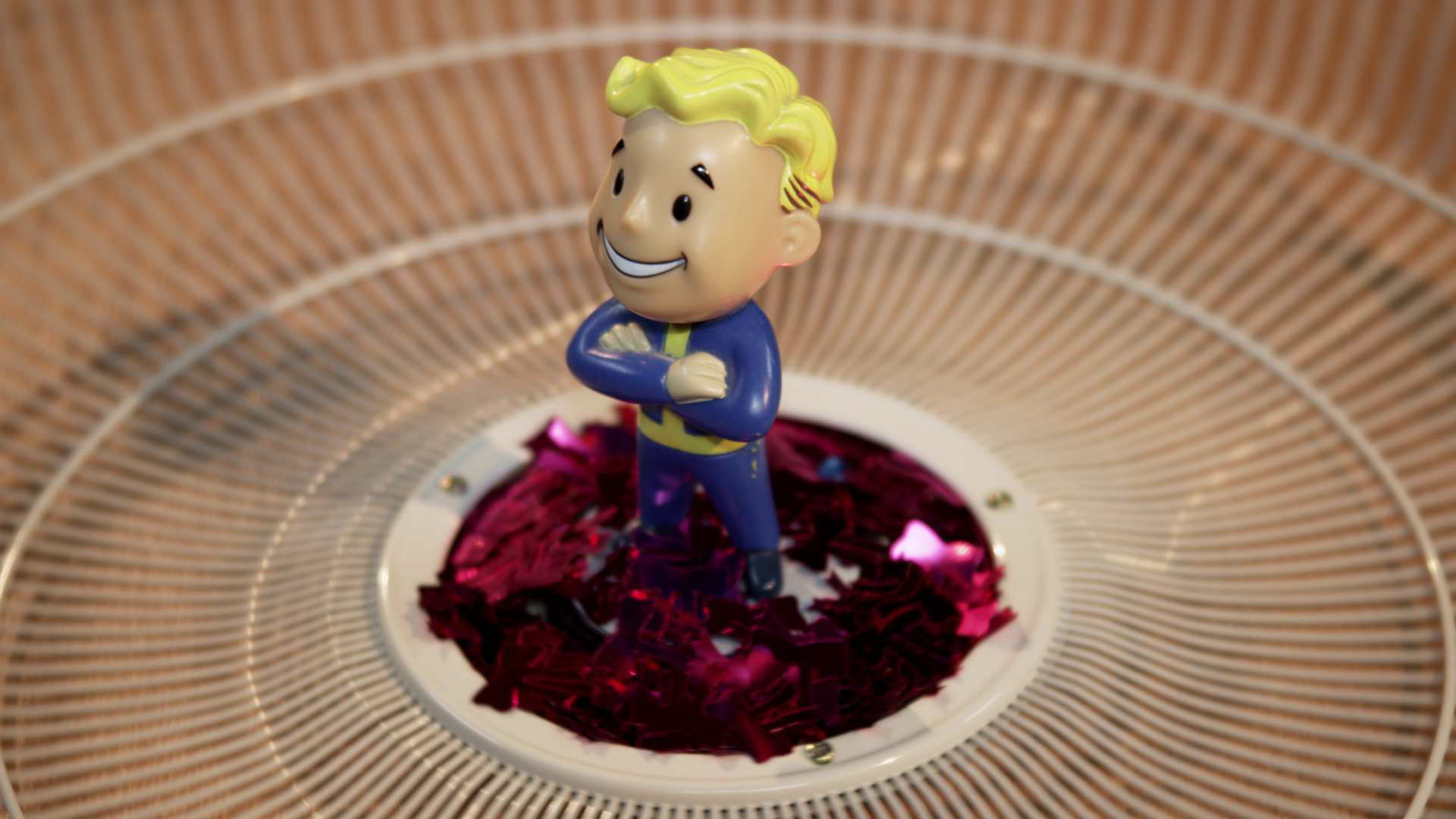}}\hfill
	\subfloat[books]{\includegraphics[width=0.24\textwidth]{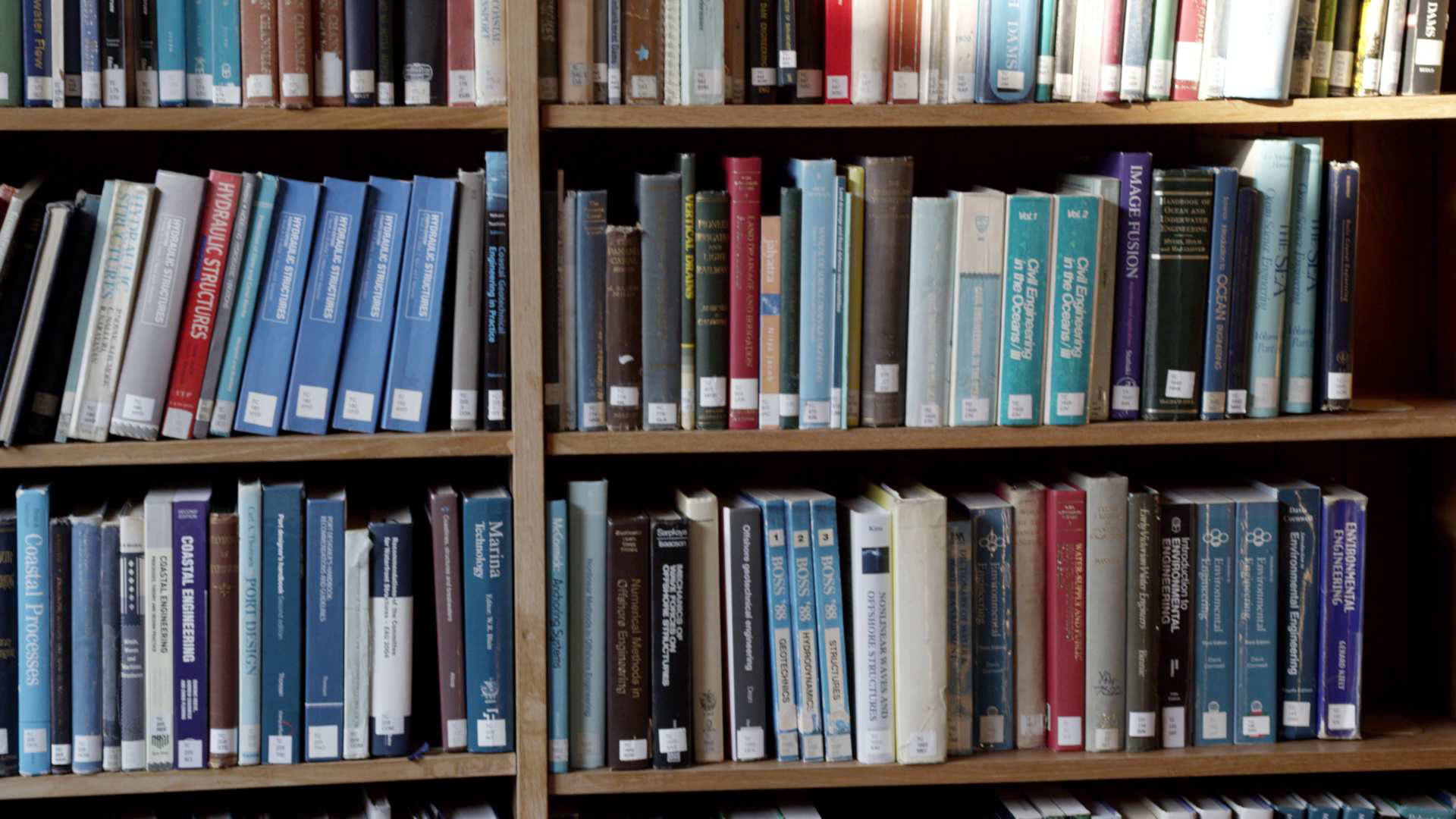}} \hfill
	\subfloat[bouncyball]{\includegraphics[width=0.24\textwidth]{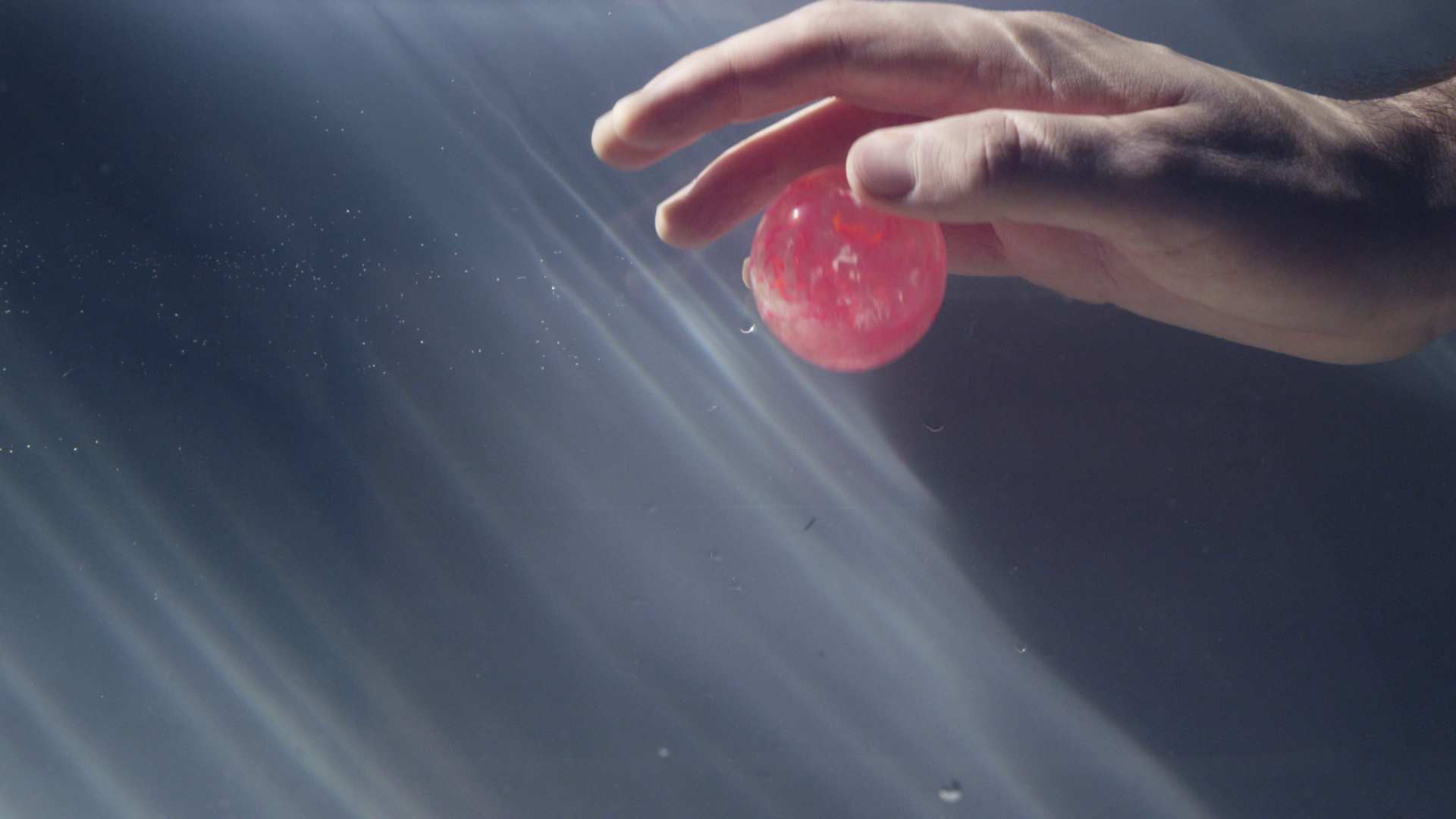}}\\
	
	\subfloat[catch-track]{\includegraphics[width=0.24\textwidth]{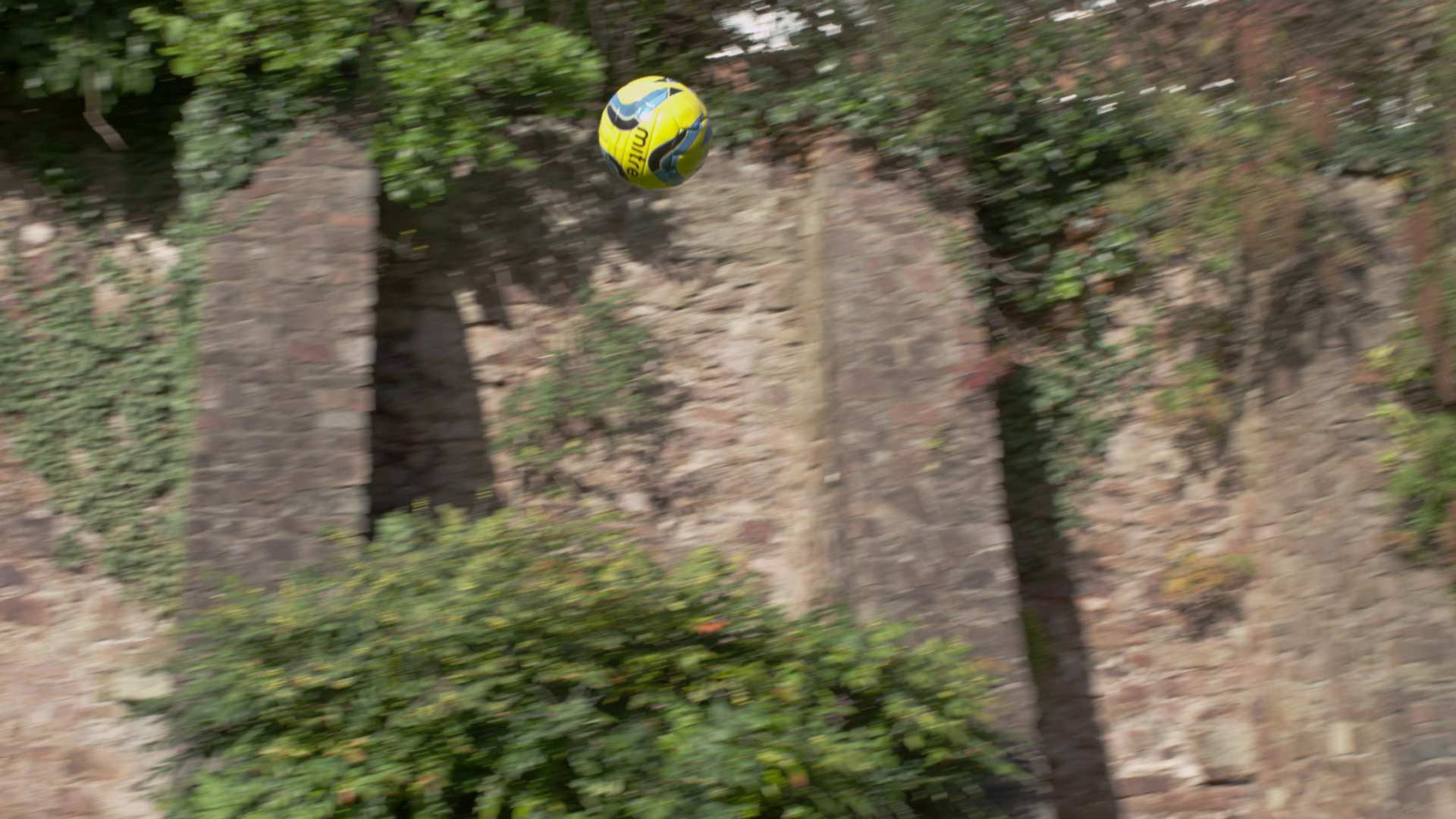}} \hfill
	\subfloat[cyclist]{\includegraphics[width=0.24\textwidth]{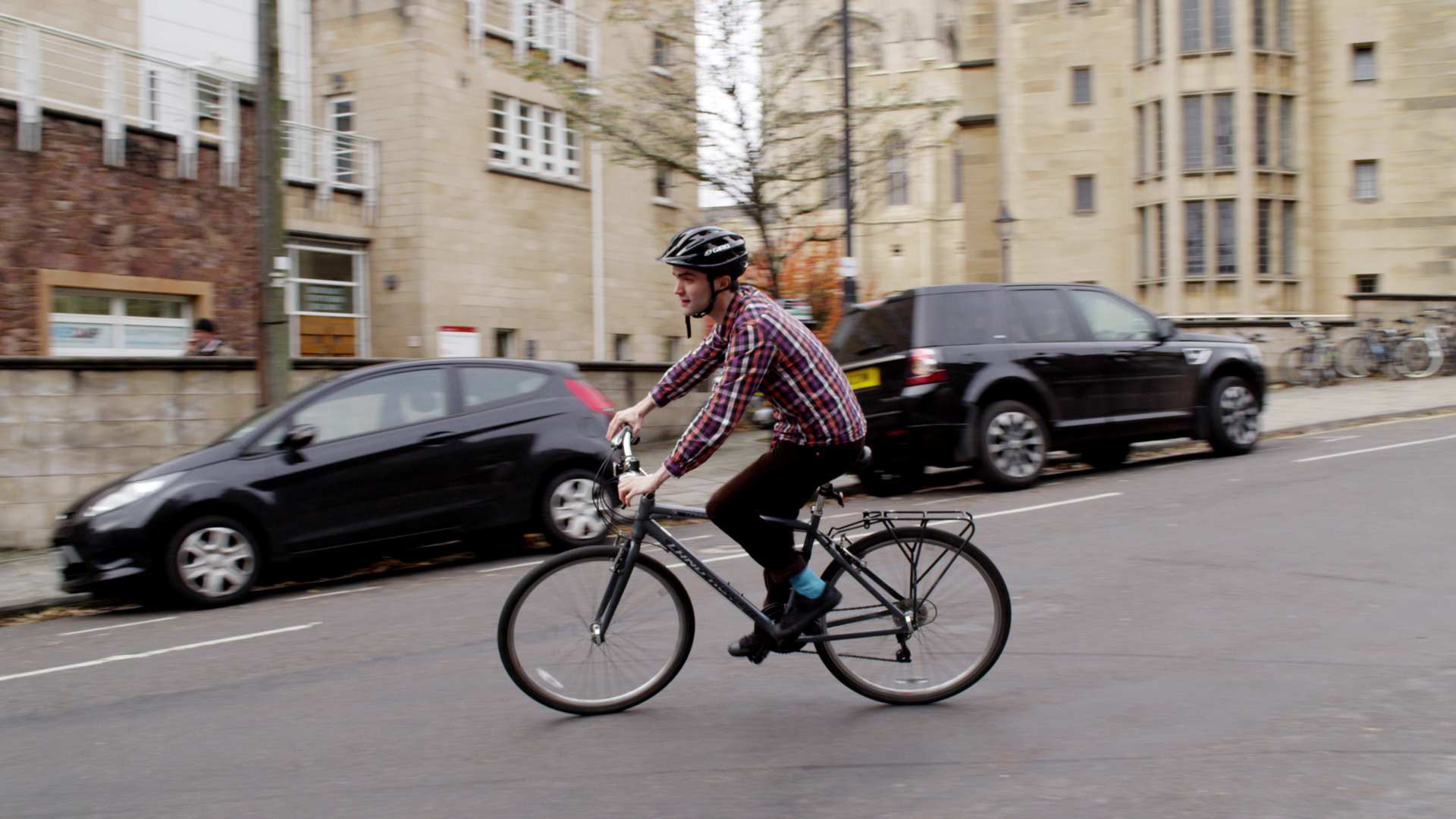}} \hfill
	\subfloat[hamster]{\includegraphics[width=0.24\textwidth]{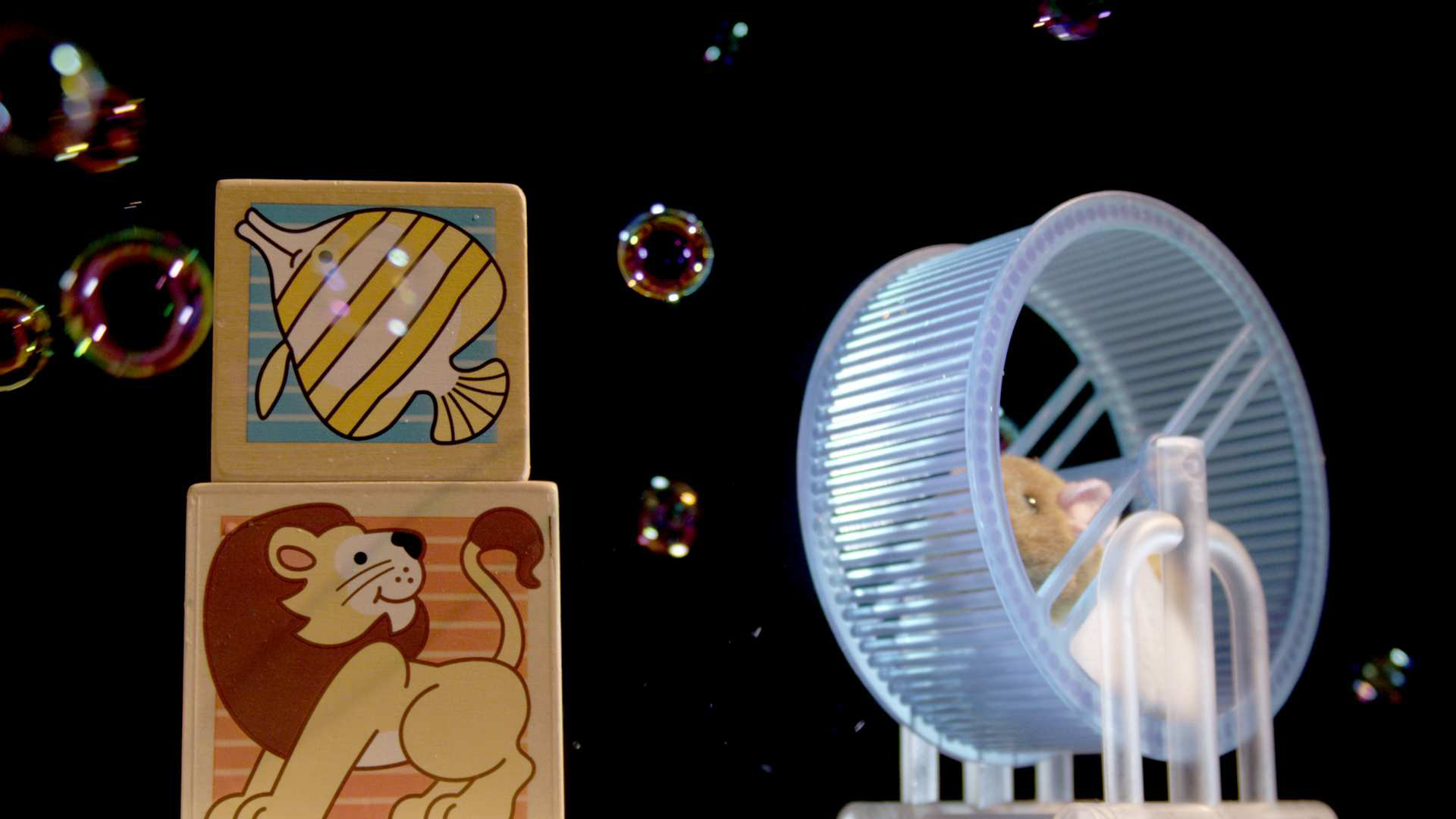}}\hfill
	\subfloat[lamppost]{\includegraphics[width=0.24\textwidth]{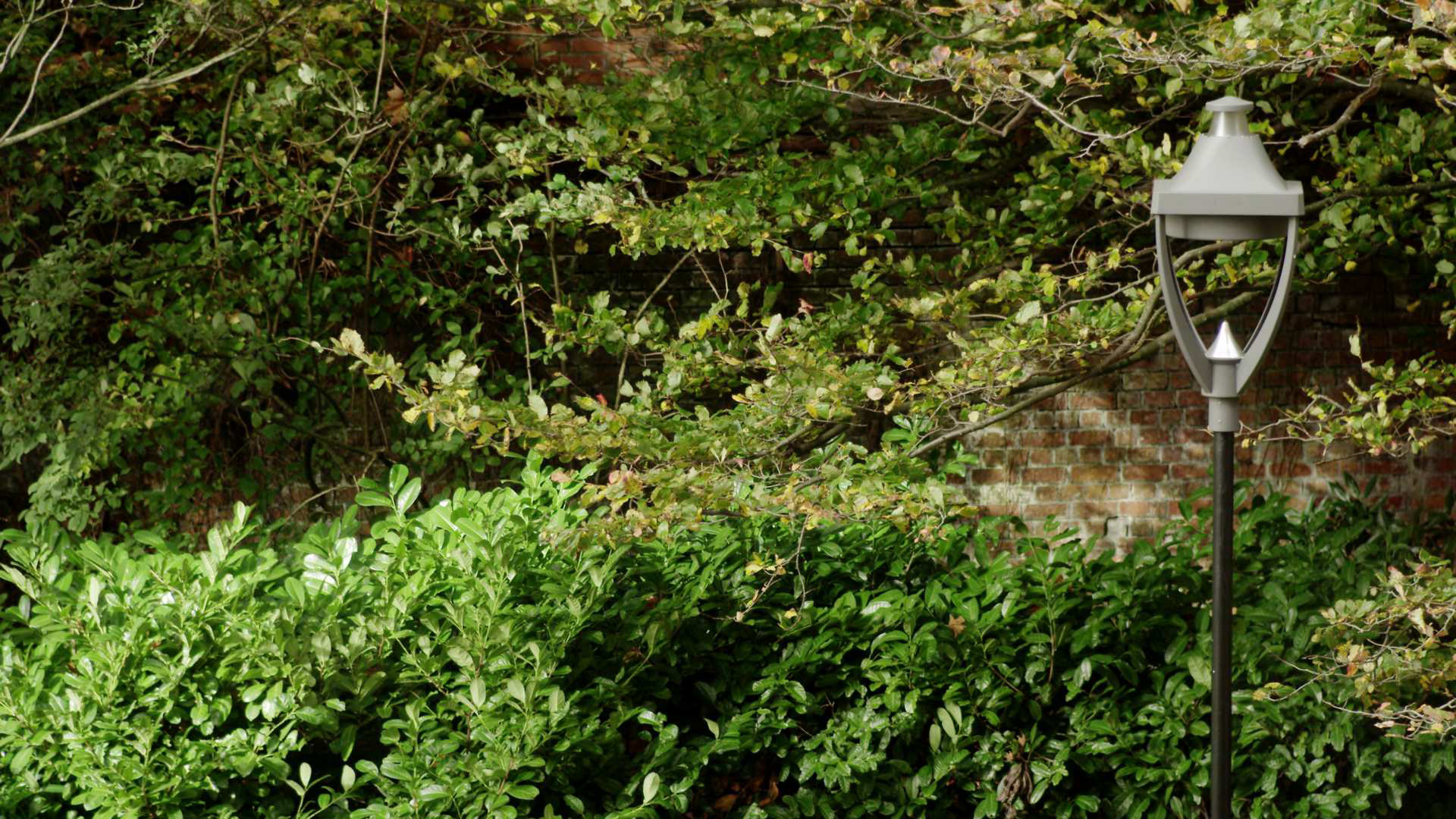}}\\
	
	\subfloat[leaves-wall]{\includegraphics[width=0.24\textwidth]{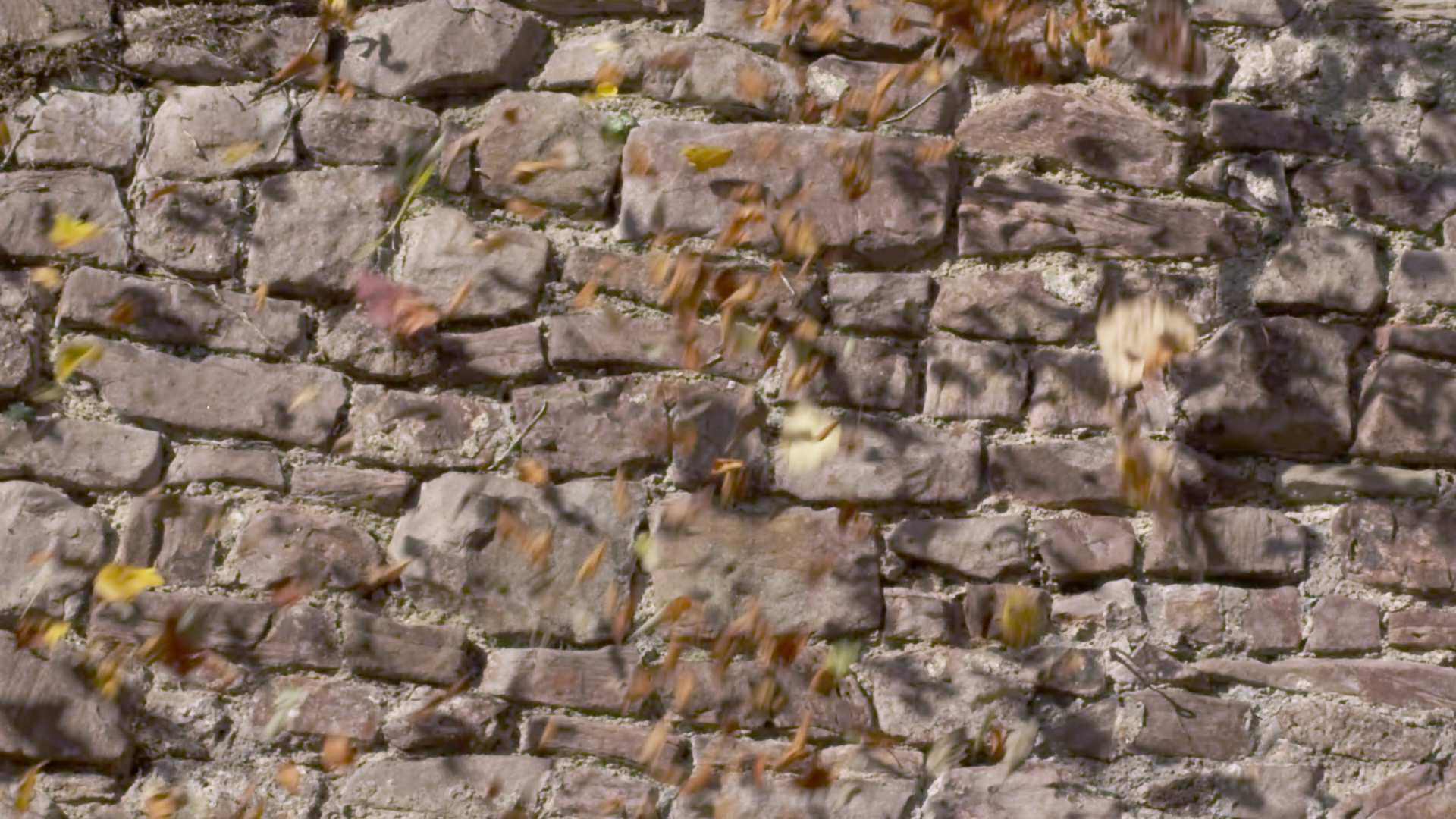}}\hfill
	\subfloat[library]{\includegraphics[width=0.24\textwidth]{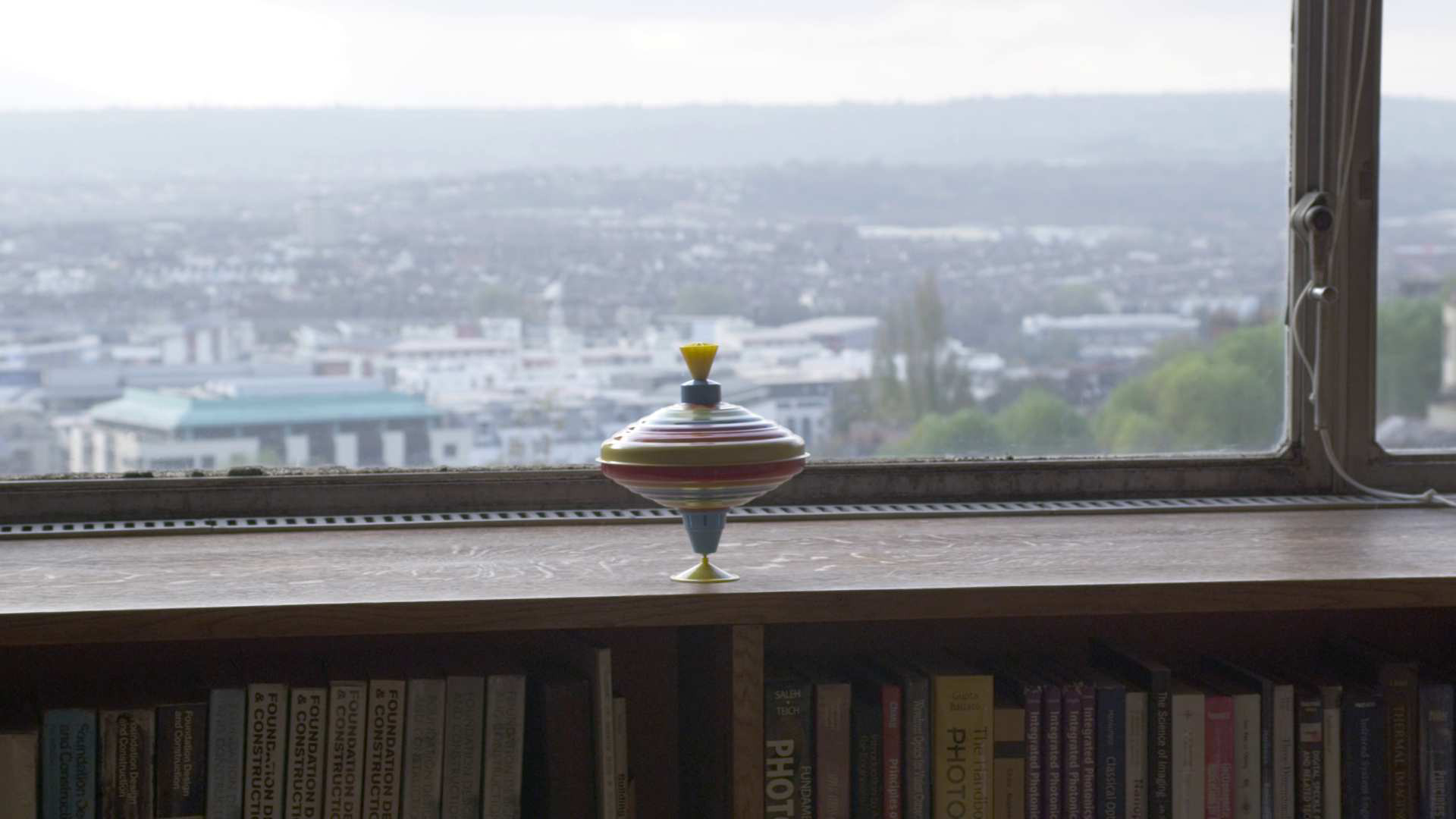}} \hfill
	\subfloat[pour]{\includegraphics[width=0.24\textwidth]{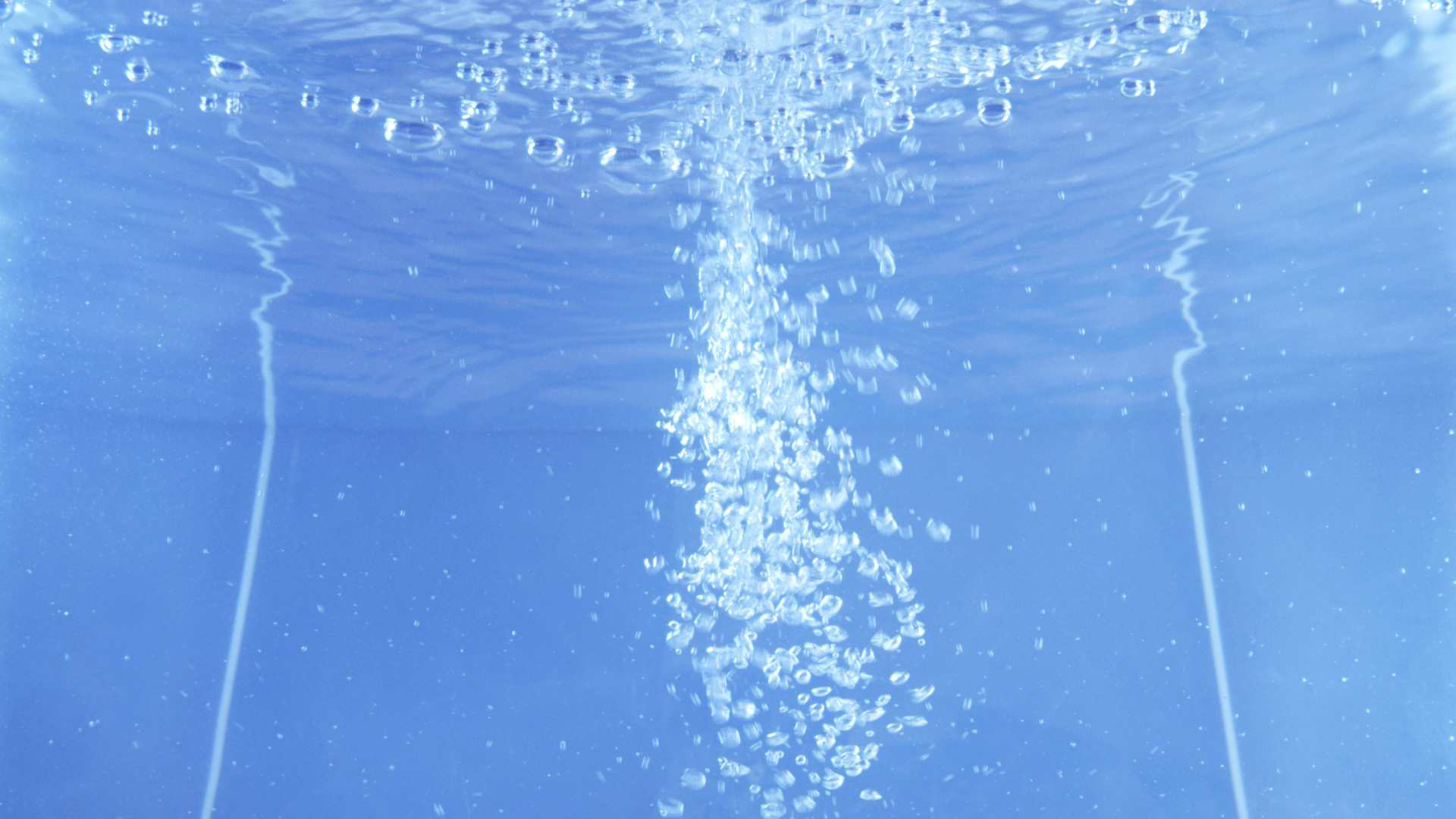}} \hfill
	\subfloat[water-splashing]{\includegraphics[width=0.24\textwidth]{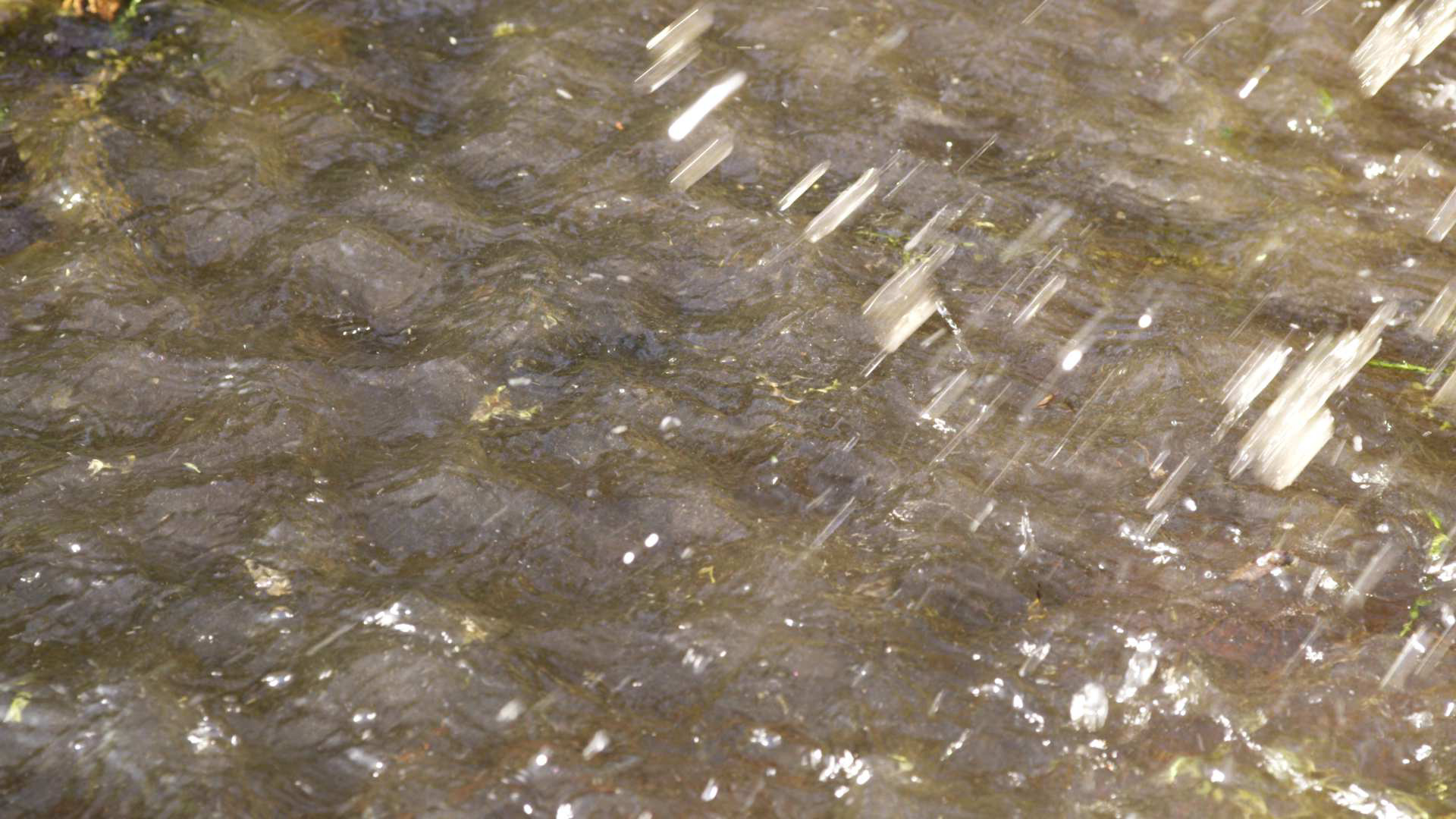}}\\
	
	\caption{Sample frames from source sequences in the LIVE-YT-HFR Database. (a) - (e): Sequences contributed by the Fox Media Group and (f) - (p): sequences from the BVI-HFR dataset.}
	\label{fig:screenshot}
\end{figure*}

Although FR Image Quality Assessment (IQA) models \cite{wang2004image,wang2003multiscale,zhang2011fsim} can be easily extended to VQA by applying them on a frame-by-frame basis, in combination with a suitable temporal pooling strategy, their performance is often limited, since temporal information is not effectively used. An early VQA model, Video Quality Metric (VQM) \cite{pinson2004new}, employs 3D spatio-temporal video blocks to compute certain features, and frame differencing to capture temporal variations. A modified SSIM algorithm \cite{seshadrinathan2007structural}, and the later MOVIE \cite{seshadrinathan2009motion} index both use a model of human visual motion processing in extra-cortical area MT to capture motion distortions. The ST-MAD \cite{vu2011spatiotemporal} index uses a ``most apparent distortion" concept \cite{larson2010most} to quantify quality. Natural Scene Statistics (NSS) based VQA models, such as ST-RRED \cite{soundararajan2012video} and SpEED-VQA \cite{bampis2017speed}, compute statistical measurements such as spatial and temporal entropic differences in the band-pass domain, to measure quality deviations. Recently, learning-based FR-VQA frameworks have gained popularity due to their superior performance. The Video Multi-method Fusion (VMAF) algorithm \cite{VMAF2016} is a highly successful and widely used method, which uses a set of features derived from VIF \cite{sheikh2006image}, a frame-difference feature, and a detail feature \cite{li2011image}, fusing them using a trained Support Vector Regressor (SVR). Kim \etal \cite{kim2018deep} proposed a model called DeepVQA, based on a CNN model in combination with a convolutional neural aggregation network (CNAN) for temporal pooling, achieving competitive performance on the LIVE-VQA and CSIQ-VQA datasets. 

VQA models relevant to HFR quality prediction are uncommon. Nasiri \etal \cite{nasiri2017perceptual} proposed an early model that measures the degree of aliasing of the temporal frequency spectrum. In \cite{nasiri2018temporal}, motion smoothness is used as a measure of cross-frame rate quality assessment. Zhang \etal \cite{zhang2017frame} proposed a wavelet domain Frame Rate Quality Metric (FRQM), whereby absolute differences between temporally wavelet filtered sequences are used to quantify quality. Although FRQM achieves competitive performance on the BVI-HFR dataset, it cannot be used when both the reference and distorted videos have the same frame rate, thus limiting its generalizability. 

The VQA models just discussed only address artifacts arising from frame rate variations, without accounting for the joint perceptual impacts of compression and frame rate. Recently a model called GSTI \cite{pavan2020gsti} was proposed, where entropic differences between temporally band-pass filtered responses were found to achieve better correlations against human judgments of quality, even when tested in the presence of both compression and frame rate.

The absence of reference information makes NR video quality prediction quite challenging. Most existing models involve some kind of learning based procedure to find mappings between features (or pixels) and human subjective judgments of quality. Good examples are \cite{mittal2012no,saad2014blind,li2016spatiotemporal,korhonen2019two}, which use NSS or other quality-aware features on which an SVR or Random Forest learner is trained to predict quality. Recent interest in assessing UGC video quality has resulted in several successful methods based on deep learning \cite{zhang2018blind,ahn2018deep,li2019quality}. Although UGC videos contain a wide variety of interesting authentic distortions, they are less topical for understanding the effects of frame rate, since usually, only very high quality source videos are subjected to frame-rate reductions (during streaming), hence UGC datasets contain only one version of each video content.  Nevertheless, frame-rate variations of UGC content may become a more important topic in the future, creating interesting research possibilities.



\begin{figure}[!t]
	\centering
	\subfloat[]{\includegraphics[width=0.5\linewidth]{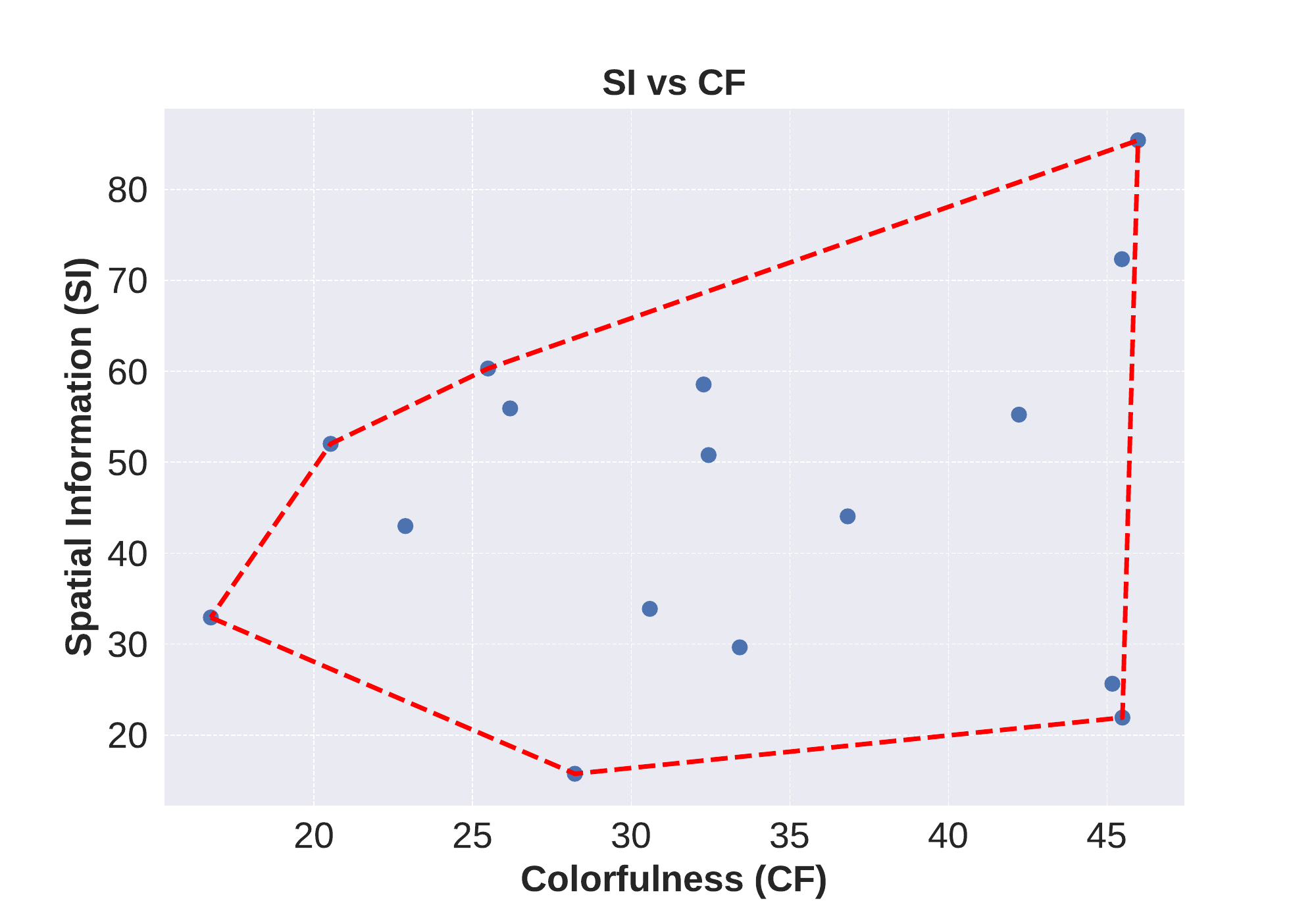}}\hfill
	\subfloat[]{\includegraphics[width=0.5\linewidth]{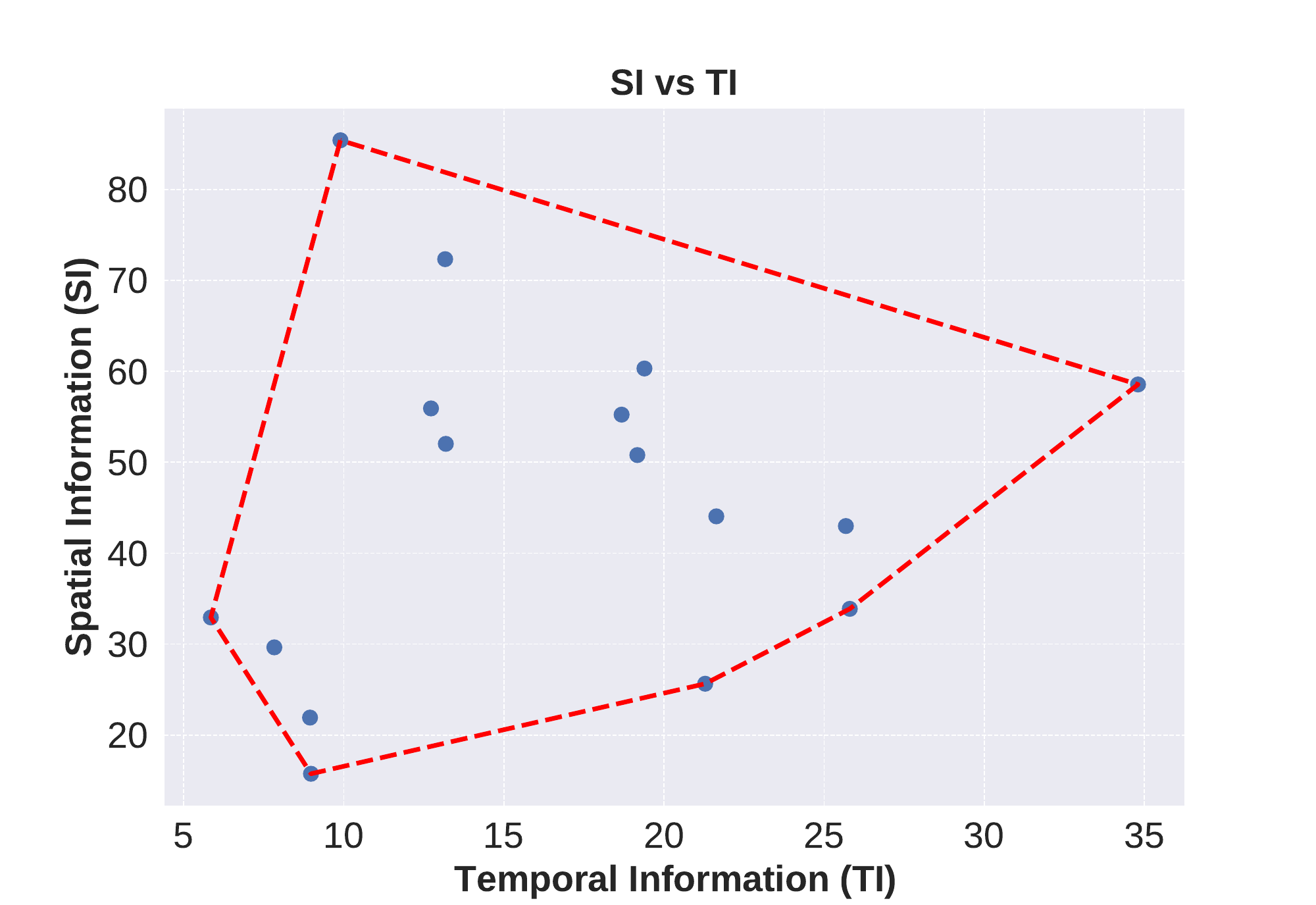}}
	\caption{(a) Spatial Information (SI) versus colorfulness (CF) and (b) Temporal Information (TI), measured on the source sequences in the LIVE-YT-HFR database respectively. The corresponding convex hulls are indicated by red lines.}
	\label{fig:SI_CF_TI}
\end{figure}

\section{LIVE-YouTube-HFR Database}
\label{sec:LIVE_HFR}
A detailed description of the new LIVE-YT-HFR database is presented in this section. Our main objective in creating this database is to provide a tool for the video quality research community to have access to when analyzing the impact of frame rates on perceptual video quality. We believe that studying the perception of artifacts arising from frame rate variations will prove to be beneficial when designing future VQA models.

\begin{table}[!t]
	\caption{Characterization of Source Sequences (120 Hz)}
	\label{table:coverage_table}
	\centering
	\footnotesize
	\begin{tabular}{|c|c|c|c|} 
		\hline
		& SI & CF & TI \\
		\hline \hline
		Range & 69.65 & 29.22 & 31.07 \\ \hline
		Uniformity of Coverage & 0.87 & 0.94 & 0.8 \\ \hline 
	\end{tabular}
\end{table}

\subsection{Source Sequences}
\label{sec:source_sequences}
We used 16 uncompressed source videos of natural scenes captured at a frame rate of 120 fps that are currently available in the public domain. Of these 16 videos, 11 sequences were borrowed from the Bristol Vision Institute High Frame Rate (BVI-HFR)
video database \cite{mackin2015study}. These were captured using a RED Epic-X video camera with a spatial resolution of $3840 \times 2160$ (UHD-1) at a frame rate of 120 fps. The publicly available version of the database contains sequences that were spatially downsampled to $1920 \times 1080$ (HD) YUV 4:2:0 8 bit format, of each 10 seconds duration. The remaining 5 videos contain high-motion sports content captured by the Fox Media Group in $3840 \times 2160$ (UHD-1) YUV 4:2:0 10 bit format, each of 6-8 seconds duration. Sample frames drawn from the source sequences, along with their IDs are shown in Fig. \ref{fig:screenshot}. This database was restricted to contain only progressively scanned videos, to avoid separate issues associated with video de-interlacing artifacts. 

\subsection{Content Description and Coverage}
Similar to \cite{mackin2018study}, we computed three low level descriptors on each source sequence: (i) Spatial Information (SI), indicating the amount of local spatial variation in each frame, (ii) Temporal Information (TI), which captures change across frames, and the (iii) Colorfulness (CF) measure \cite{hasler2003measuring}. SI is a Sobel magnitude measure, whereas TI uses the average squared luminance difference between successive frames:
\begin{align}
	TI = \sqrt{\frac{1}{N - 1} \sum_{t = 1}^{N - 1}\sum_{i,j}^{P} \frac{(I(i,j,t+1) - I(i,j,t))^2}{P}}
\end{align} 
where $I(i,j,t)$ is luminance at co-ordinate $i,j$ in frame $t$, $P$ is the total number of pixels in each frame, and $N$ is the number of frames in the video. Table \ref{table:coverage_table} shows the range and uniformity characteristics of the source sequences, while the raw SI, CF and TI values are plotted in Fig. \ref{fig:SI_CF_TI}. These plots illustrate a diverse span of scenes and motions among the selected source sequences. 

\subsection{Temporal Downsampling}
Simultaneously capturing a same scene across multiple frame rates without downsampling is impractical, as it would either require a specialized camera with concurrent multi-frame rate capture capability, or a careful configuration of a multi-camera system. Thus, lower frame rate versions were generated by employing temporal downsampling of original high frame rate (120fps) source videos. In prior studies, two methods of downsampling have been used: frame dropping and frame averaging \cite{mackin2018study}. Dropping frames is similar to native capture at a lower frame rate with a reduced shutter angle \cite{watson2013high}. However, while frame dropping is simple and computationally inexpensive, it can introduce judder/strobing artifacts, especially on videos captured with significant camera motion. Conversely, frame averaging alleviates the problem of judder/strobing distortions, but can introduce motion blur, resulting in the attenuation of visually important high spatio-temporal frequencies. The degree of high-frequency attenuation increases with increasing downsampling factor, making videos subsampled to low frame rates, such as 24, 30 fps, strikingly blurred. Of course, motion compensation methods of frame averaging might be considered, but these can create other kinds of artifacts, and are not commonly used \cite{choi2007motion}. Since the choice of temporal downsampling method influences the perception of video quality, we decided to use the frame dropping method, in order to avoid the introduction of motion blur, and to obtain low frame-rate videos closer to natively captured ones. Frame dropping was performed by suitably modifying the \textit{fps} filter available in FFmpeg \cite{ffmpeg}.

\begin{figure}[t]
	\centering
	\includegraphics[width=0.5\textwidth]{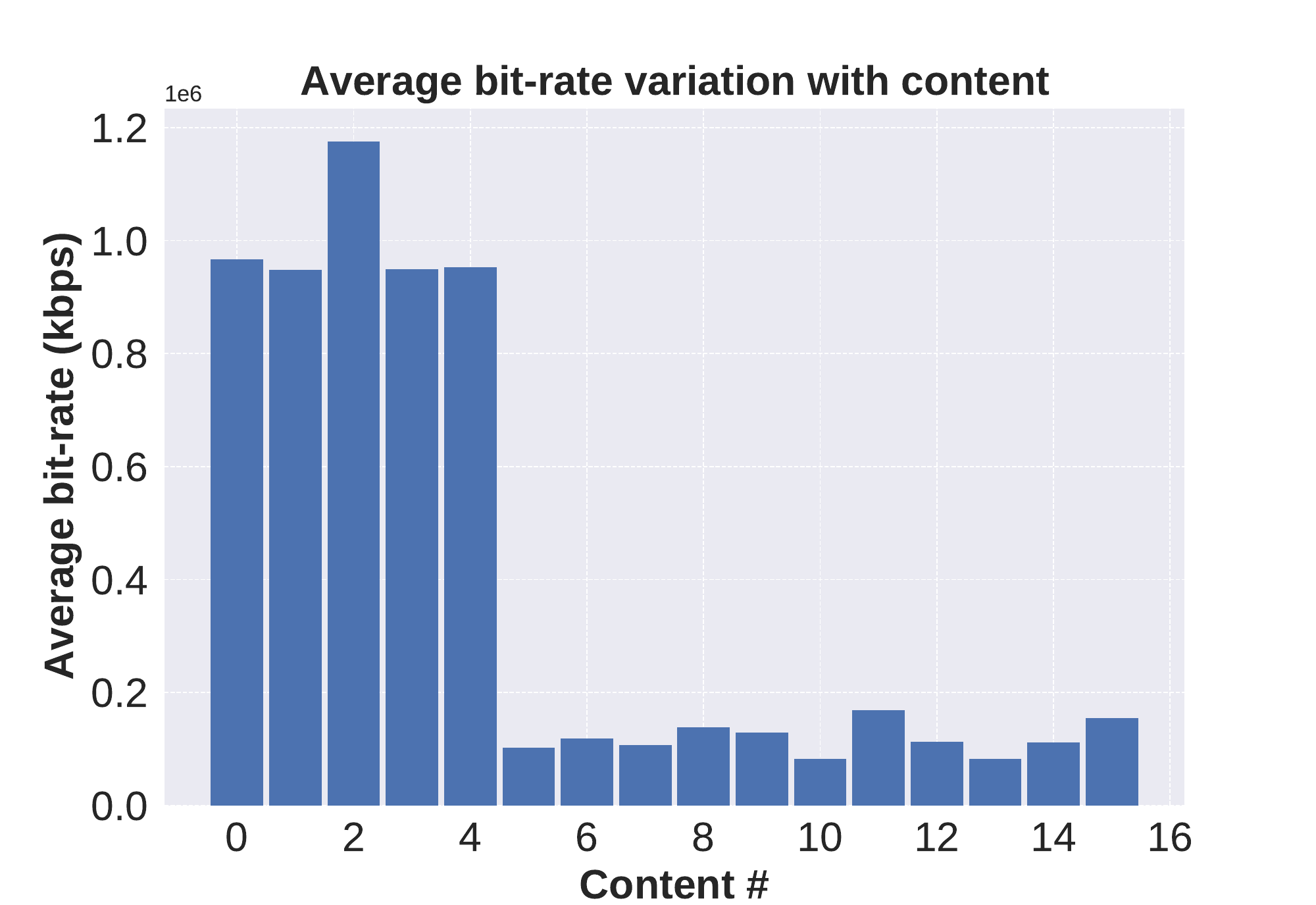}
	\caption{Variation of average bit-rate with content in the LIVE-YT-HFR Database. The horizontal axis indexes video contents ordered in the same manner as shown in Fig. \ref{fig:screenshot}}
	\label{fig:bit_rate_bar}
\end{figure}

\subsection{Test Sequences}
\label{sec:test_sequences}
We created 30 test sequences from each source sequence, by subsampling them to 6 different frame rates: 24, 30, 60, 82, 98 fps and 120 fps. Each of these were subsequently subjected to 5 levels of VP9 compression. Our goal in this work is to create a video quality database containing varying frame rates that can be used to enable the development of general purpose space-time video quality models applicable to any frame rate up to 120 fps. The frame rate values 82 and 98, although currently not as popular as the others, were included to obtain more fine-grained video quality ratings, particularly in the range of 60 fps and 120 fps, which allows for better model-building. These frame rates were chosen because they were natively supported by the display device (Acer Predator X27 \cite{Predator}) that was employed in the human study. This removed the need of any device-specific interpolation that could introduce distortions. In order to ensure that non-uniform sampling would not introduce additional artifacts or alter perceived visual quality, we performed a pilot study involving a small set of expert subjects to ensure that non-uniform subsampling did not alter perceived quality. We found that reported visual quality increased with frame rates, including 82 fps and 98 fps, over the entire range. All of the sequences were compressed using FFmpeg VP9 compression \cite{VP9} with single pass encoding by varying the Constant Rate Factor (CRF) values, resulting in bit-rates $R_i, \text{\hspace{2pt}} i \in \{1,\ldots,5\}$, where $R_i < R_j,\text{\hspace{2pt}} \forall i < j$. The strategy used to choose the 5 compression levels for a given source sequence was done as follows: two of the levels $R_1,R_5$ correspond, respectively, to the lossless and highest (CRF=63) possible compression levels in VP9. The other three bit-rates $R_2,R_3$, and $R_4$ were chosen such that compression resulted in approximately the same bit-rates across all frame rates (for a given bit-rate, the CRF values vary across frame rate, with higher frame rates having larger CRF values than corresponding lower frame rate sequences). Thus, for a given source sequence, bit-rates $R_2,R_3$ and $R_4$ remained constant and were selected to ensure that there was adequate perceptual separation between them. The CRF values of the remaining videos derived from the source sequence were determined to approximately match these bit-rates. Thus, for each source content, there are 6 (Frame rates) $\times$ 5 (lossless + 4 CRF values) = 30 test sequences. The above procedure was repeated on every source sequence present in the database. Although extreme compression levels (lossless and CRF=63) are rarely employed in practical applications, the objective was to capture and model human behavioral responses inclusive of, but not limited to, realistic ranges of distortion. These more general data can result in robust models that are applicable across wide operating conditions. Since bit-rates depend on content, there is significant variation of bit-rates across the compressed source sequences. This is illustrated in Fig. \ref{fig:bit_rate_bar}, where average bit-rates are plotted against content indices, and where the initial contents were 4K videos having higher bit-rate values. Given the 16 source videos described in Sec. \ref{sec:source_sequences}, we arrived at 16$\times$30 = 480 videos in the database.

\begin{table}[!t]
	\caption{Display parameters and viewing conditions of subjective study}
	\label{table:Monitor_specs}
	\centering
	\footnotesize
	\begin{tabular}{c|c} 
		\hline
		Parameter & Value\\ 
		\hline \hline
		Screen Resolution  & 3840 $\times$ 2160\\ 
		Screen Size & 27 inch\\
		Screen Width & 23.4 inch\\
		Screen Height & 13.2 inch\\ 
		Aspect ratio & 16:9\\
		Viewing distance & 30 inch\\
		\hline
	\end{tabular}
\end{table}

\begin{figure}[!t]
	\centering
	\includegraphics[width=0.8\linewidth]{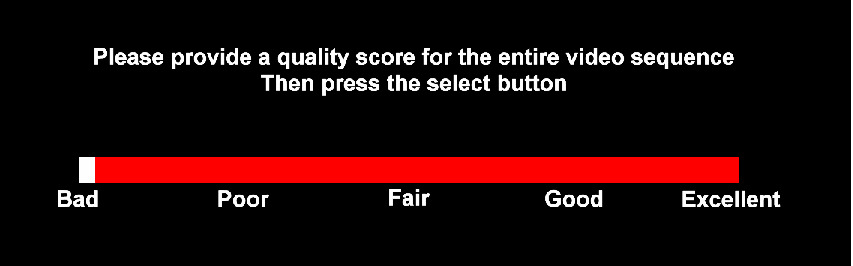}
	\caption{Screenshot of the scoring slider used in the human study, prompting the user to enter a quality score.}
	\label{fig:vote}
\end{figure}

\subsection{Significance of LIVE-YT-HFR Database}
The LIVE-YT-HFR database possesses some important and unique characteristics that distinguishes it from both existing HFR and standard VQA databases. First, it contains sequences corresponding to six different frame rates, spanning the range 24 fps to 120 fps. Prior HFR datasets have either limited the content to be less than 60 fps \cite{nasiri2015perceptual}, or have contained only a few frame rates \cite{mackin2018study}. Standard VQA databases generally restrict all of the reference and distorted videos to the same frame rate. We believe that having a more fine-grained sampling of frame rates will make it possible to create better models of the impact of frame rate on perceptual video quality. Second, the database contains a mixture of contents at spatial resolutions 1080p and 4K. The inclusion of 4K contents increases the relevancy of the database, given strong trends in video streaming towards 4K standards. Lastly, the LIVE-YT-HFR Database includes VP9 compression artifacts, enabling the study of the joint effects of compression and frame rate on video quality. Note that, although the database contains a mix of 1080p and 4K sequences, the quality score on each video was obtained at a single spatial resolution. In other words, we did not attempt to capture video quality variations over multiple resolutions. VP9 is a widely used alternative to MPEG compression, and it is heavily used by YouTube. The principles that can be learned will likely be applicable to other codecs as well, such as HEVC and AV-1. Overall, the new database comprises 480 videos, making it one of the largest VQA databases currently available. 

\begin{figure}[t]
	\centering
	\includegraphics[width=1\linewidth]{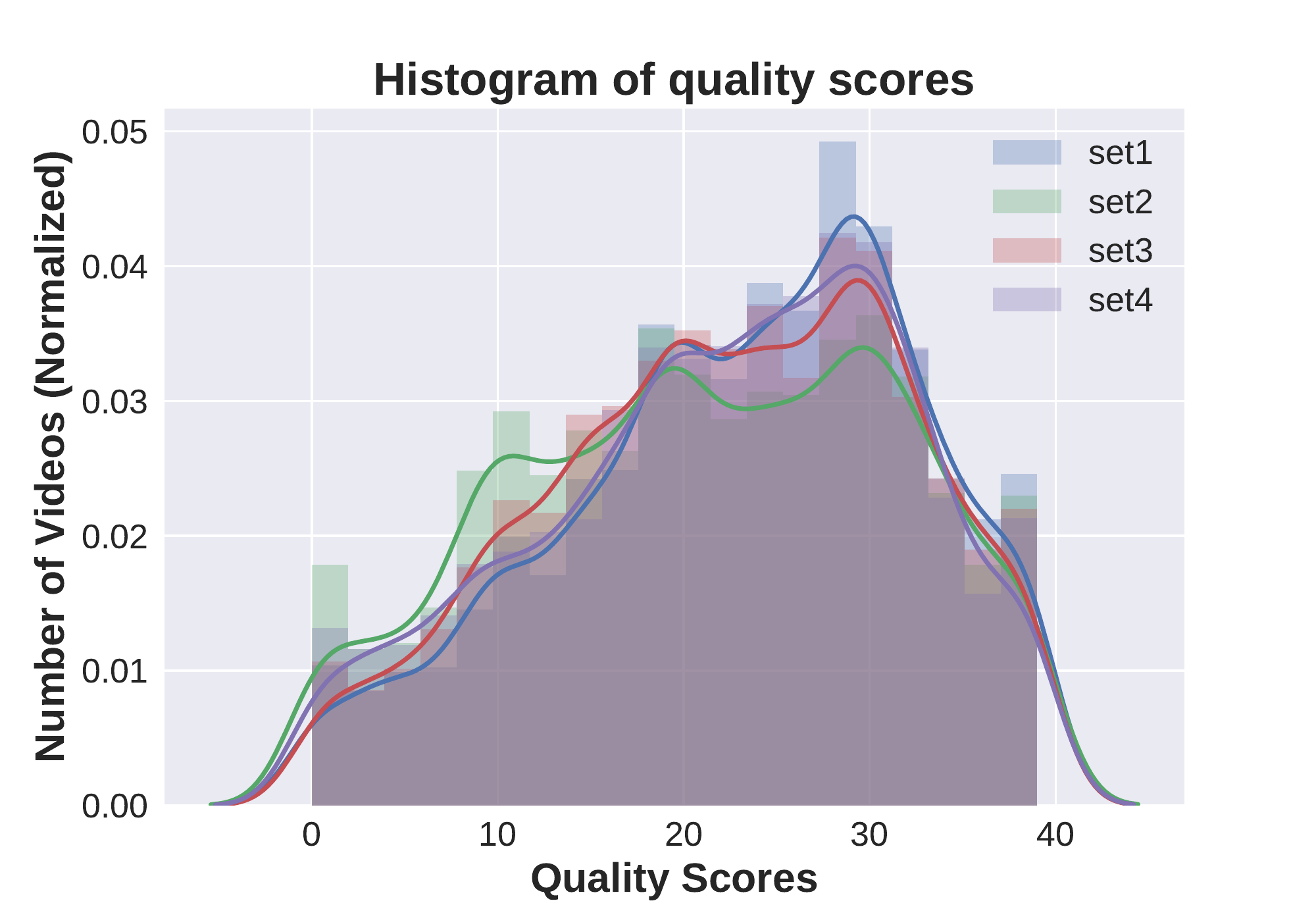}
	\caption{Histogram of raw scores across all four subsets of the LIVE-YT-HFR Database}
	\label{fig:histogram_sets}
\end{figure}

\section{Subjective Quality Assessment}
\label{sec:subjective_quality}
\subsection{Subjective Testing Design}
We employed a Single-Stimulus Continuous Quality Evaluation (SSCQE) \cite{recommendation2002500} procedure to obtain subjective quality ratings on the videos in the LIVE-YT-HFR database. By ``continuous," we refer to a continuous quality scale, as opposed to continuous quality collection over the duration of each video. The display parameters and viewing conditions employed in the subjective study are shown in Table \ref{table:Monitor_specs}. Since the screen resolution of the display device is 4K, the 1080p sequences were spatially upsampled to 4K using Lanczos interpolation, while the 4K videos were shown at their native resolution. This is how 1080p videos are commonly reformatted for display in practice. Lanczos interpolation was used as it is the most popular method, and this gave us control over the upsampling scheme, thereby avoiding device dependent proprietary upsampling methods. During the study, the videos were played out using the Venueplayer \cite{videoclarity} application developed by VideoClarity, which supports high frame rates and does not introduce artifacts that could impact the perception of video quality. To ensure perfect playback, all of the distorted sequences were processed and stored as raw YUV 4:2:0 files. Note that each video was viewed at a single spatial resolution, hence the subjects were never asked to rate any videos at different scales.

The LIVE-YT-HFR database was divided into 4 subsets of 120 videos each, such that every subject viewed only 2 of the 4 sets. Thus, each subject rated 240 videos across 2 sessions, where 120 videos were viewed in each session. We prepared playlists for each subject by randomly re-ordering the 120 videos. Care was taken to ensure that successive videos were obtained from different source sequences as well as different frame rates. This was done in order to inhibit any contextual and memory biases that could affect subjective quality judgments. Distinct playlists were created for every subject across every session, to avoid any prejudice arising from playing videos in any specific order. To avoid latency issues due to slow hard disk access, the entire playlist was loaded into memory before playback in each session. The monitor refresh rates were altered to exactly match each video's frame rate before it was played back. 

After each video plays, an interactive continuous quality rating scale was displayed on the screen as shown in Fig. \ref{fig:vote}. The initial position of the cursor was randomized for every video. The quality bar was labeled with 5 Likert indicators, to assist the subjects in their rating task, ranging from ``Bad" to ``Excellent." The subjects could move the cursor using a Palette gear console \cite{palette}, then press a key on the console to enter each quality score. The subject was provided as much time as needed to enter each score, but could not modify the score once entered. After the score was received, the next video was presented.

\begin{figure}[!t]
	\centering
	\includegraphics[width=1\linewidth]{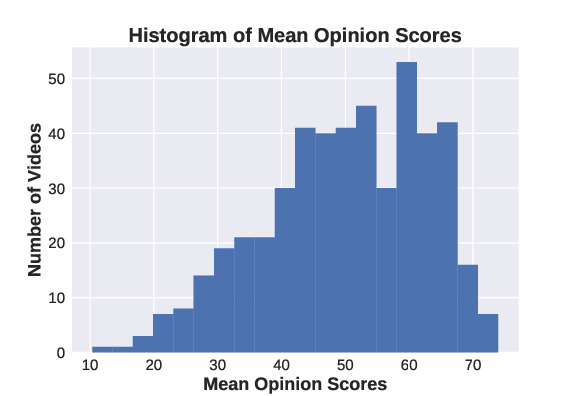}
	\caption{Histogram of MOS in 20 equally spaced bins}
	\label{fig:MOS_hist}
\end{figure}

\subsection{Subjects and Training}
A total of 85 volunteer undergraduate subjects were recruited at The University of Texas at Austin. The subject pool consisted of 14 female and 71 male participants, aged between 20 to 30 years. All subjects were screened for normal or corrected-to-normal color vision, and no subjects were rejected during screening. Each subject was individually informed of the purpose of the study, and a short training session was conducted to familiarize them with the rating procedure. During the training session, 6 videos were shown approximately spanning the overall quality range of test sequences, to give the subjects an idea of the video quality they could expect during the actual study. The training videos were not part of the database and contained different content, and the scores on them were not recorded or considered. Training was only performed before the start of each subject's first session. The subjects were instructed to provide ratings based on perceived quality, rather than on any preference for, or interestingness of content. To reduce subjective fatigue, a minimum of 24 hours was required between successive sessions.

No subject required more than 40 minutes to complete any session. In the end, each video was labelled by a minimum of 42 user ratings. The histograms of raw subjective scores for all four subsets of scores are shown in Fig. \ref{fig:histogram_sets}. The very similar score distributions over the four subsets indicate that they contain very similar quality distributions.

\subsection{Processing of Subjective Scores} 
Let $m_{ijk}$ denote the score provided by subject $i$ to video $j$ in session $k = \lbrace 1,2 \rbrace$. Since not all videos in the LIVE-YT-HFR Database were rated by every subject, let $\delta(i,j)$ be an indicator function such that 
\begin{equation}
	\delta(i,j) =
	\begin{cases}
		1 & \text{if subject $i$ rated video $j$} \\
		0 & \text{otherwise}. 
	\end{cases}
	\label{eqn:delta}
\end{equation}
\begin{figure}[!t]
	\centering
	\includegraphics[width=1\linewidth]{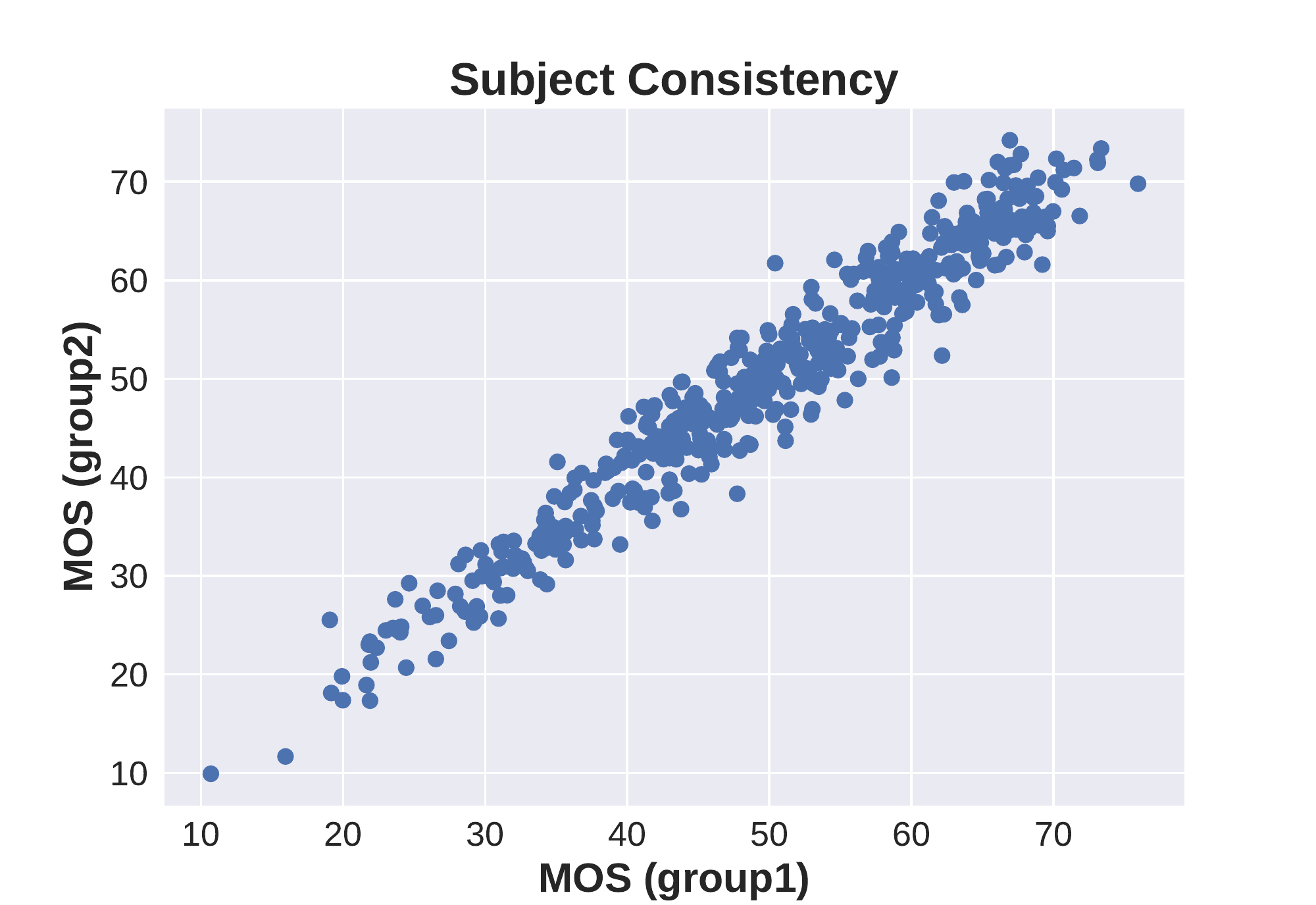}
	\caption{Scatter plot of MOS between two groups of subjects.}
	\label{fig:Subject_consistency}
\end{figure}
Then, to normalize the scores received across multiple sessions of each subject, we calculate the Z-scores per session  \cite{van1995quality} as
\begin{align*}
	\mu_{ik} &= \frac{1}{N_{ik}}\sum_{j=1} ^{N_{ik}} m_{ijk}\\
	\sigma_{ik} &= \sqrt{\frac{1}{N_{ik}-1} \sum_{j=1} ^{N_{ik}} (m_{ijk} - \mu_{ik})^2} \\
	z_{ijk} &= \frac{m_{ijk} - \mu_{ik}}{\sigma_{ik}},
\end{align*}
where $N_{ik}$ is the number of videos seen by subject $i$ in session $k$. The Z-scores from all sessions were concatenated to form the matrix $\lbrace z_{ij} \rbrace$ denoting the Z-score assigned by subject $i$ to the videos indexed by $j$ with $j \in \lbrace 1,2 \ldots 480 \rbrace$, where the entries of $\lbrace z_{ij} \rbrace$ are empty at locations $(i,j)$ where $\delta(i,j)=0$. We elected not to enforce any subject rejection procedure, as we observed that the inter-subject correlation was very high (inter-subject consistency is discussed in Sec. \ref{sec:subject_consistency}). A high inter-subject correlation suggests that there is a high degree of agreement among subjects regarding the video quality, and a reduced probability of a subject lying outside the population of honest subjects. 
Assuming $z_{ij}$ to have a standard normal distribution, $99\%$ of the Z-scores were found to lie in $[-3,3]$. A linear rescaling was used to map scores to the range $[0,100]$ as
\begin{align}
	z_{ij}' = \frac{100(z_{ij}+3)}{6}.
\end{align}
Finally the Mean Opinion Score (MOS) of each video was calculated by averaging the scores received for that video as
\begin{align}
	MOS_j = \frac{1}{N_j} \sum_{i=1} ^N z_{ij}'\delta(i,j),
\end{align}
where $N_j = \sum_{i=1} ^N \delta(i,j)$ and $N = 480$. The MOS were found to lie in the range $[10.26,73.15]$, and the mean of standard deviations of the rescaled Z-scores obtained from all subjects across all images was found to be $10.26$. The histogram of MOS is shown in Fig. \ref{fig:MOS_hist} indicating a relatively broad MOS variation.  

\begin{figure}[t]
	\centering
	\includegraphics[width=1\linewidth]{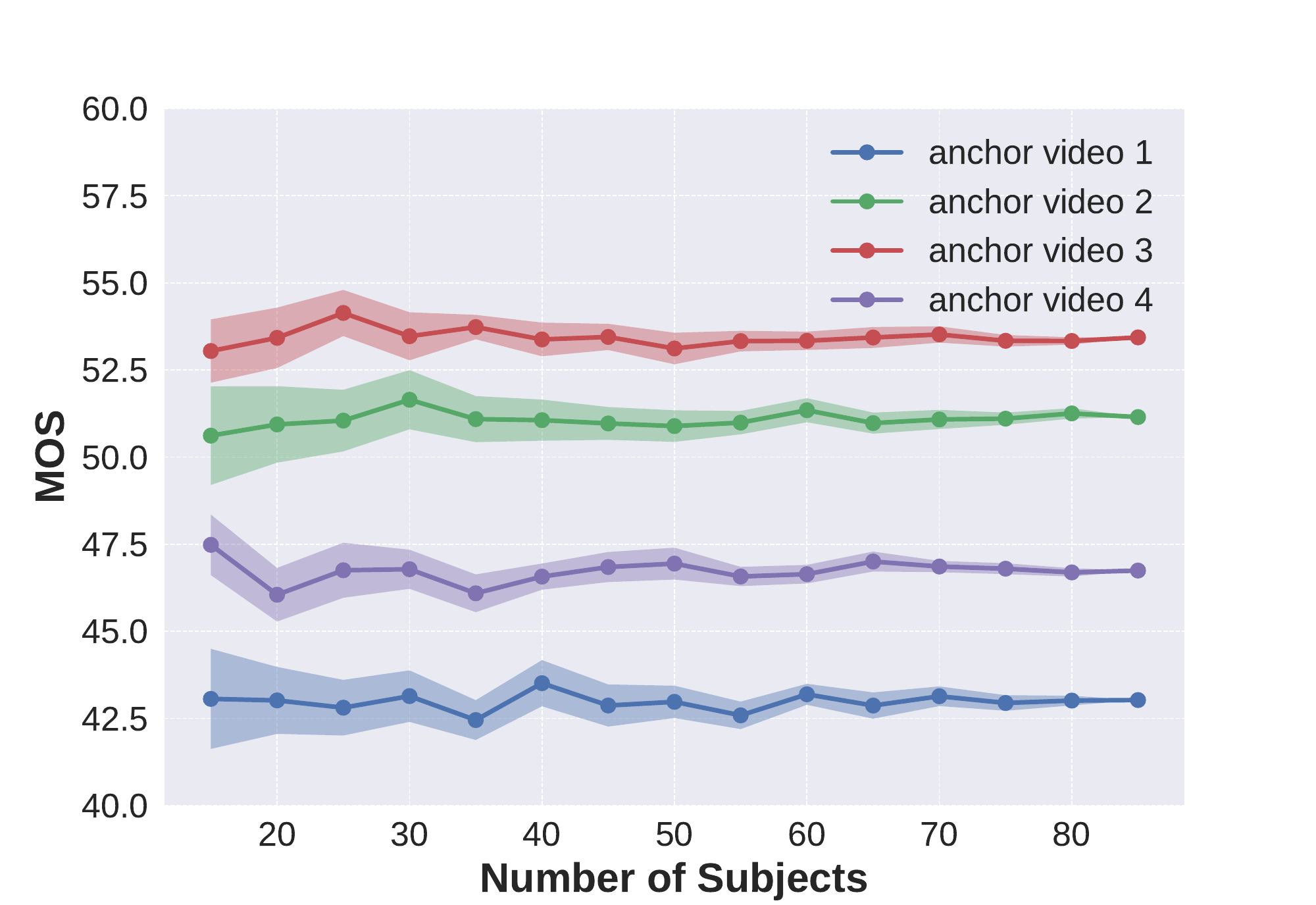}
	\caption{MOS of anchor videos plotted against number of subjects along with 95\% confidence intervals.}
	\label{fig:anchor_MOS}
\end{figure}

\begin{figure}[t]
	\centering
	\subfloat{\includegraphics[width=0.5\linewidth]{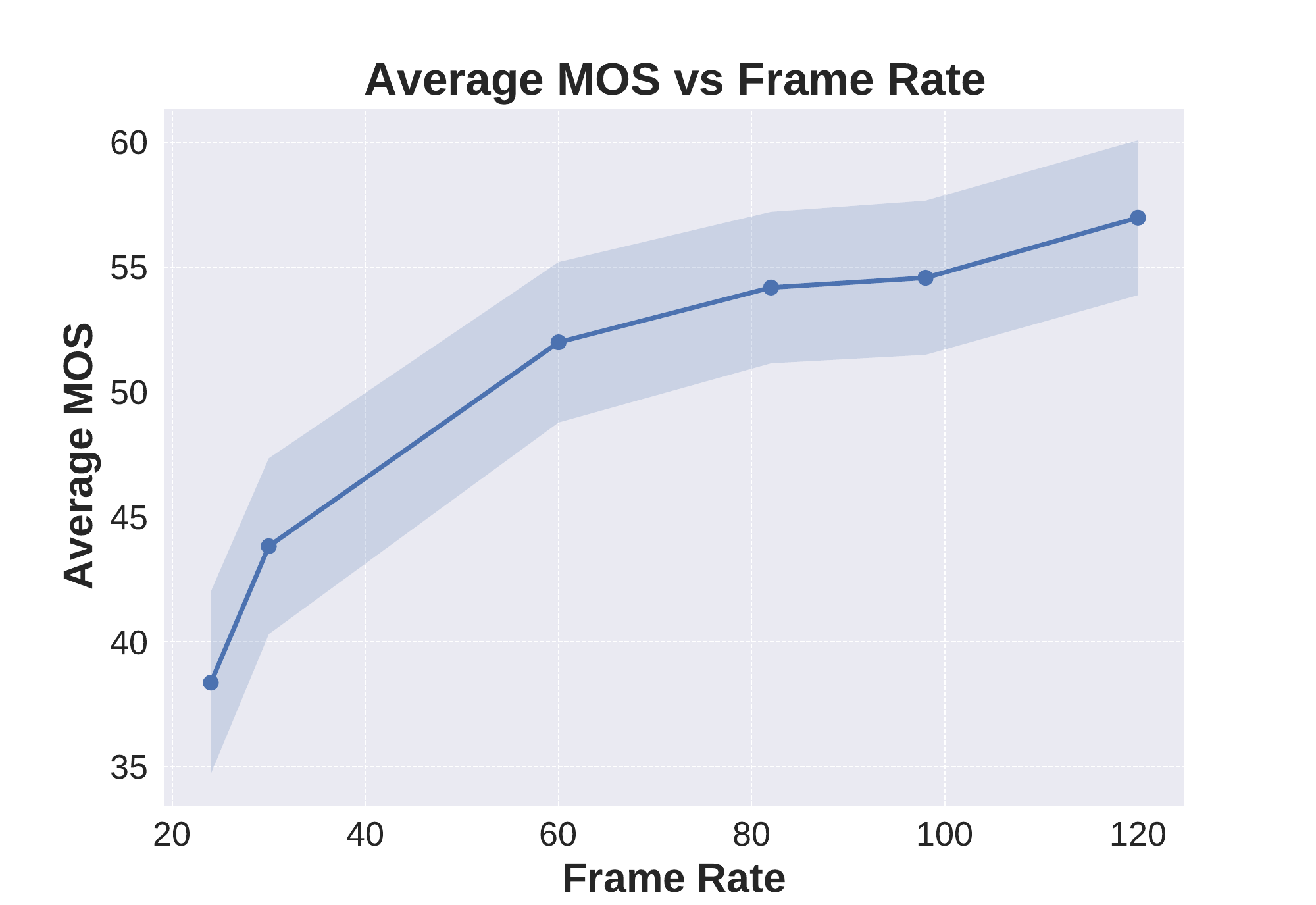}}\hfill
	\subfloat{\includegraphics[width=0.5\linewidth]{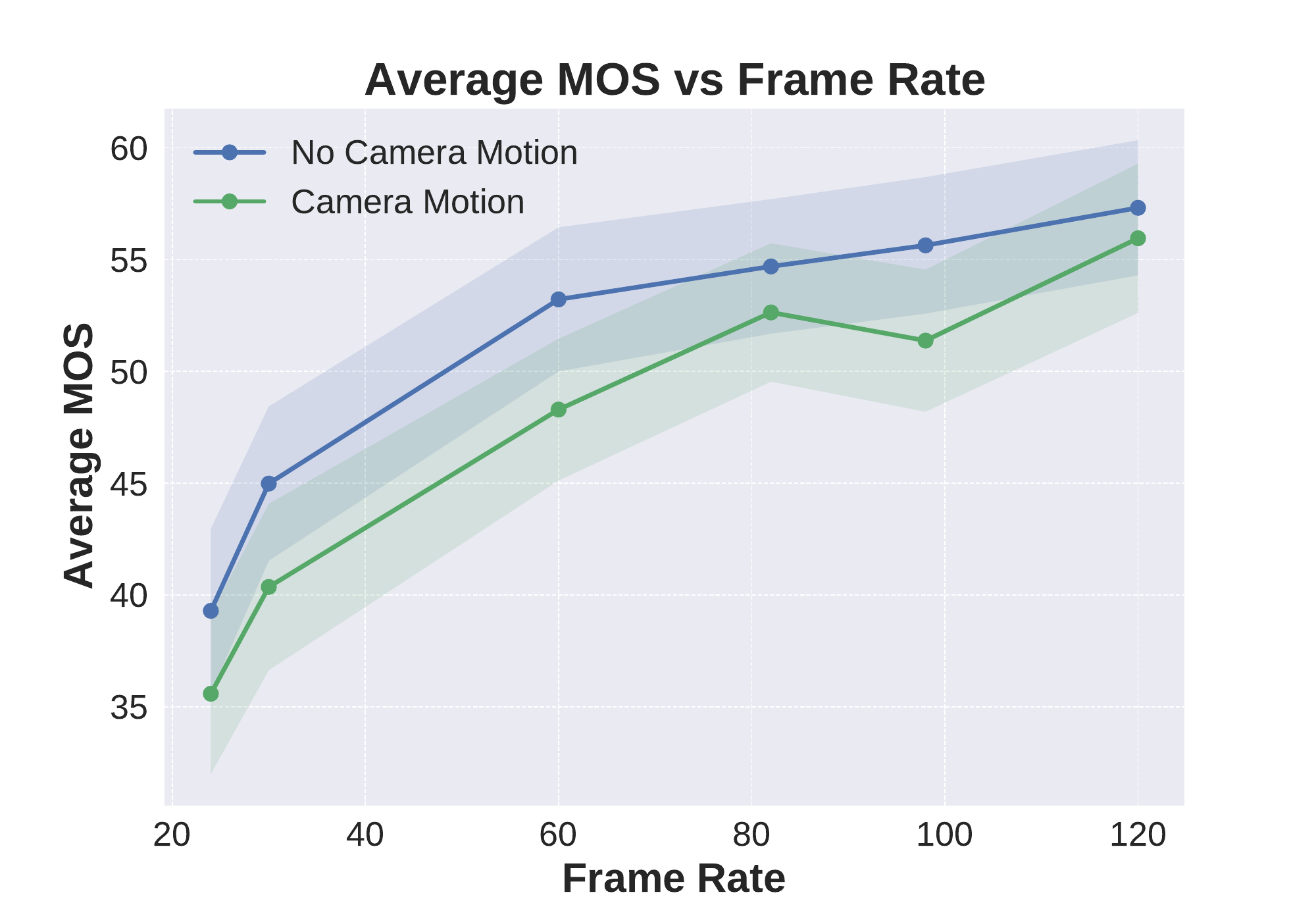}}
	\caption{(Left) Relationship between average MOS and frame rate, and (Right) The effect of camera motion. Shaded regions represent 95\% confidence intervals.}
	\label{fig:average_MOS}
\end{figure}

We calculated Difference MOS (DMOS) by subtracting the MOS of each video from the MOS of its corresponding reference as:
\begin{align}
	DMOS_j = MOS_j ^{ref} - MOS_j.
	\label{eqn:DMOS}
\end{align}
DMOS is particularly useful for FR-VQA problems to reduce content dependence.

\subsection{Subject-Consistency Analysis}
\label{sec:subject_consistency}
To ensure that the subjects' ratings were reliable, we performed additional analysis to evaluate the inter and intra subject reliability. 
\paragraph{\textit{Inter-Subject Consistency}} To check inter-subject consistency we split the scores received for every video into two disjoint equal groups, and measured the correlation of MOS between these two groups. The random splits were performed over 100 trials and the medians of both the Spearman rank order correlation coefficient (SROCC) and the Pearson linear correlation coefficient (PLCC) between the two groups were found to be \textbf{0.96}. Fig. \ref{fig:Subject_consistency} shows a scatter plot of MOS between the two randomly divided groups. It may be observed that the majority of the scores are concentrated near a line of unit slope passing through the origin, indicating a high degree of consistency between the groups.

\begin{figure}[t]
	\centering
	\includegraphics[width=1\linewidth]{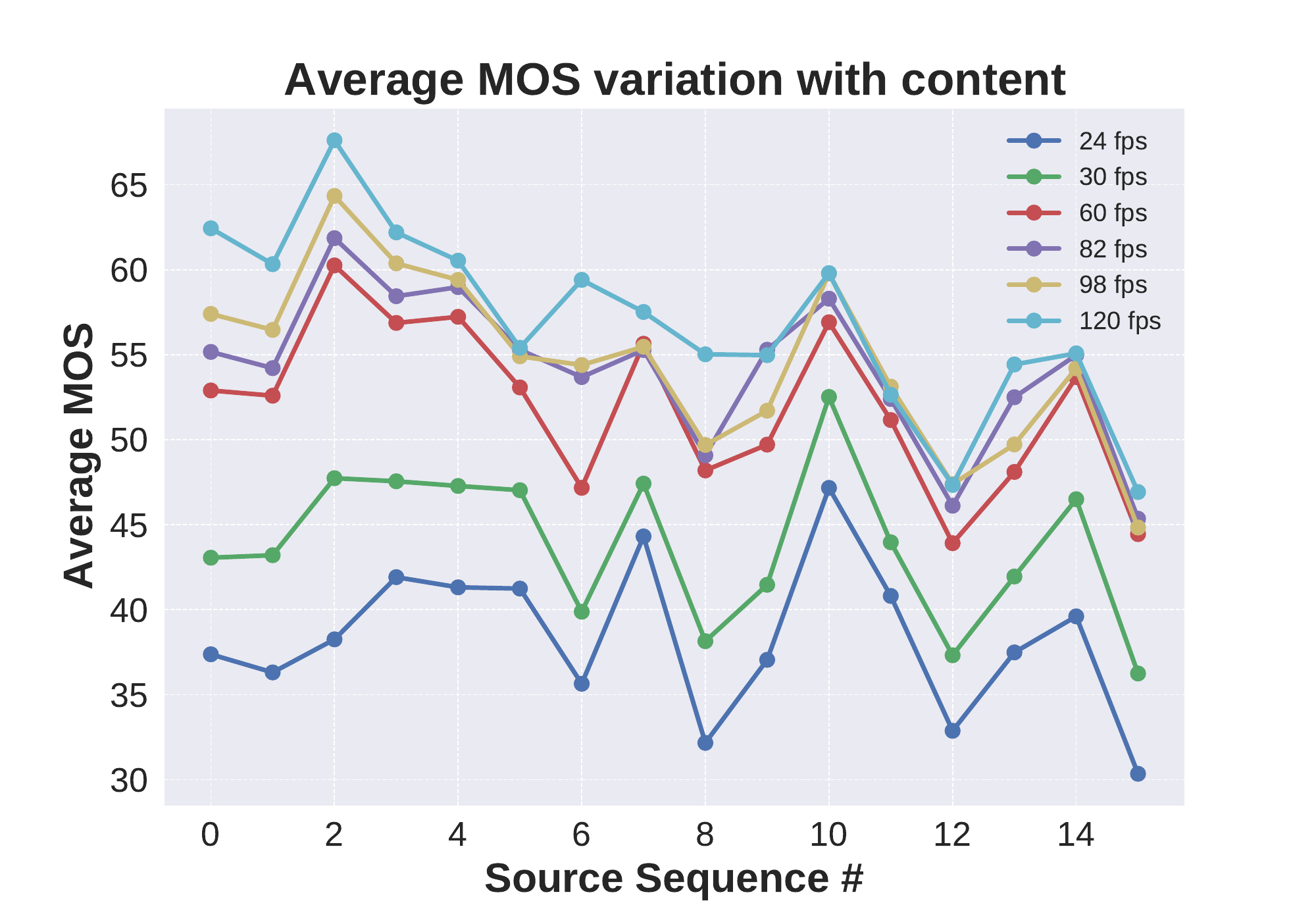}
	\caption{Variation of average MOS with content across frame rates. The horizontal axis indexes videos ordered in the same manner as shown in Fig. \ref{fig:screenshot}}
	\label{fig:average_MOS_content}
\end{figure}

\begin{figure}[t]
	\centering
	\subfloat{\includegraphics[width=0.5\linewidth]{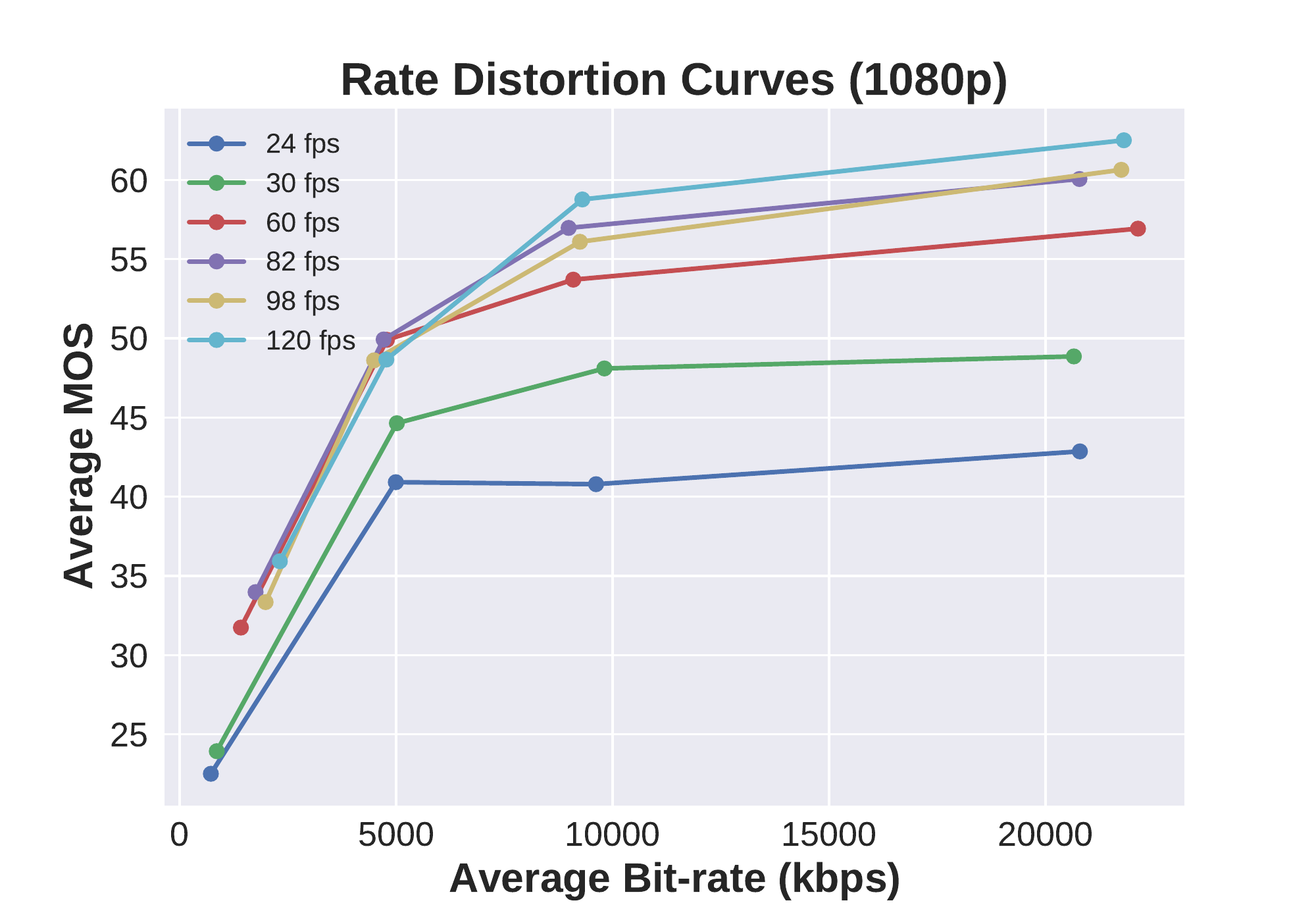}}\hfill
	\subfloat{\includegraphics[width=0.5\linewidth]{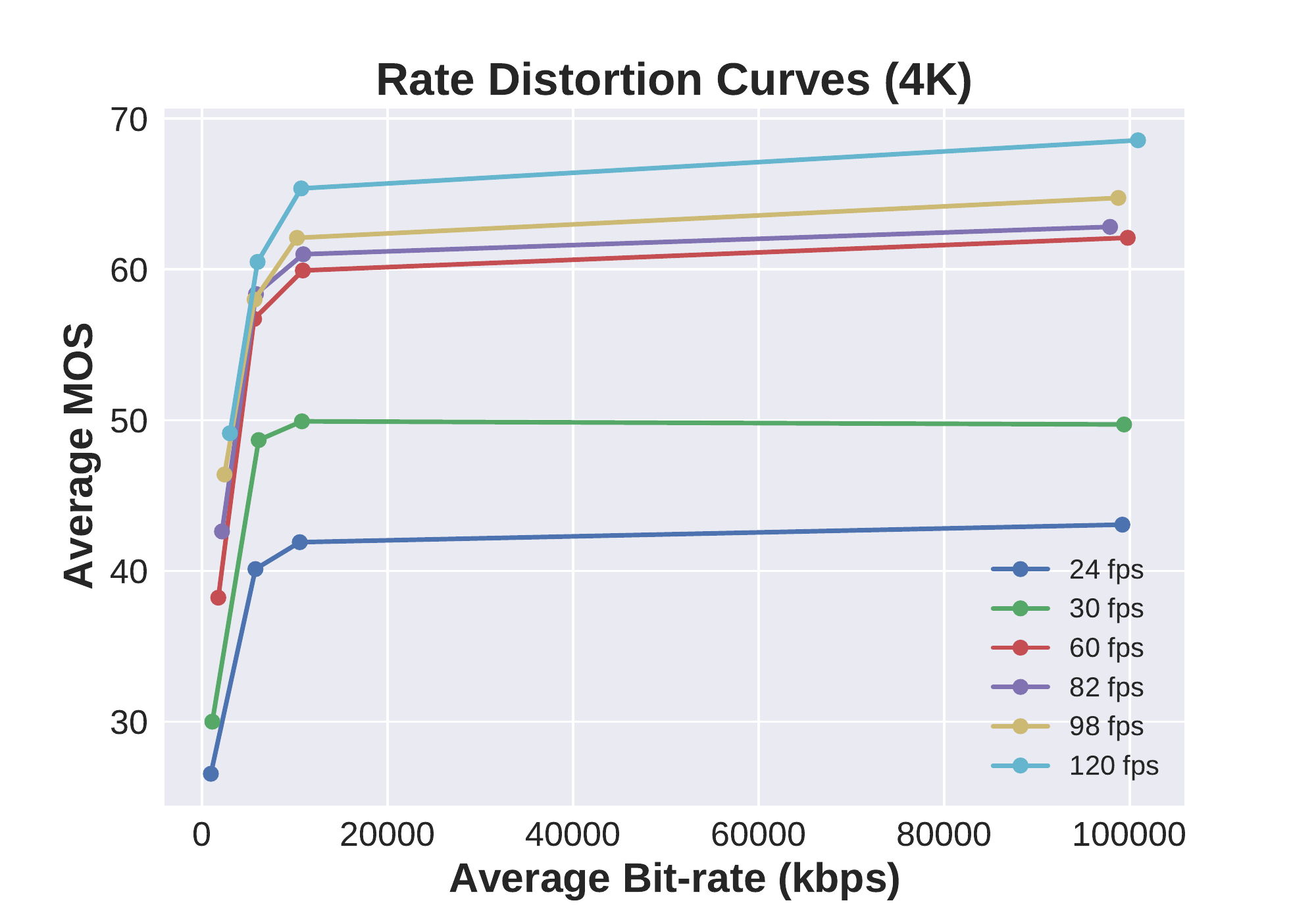}}
	\caption{Rate distortion curves for different frame rates with 1080p (left) and 4K (right) resolutions.}
	\label{fig:rd_plots}
\end{figure}

\begin{table*}[t]
	\caption{Results of t-test between videos at various frame rates. A value of `1' indicates that the row is statistically superior (better visual quality) than the column, while a value of `0' indicates that the column is statistically superior than the row. A value of `-' indicates that the row and column are statistically similar. Each sub-entry in row/column corresponds to 16 contents arranged in the same order, as shown in Fig. \ref{fig:screenshot}}
	\label{table:fps_significance}
	\centering
	\scriptsize
	\scalebox{0.92}{
		\begin{tabular}{|c||c|c|c|c|c|c|}
			\hline
			~ & 24 fps & 30 fps & 60 fps & 82 fps & 98 fps & 120 fps \\ \hline \hline
			24 fps & -\hspace{1pt}-\hspace{1pt}-\hspace{1pt}-\hspace{1pt}-\hspace{1pt}-\hspace{1pt}-\hspace{1pt}-\hspace{1pt}-\hspace{1pt}-\hspace{1pt}-\hspace{1pt}-\hspace{1pt}-\hspace{1pt}-\hspace{1pt}-\hspace{1pt}- & 0\hspace{1pt}0\hspace{1pt}0\hspace{1pt}0\hspace{1pt}0\hspace{1pt}0\hspace{1pt}0\hspace{1pt}-\hspace{1pt}0\hspace{1pt}0\hspace{1pt}0\hspace{1pt}0\hspace{1pt}0\hspace{1pt}0\hspace{1pt}0\hspace{1pt}0\hspace{1pt} & 0\hspace{1pt}0\hspace{1pt}0\hspace{1pt}0\hspace{1pt}0\hspace{1pt}0\hspace{1pt}0\hspace{1pt}0\hspace{1pt}0\hspace{1pt}0\hspace{1pt}0\hspace{1pt}0\hspace{1pt}0\hspace{1pt}0\hspace{1pt}0\hspace{1pt}0\hspace{1pt} & 0\hspace{1pt}0\hspace{1pt}0\hspace{1pt}0\hspace{1pt}0\hspace{1pt}0\hspace{1pt}0\hspace{1pt}0\hspace{1pt}0\hspace{1pt}0\hspace{1pt}0\hspace{1pt}0\hspace{1pt}0\hspace{1pt}0\hspace{1pt}0\hspace{1pt}0\hspace{1pt} & 0\hspace{1pt}0\hspace{1pt}0\hspace{1pt}0\hspace{1pt}0\hspace{1pt}0\hspace{1pt}0\hspace{1pt}0\hspace{1pt}0\hspace{1pt}0\hspace{1pt}0\hspace{1pt}0\hspace{1pt}0\hspace{1pt}0\hspace{1pt}0\hspace{1pt}0\hspace{1pt} & 0\hspace{1pt}0\hspace{1pt}0\hspace{1pt}0\hspace{1pt}0\hspace{1pt}0\hspace{1pt}0\hspace{1pt}0\hspace{1pt}0\hspace{1pt}0\hspace{1pt}0\hspace{1pt}0\hspace{1pt}0\hspace{1pt}0\hspace{1pt}0\hspace{1pt}0\hspace{1pt} \\ 
			30 fps & 1\hspace{1pt}1\hspace{1pt}1\hspace{1pt}1\hspace{1pt}1\hspace{1pt}1\hspace{1pt}1\hspace{1pt}-\hspace{1pt}1\hspace{1pt}1\hspace{1pt}1\hspace{1pt}1\hspace{1pt}1\hspace{1pt}1\hspace{1pt}1\hspace{1pt}1\hspace{1pt} & -\hspace{1pt}-\hspace{1pt}-\hspace{1pt}-\hspace{1pt}-\hspace{1pt}-\hspace{1pt}-\hspace{1pt}-\hspace{1pt}-\hspace{1pt}-\hspace{1pt}-\hspace{1pt}-\hspace{1pt}-\hspace{1pt}-\hspace{1pt}-\hspace{1pt}- & 0\hspace{1pt}0\hspace{1pt}0\hspace{1pt}0\hspace{1pt}0\hspace{1pt}0\hspace{1pt}0\hspace{1pt}0\hspace{1pt}0\hspace{1pt}0\hspace{1pt}0\hspace{1pt}0\hspace{1pt}0\hspace{1pt}-\hspace{1pt}0\hspace{1pt}0\hspace{1pt} & 0\hspace{1pt}0\hspace{1pt}0\hspace{1pt}0\hspace{1pt}0\hspace{1pt}0\hspace{1pt}0\hspace{1pt}0\hspace{1pt}0\hspace{1pt}0\hspace{1pt}0\hspace{1pt}0\hspace{1pt}0\hspace{1pt}0\hspace{1pt}0\hspace{1pt}0\hspace{1pt} & 0\hspace{1pt}0\hspace{1pt}0\hspace{1pt}0\hspace{1pt}0\hspace{1pt}0\hspace{1pt}0\hspace{1pt}0\hspace{1pt}0\hspace{1pt}0\hspace{1pt}0\hspace{1pt}0\hspace{1pt}0\hspace{1pt}0\hspace{1pt}0\hspace{1pt}0\hspace{1pt} & 0\hspace{1pt}0\hspace{1pt}0\hspace{1pt}0\hspace{1pt}0\hspace{1pt}0\hspace{1pt}0\hspace{1pt}0\hspace{1pt}0\hspace{1pt}0\hspace{1pt}0\hspace{1pt}0\hspace{1pt}0\hspace{1pt}0\hspace{1pt}0\hspace{1pt}0\hspace{1pt} \\ 
			60 fps & 1\hspace{1pt}1\hspace{1pt}1\hspace{1pt}1\hspace{1pt}1\hspace{1pt}1\hspace{1pt}1\hspace{1pt}1\hspace{1pt}1\hspace{1pt}1\hspace{1pt}1\hspace{1pt}1\hspace{1pt}1\hspace{1pt}1\hspace{1pt}1\hspace{1pt}1\hspace{1pt} & 1\hspace{1pt}1\hspace{1pt}1\hspace{1pt}1\hspace{1pt}1\hspace{1pt}1\hspace{1pt}1\hspace{1pt}1\hspace{1pt}1\hspace{1pt}1\hspace{1pt}1\hspace{1pt}1\hspace{1pt}1\hspace{1pt}-\hspace{1pt}1\hspace{1pt}1\hspace{1pt} & -\hspace{1pt}-\hspace{1pt}-\hspace{1pt}-\hspace{1pt}-\hspace{1pt}-\hspace{1pt}-\hspace{1pt}-\hspace{1pt}-\hspace{1pt}-\hspace{1pt}-\hspace{1pt}-\hspace{1pt}-\hspace{1pt}-\hspace{1pt}-\hspace{1pt}- & -\hspace{1pt}-\hspace{1pt}-\hspace{1pt}-\hspace{1pt}0\hspace{1pt}-\hspace{1pt}0\hspace{1pt}-\hspace{1pt}1\hspace{1pt}0\hspace{1pt}-\hspace{1pt}-\hspace{1pt}0\hspace{1pt}0\hspace{1pt}0\hspace{1pt}-\hspace{1pt} & 0\hspace{1pt}0\hspace{1pt}0\hspace{1pt}0\hspace{1pt}-\hspace{1pt}-\hspace{1pt}0\hspace{1pt}-\hspace{1pt}-\hspace{1pt}-\hspace{1pt}-\hspace{1pt}-\hspace{1pt}0\hspace{1pt}0\hspace{1pt}-\hspace{1pt}-\hspace{1pt} & 0\hspace{1pt}0\hspace{1pt}0\hspace{1pt}0\hspace{1pt}0\hspace{1pt}0\hspace{1pt}0\hspace{1pt}-\hspace{1pt}0\hspace{1pt}0\hspace{1pt}0\hspace{1pt}-\hspace{1pt}0\hspace{1pt}0\hspace{1pt}-\hspace{1pt}-\hspace{1pt} \\ 
			82 fps & 1\hspace{1pt}1\hspace{1pt}1\hspace{1pt}1\hspace{1pt}1\hspace{1pt}1\hspace{1pt}1\hspace{1pt}1\hspace{1pt}1\hspace{1pt}1\hspace{1pt}1\hspace{1pt}1\hspace{1pt}1\hspace{1pt}1\hspace{1pt}1\hspace{1pt}1\hspace{1pt} & 1\hspace{1pt}1\hspace{1pt}1\hspace{1pt}1\hspace{1pt}1\hspace{1pt}1\hspace{1pt}1\hspace{1pt}1\hspace{1pt}1\hspace{1pt}1\hspace{1pt}1\hspace{1pt}1\hspace{1pt}1\hspace{1pt}1\hspace{1pt}1\hspace{1pt}1\hspace{1pt} & -\hspace{1pt}-\hspace{1pt}-\hspace{1pt}-\hspace{1pt}1\hspace{1pt}-\hspace{1pt}1\hspace{1pt}-\hspace{1pt}0\hspace{1pt}1\hspace{1pt}-\hspace{1pt}-\hspace{1pt}1\hspace{1pt}1\hspace{1pt}1\hspace{1pt}-\hspace{1pt} & -\hspace{1pt}-\hspace{1pt}-\hspace{1pt}-\hspace{1pt}-\hspace{1pt}-\hspace{1pt}-\hspace{1pt}-\hspace{1pt}-\hspace{1pt}-\hspace{1pt}-\hspace{1pt}-\hspace{1pt}-\hspace{1pt}-\hspace{1pt}-\hspace{1pt}- & -\hspace{1pt}0\hspace{1pt}0\hspace{1pt}-\hspace{1pt}-\hspace{1pt}-\hspace{1pt}-\hspace{1pt}-\hspace{1pt}0\hspace{1pt}1\hspace{1pt}-\hspace{1pt}-\hspace{1pt}-\hspace{1pt}-\hspace{1pt}-\hspace{1pt}-\hspace{1pt} & 0\hspace{1pt}0\hspace{1pt}0\hspace{1pt}0\hspace{1pt}-\hspace{1pt}-\hspace{1pt}0\hspace{1pt}-\hspace{1pt}0\hspace{1pt}-\hspace{1pt}-\hspace{1pt}-\hspace{1pt}-\hspace{1pt}-\hspace{1pt}-\hspace{1pt}-\hspace{1pt} \\ 
			98 fps & 1\hspace{1pt}1\hspace{1pt}1\hspace{1pt}1\hspace{1pt}1\hspace{1pt}1\hspace{1pt}1\hspace{1pt}1\hspace{1pt}1\hspace{1pt}1\hspace{1pt}1\hspace{1pt}1\hspace{1pt}1\hspace{1pt}1\hspace{1pt}1\hspace{1pt}1\hspace{1pt} & 1\hspace{1pt}1\hspace{1pt}1\hspace{1pt}1\hspace{1pt}1\hspace{1pt}1\hspace{1pt}1\hspace{1pt}1\hspace{1pt}1\hspace{1pt}1\hspace{1pt}1\hspace{1pt}1\hspace{1pt}1\hspace{1pt}1\hspace{1pt}1\hspace{1pt}1\hspace{1pt} & 1\hspace{1pt}1\hspace{1pt}1\hspace{1pt}1\hspace{1pt}-\hspace{1pt}-\hspace{1pt}1\hspace{1pt}-\hspace{1pt}-\hspace{1pt}-\hspace{1pt}-\hspace{1pt}-\hspace{1pt}1\hspace{1pt}1\hspace{1pt}-\hspace{1pt}-\hspace{1pt} & -\hspace{1pt}1\hspace{1pt}1\hspace{1pt}-\hspace{1pt}-\hspace{1pt}-\hspace{1pt}-\hspace{1pt}-\hspace{1pt}1\hspace{1pt}0\hspace{1pt}-\hspace{1pt}-\hspace{1pt}-\hspace{1pt}-\hspace{1pt}-\hspace{1pt}-\hspace{1pt} & -\hspace{1pt}-\hspace{1pt}-\hspace{1pt}-\hspace{1pt}-\hspace{1pt}-\hspace{1pt}-\hspace{1pt}-\hspace{1pt}-\hspace{1pt}-\hspace{1pt}-\hspace{1pt}-\hspace{1pt}-\hspace{1pt}-\hspace{1pt}-\hspace{1pt}- & 0\hspace{1pt}0\hspace{1pt}0\hspace{1pt}0\hspace{1pt}-\hspace{1pt}-\hspace{1pt}0\hspace{1pt}-\hspace{1pt}0\hspace{1pt}-\hspace{1pt}-\hspace{1pt}-\hspace{1pt}-\hspace{1pt}0\hspace{1pt}-\hspace{1pt}-\hspace{1pt} \\ 
			120 fps & 1\hspace{1pt}1\hspace{1pt}1\hspace{1pt}1\hspace{1pt}1\hspace{1pt}1\hspace{1pt}1\hspace{1pt}1\hspace{1pt}1\hspace{1pt}1\hspace{1pt}1\hspace{1pt}1\hspace{1pt}1\hspace{1pt}1\hspace{1pt}1\hspace{1pt}1\hspace{1pt} & 1\hspace{1pt}1\hspace{1pt}1\hspace{1pt}1\hspace{1pt}1\hspace{1pt}1\hspace{1pt}1\hspace{1pt}1\hspace{1pt}1\hspace{1pt}1\hspace{1pt}1\hspace{1pt}1\hspace{1pt}1\hspace{1pt}1\hspace{1pt}1\hspace{1pt}1\hspace{1pt} & 1\hspace{1pt}1\hspace{1pt}1\hspace{1pt}1\hspace{1pt}1\hspace{1pt}1\hspace{1pt}1\hspace{1pt}-\hspace{1pt}1\hspace{1pt}1\hspace{1pt}1\hspace{1pt}-\hspace{1pt}1\hspace{1pt}1\hspace{1pt}-\hspace{1pt}-\hspace{1pt} & 1\hspace{1pt}1\hspace{1pt}1\hspace{1pt}1\hspace{1pt}-\hspace{1pt}-\hspace{1pt}1\hspace{1pt}-\hspace{1pt}1\hspace{1pt}-\hspace{1pt}-\hspace{1pt}-\hspace{1pt}-\hspace{1pt}-\hspace{1pt}-\hspace{1pt}-\hspace{1pt} & 1\hspace{1pt}1\hspace{1pt}1\hspace{1pt}1\hspace{1pt}-\hspace{1pt}-\hspace{1pt}1\hspace{1pt}-\hspace{1pt}1\hspace{1pt}-\hspace{1pt}-\hspace{1pt}-\hspace{1pt}-\hspace{1pt}1\hspace{1pt}-\hspace{1pt}-\hspace{1pt} & -\hspace{1pt}-\hspace{1pt}-\hspace{1pt}-\hspace{1pt}-\hspace{1pt}-\hspace{1pt}-\hspace{1pt}-\hspace{1pt}-\hspace{1pt}-\hspace{1pt}-\hspace{1pt}-\hspace{1pt}-\hspace{1pt}-\hspace{1pt}-\hspace{1pt}- \\
			\hline
		\end{tabular}
	}
\end{table*}

\paragraph{\textit{Intra-Subject Consistency}} Measuring intra-subject reliability provides information on the level of consistency demonstrated by individual subjects \cite{hossfeld2013best} over the videos rated by them. We thus measured the SROCC and PLCC between the individual opinion scores and MOS. Both the SROCC and PLCC were observed to have a median value of \textbf{0.75} with a standard deviation of 0.09.

These additional experiments indicate that we can ascribe a high degree of confidence in the veracity of the obtained opinion scores, as well as the framework used to conduct the subjective study. While our above subject agreement analysis is deeply considered and effective, methods proposed in \cite{li2017recover} and \cite{nehme2020visual} are other choices.

\subsection{Anchor Videos}
In our study, not every video was rated by all of subjects, and each subject viewed only 50\% of the entire set of videos present in the database. Since we subscribed 85 subjects, we obtained roughly 43 ratings per video. In order to analyze the impact on MOS of having a different subset of subjects view each video as opposed to the entire population, we chose a subset of 30 \textit{anchor videos} which were present in the viewing sets of all subjects. Thus anchor videos received twice as many ratings as compared to non-anchor videos. To analyze the influence of different subject groups contributing MOS, we randomly sampled subsets of scores received for these anchor videos, and recalculated MOS on the reduced subsets, as shown in Fig. \ref{fig:anchor_MOS}. We notice that these computed MOS values remained nearly constant across the number of subjects, although the standard deviation tended to be higher when the number of subjects fell below 40. The confidence intervals were calculated based on MOS variation over 25 trials. Fig. \ref{fig:anchor_MOS} depicts the results on 4 anchor videos, but very similar observations were made on the remaining anchor videos. A key takeaway of this exercise is that MOS was relatively robust against the number of subjects.

\subsection{Analysis of Opinion Scores}
\label{sec:MOS_analysis}

\paragraph{\textbf{Impact of frame rate on MOS}}In Fig. \ref{fig:average_MOS} (left), the average MOS over all videos at each frame rate is plotted, along with their corresponding confidence intervals. Clearly, increases in frame rate led to higher perceived quality, but with diminishing returns for videos beyond 60 fps. It may also be inferred that non-uniform sampling of 120 fps to 82 and 98 frame rates dis not alter the increasing trend of quality with frame rate. In Fig. \ref{fig:average_MOS} (right) the impact of camera motion on MOS is illustrated. Videos with significant camera motion suffer from judder/strobing artifacts, particularly among lower frame rate versions. Thus videos with camera motion tended to have lower MOS values, as compared to non-camera motion videos at lower frame rates. However this gap narrowed with increases in frame rate, indicating a valuable reduction in judder/strobing distortions at higher frame rates.

\paragraph{\textbf{MOS content dependence}}In Fig. \ref{fig:average_MOS_content} the impact of source content on MOS across different frame rates is analyzed. It may be seen that for certain contents, there exists a clear demarcation between frame rates, however this separation is considerably reduced beyond 60 fps. Note that videos at lower frame rates (24 fps, 30 fps) always had lower MOS values, irrespective of content, indicating the existence of annoying temporal distortions arising from frame rate variations. A salient takeaway from these plots is that there exists high perceptual disparity in low fps regions, irrespective of the content. However, moving towards high fps, there is significant reduction in this gap, with the amount of reduction depending on the content. 

\paragraph{\textbf{Rate distortion curves}} In Fig. \ref{fig:rd_plots} rate distortion (RD) curves are plotted for various frame rates of 1080p (left) and 4K (right) videos. The horizontal axis denotes bit-rates averaged across content over 5 compression levels, as discussed in Sec. \ref{sec:test_sequences}. Note that we ignored the lossless compression level when plotting Fig. \ref{fig:rd_plots}, as bit-rates associated with those sequences are large, hence including them would make it harder to compare lower bit-rate videos. From the plots we may discern that there exists considerable overlap among the RD curves for frame rates above 60 fps in the low bit-rate region, while the amount of overlap gradually decreased as we moved towards the high bit-rate regime. Here as well, lower frame rates (24 fps, 30 fps) led to much lower MOS values across all bit-rates, reflecting the impact of temporal distortions on video quality.

\begin{figure*}[t]
	\centering
	\subfloat[PSNR]{\includegraphics[width=0.24\textwidth]{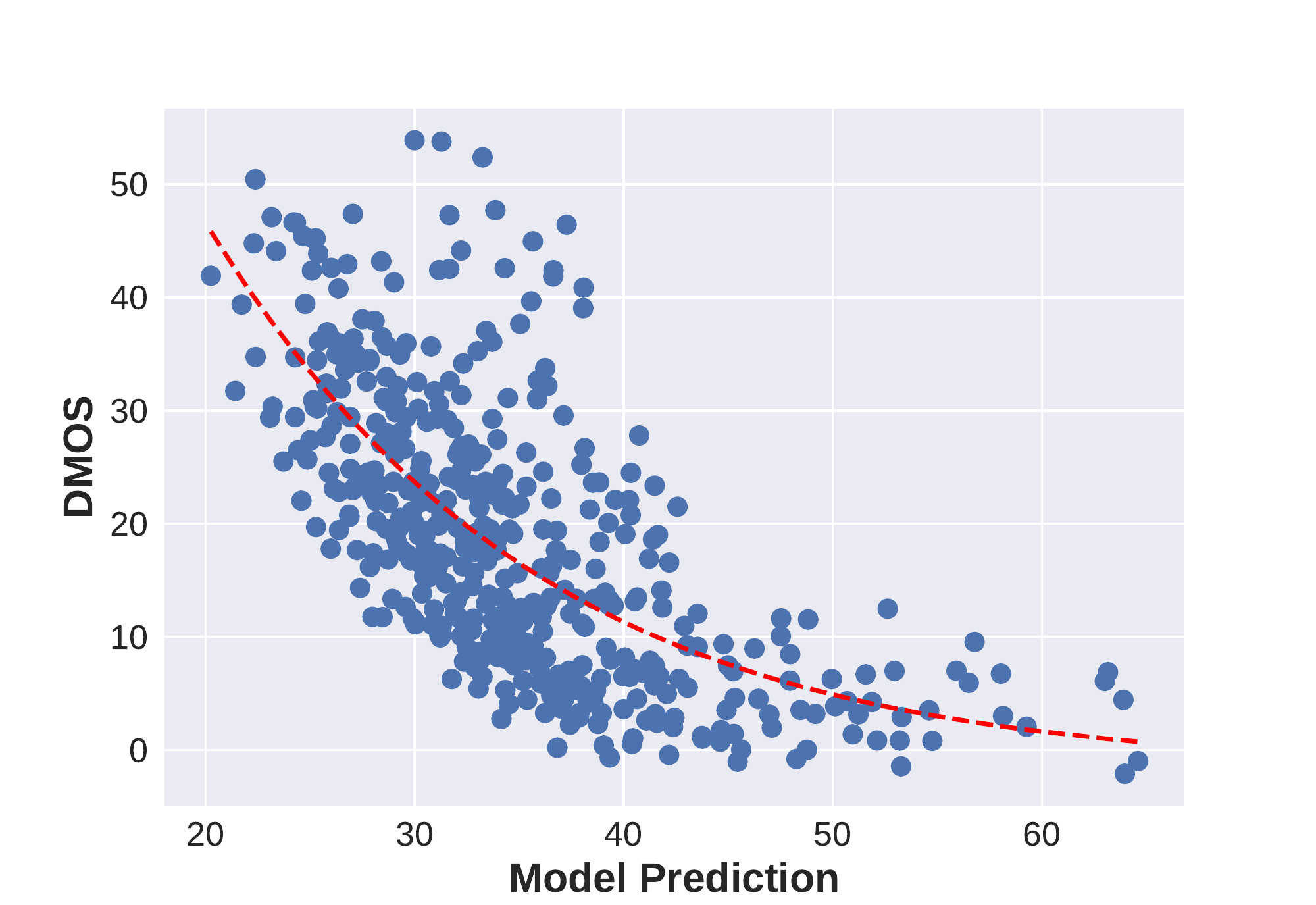}} \hfill
	\subfloat[SSIM \cite{wang2004image}]{\includegraphics[width=0.24\textwidth]{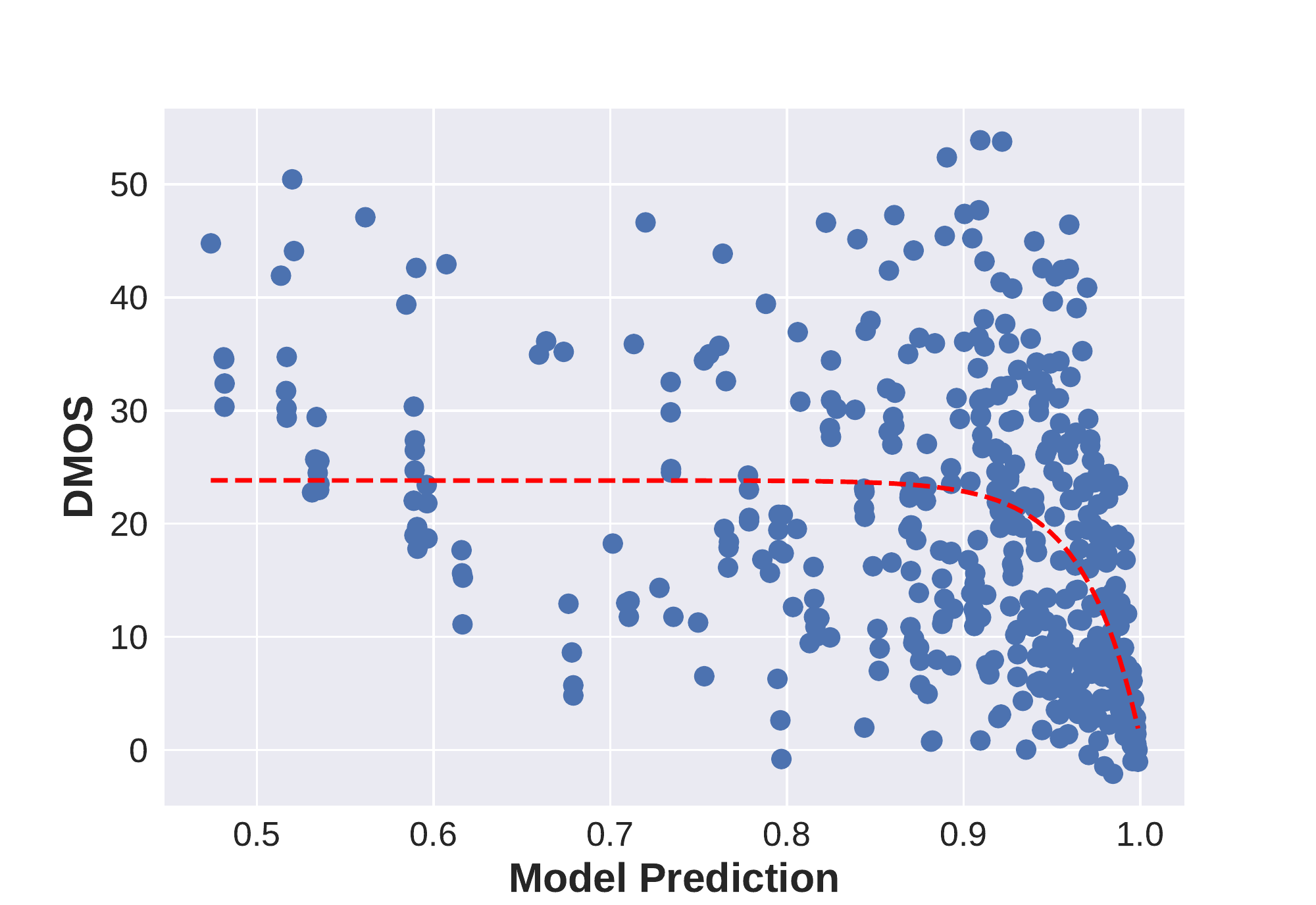}} \hfill
	\subfloat[MS-SSIM \cite{wang2003multiscale}]{\includegraphics[width=0.24\textwidth]{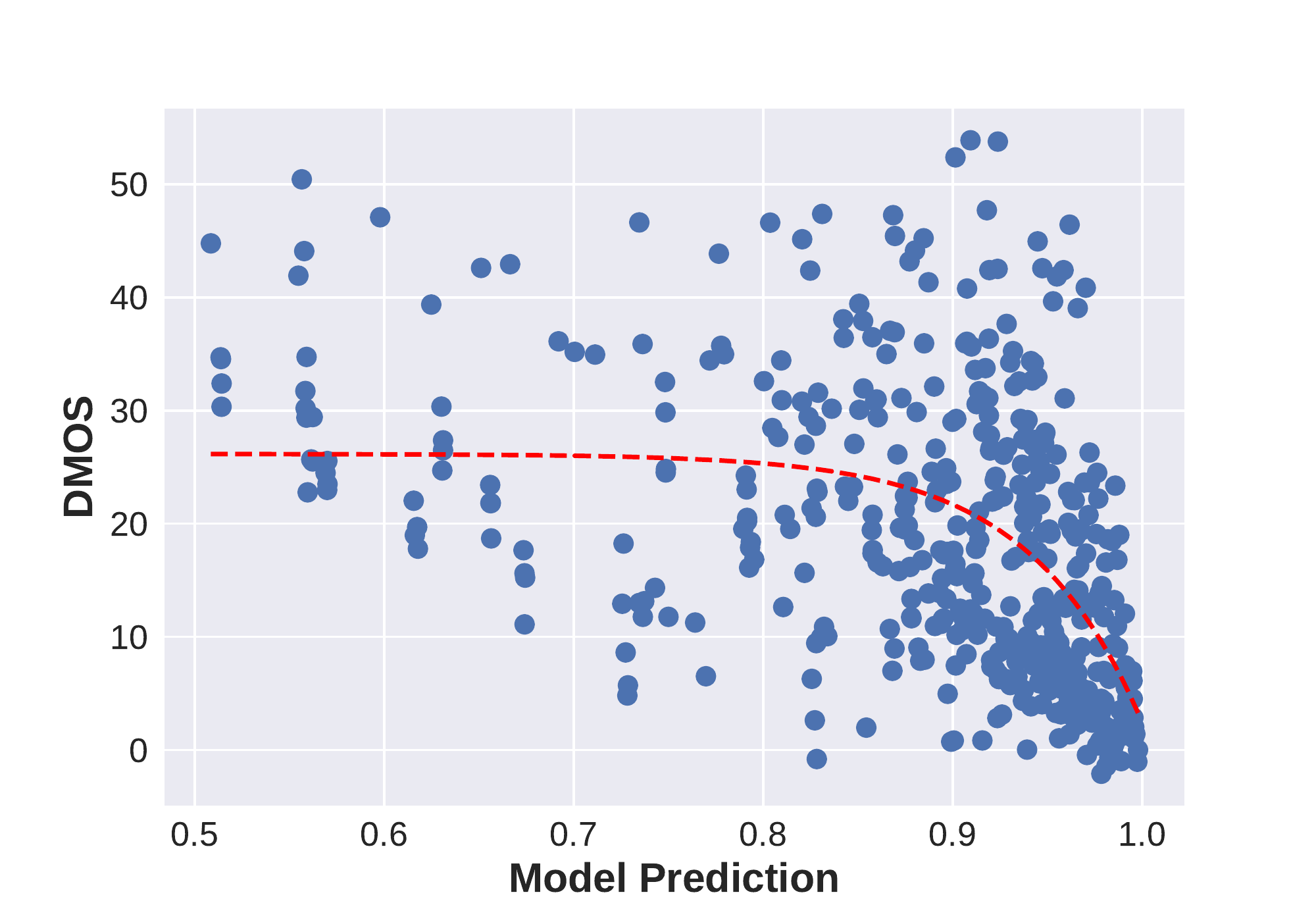}} \hfill
	\subfloat[FSIM \cite{zhang2011fsim}]{\includegraphics[width=0.24\textwidth]{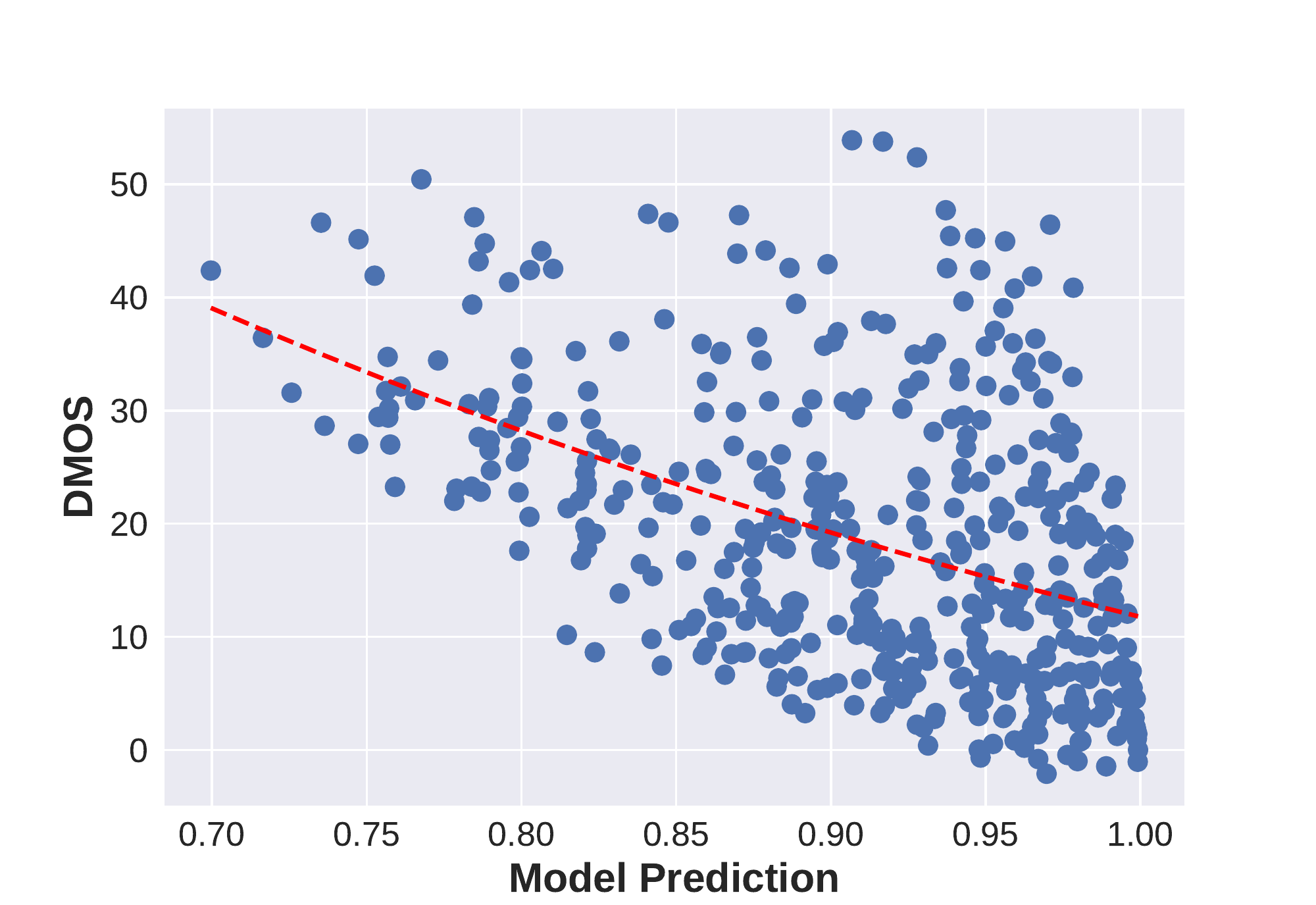}} \\
	\subfloat[ST-RRED \cite{soundararajan2012video}]{\includegraphics[width=0.24\textwidth]{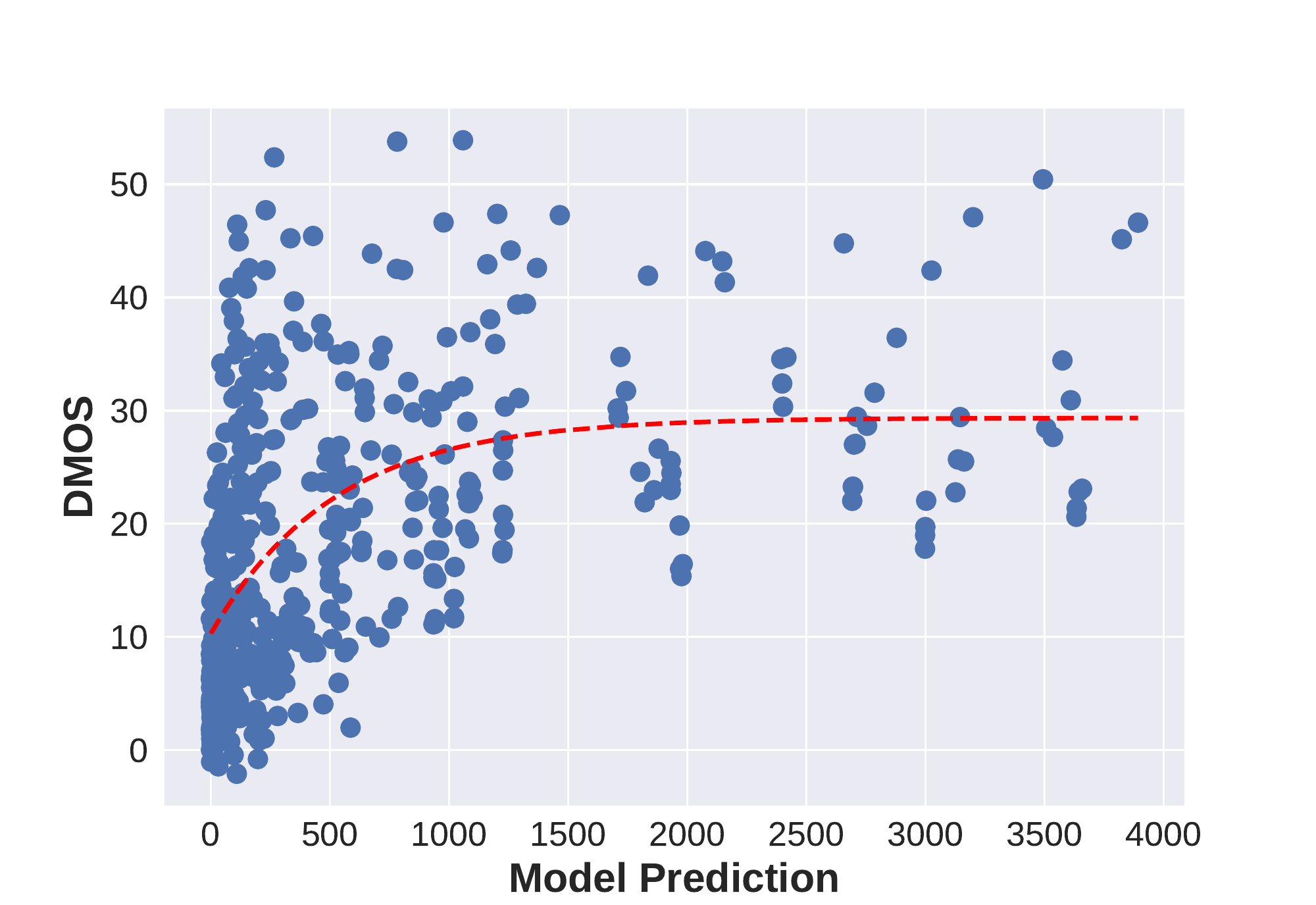}} \hfill
	\subfloat[SpEED \cite{bampis2017speed}]{\includegraphics[width=0.24\textwidth]{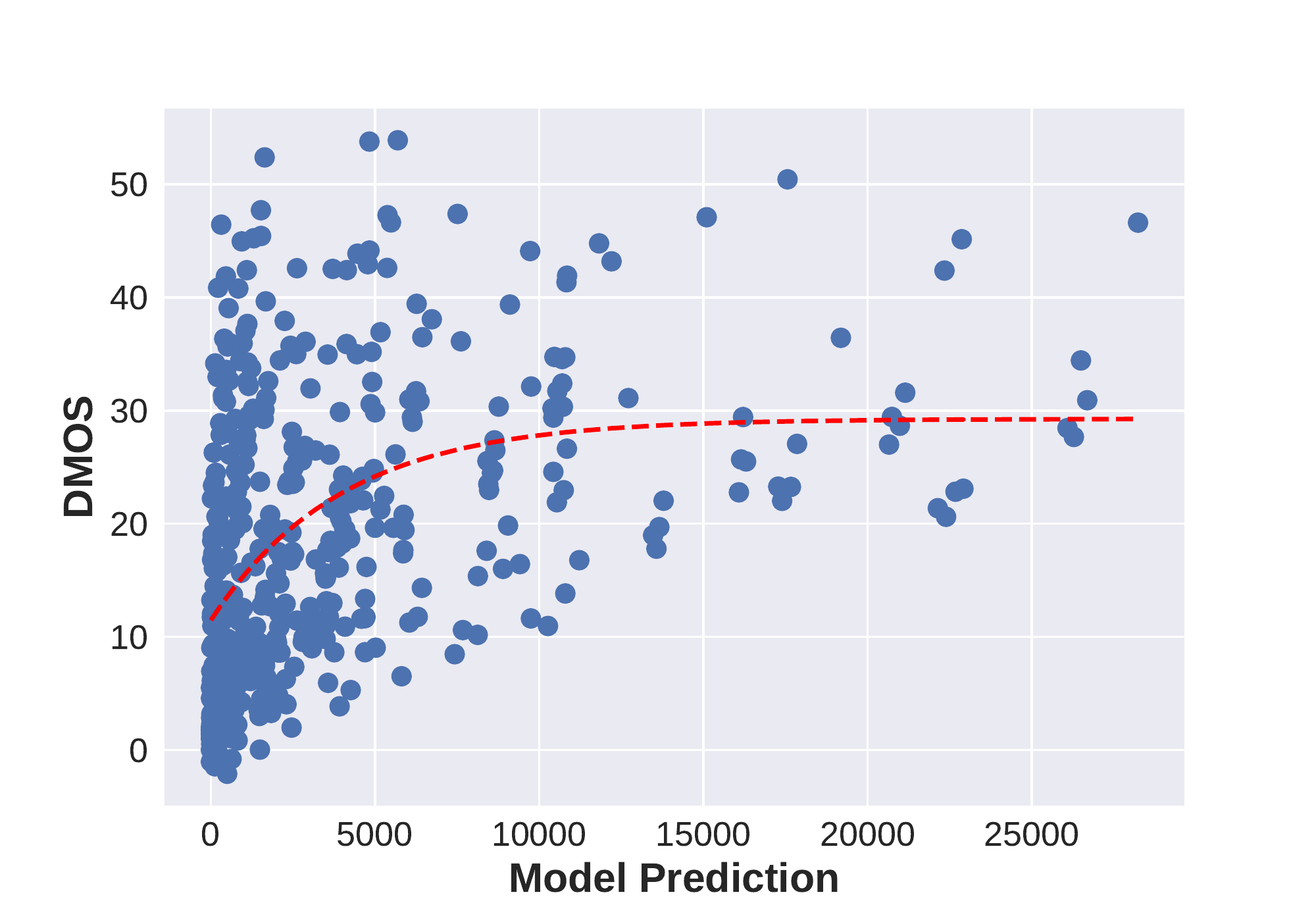}} \hfill
	\subfloat[FRQM \cite{zhang2017frame}]{\includegraphics[width=0.24\textwidth]{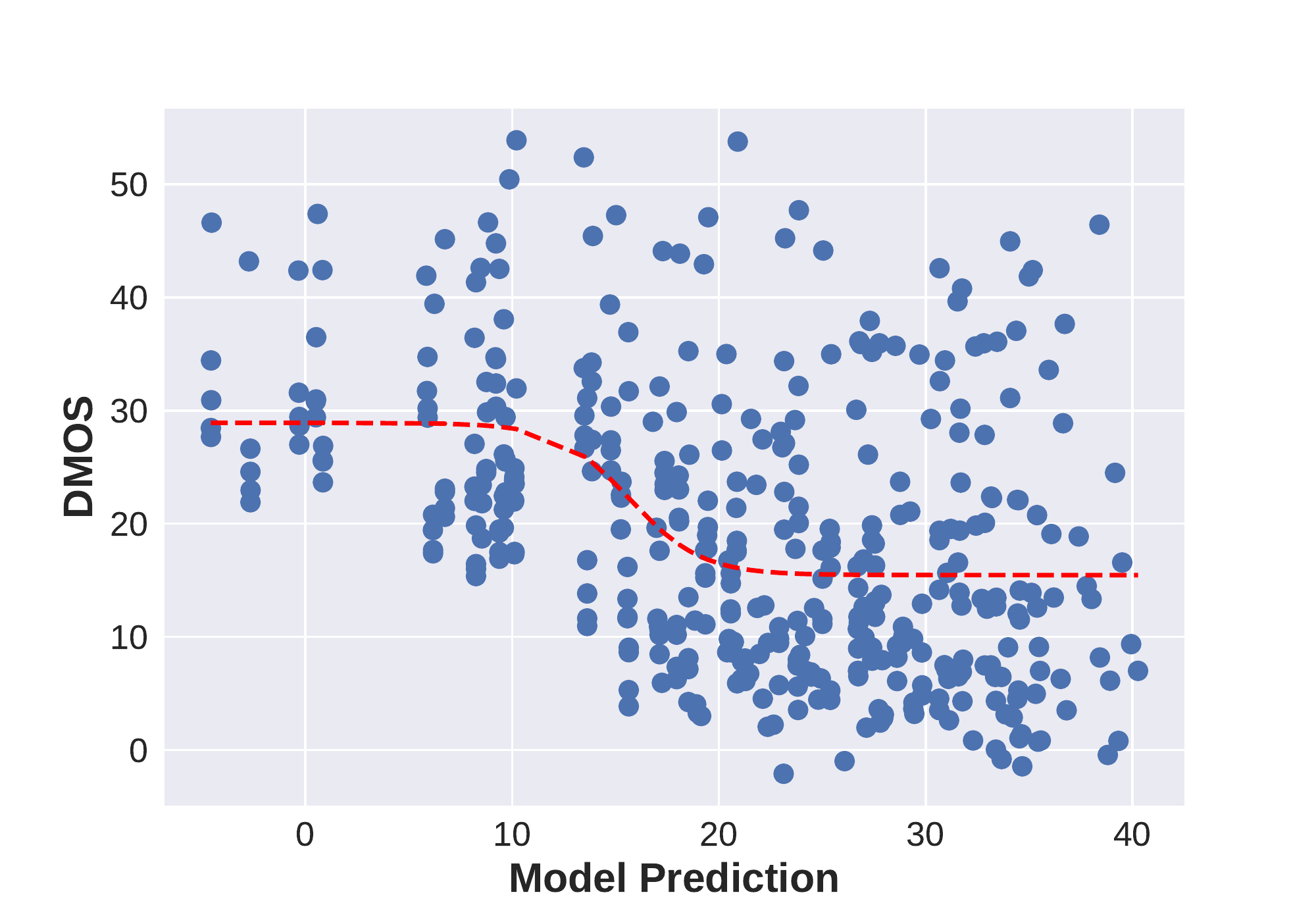}} \hfill
	\subfloat[VMAF \cite{VMAF2016}]{\includegraphics[width=0.24\textwidth]{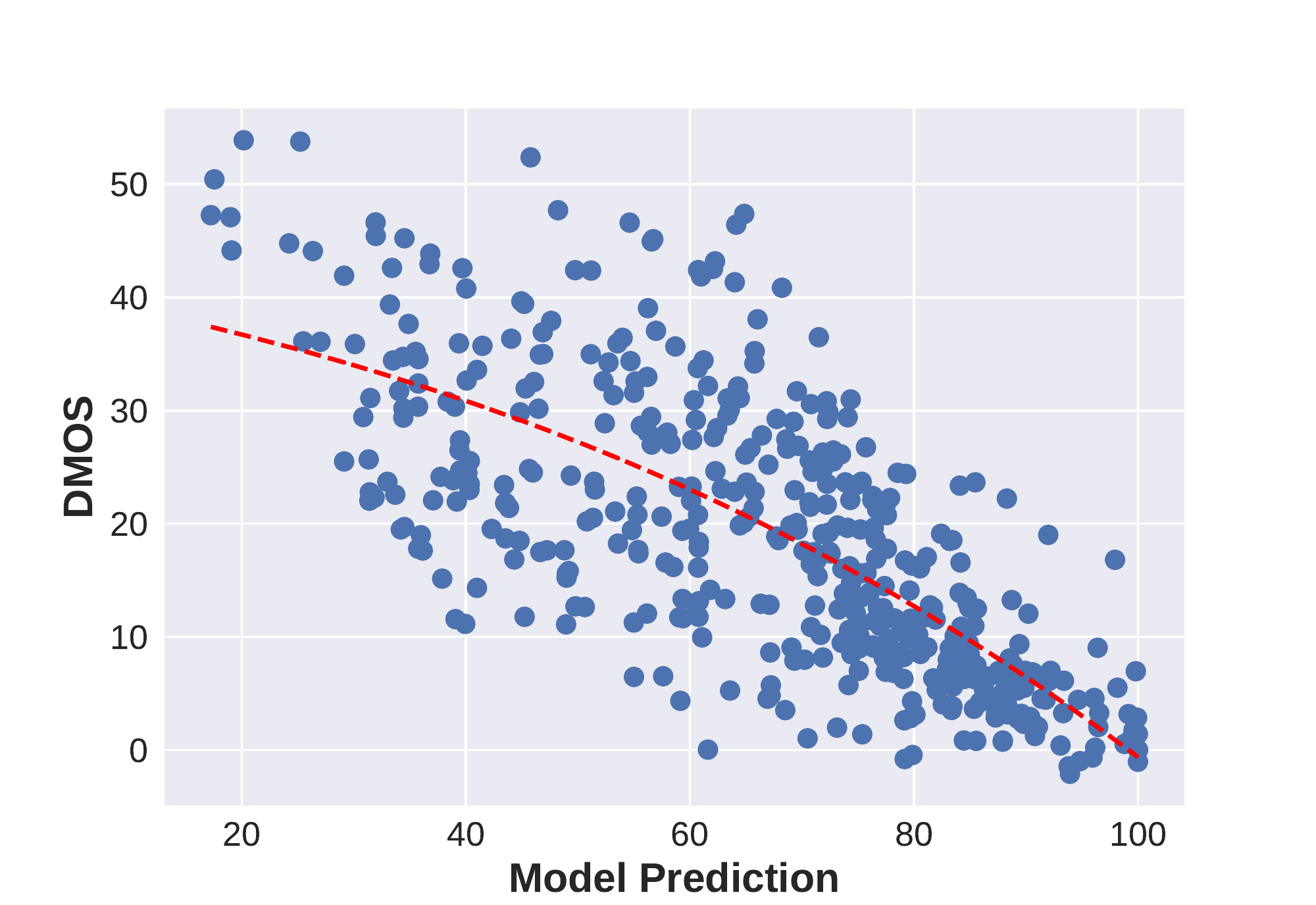}}\\
	\subfloat[deepVQA \cite{kim2018deep}]{\includegraphics[width=0.24\textwidth]{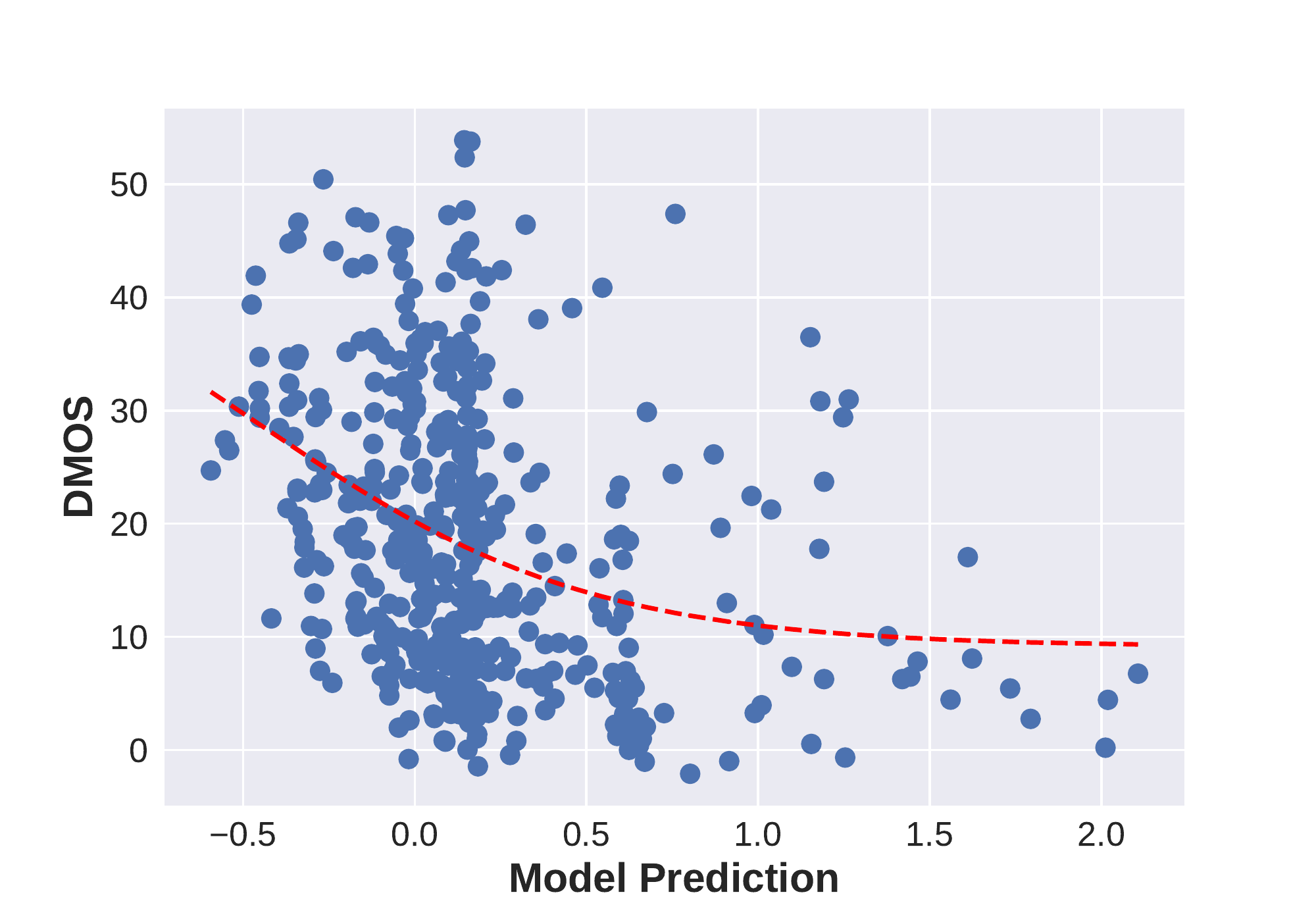}} \qquad
	\subfloat[GSTI \cite{pavan2020gsti}]{\includegraphics[width=0.24\textwidth]{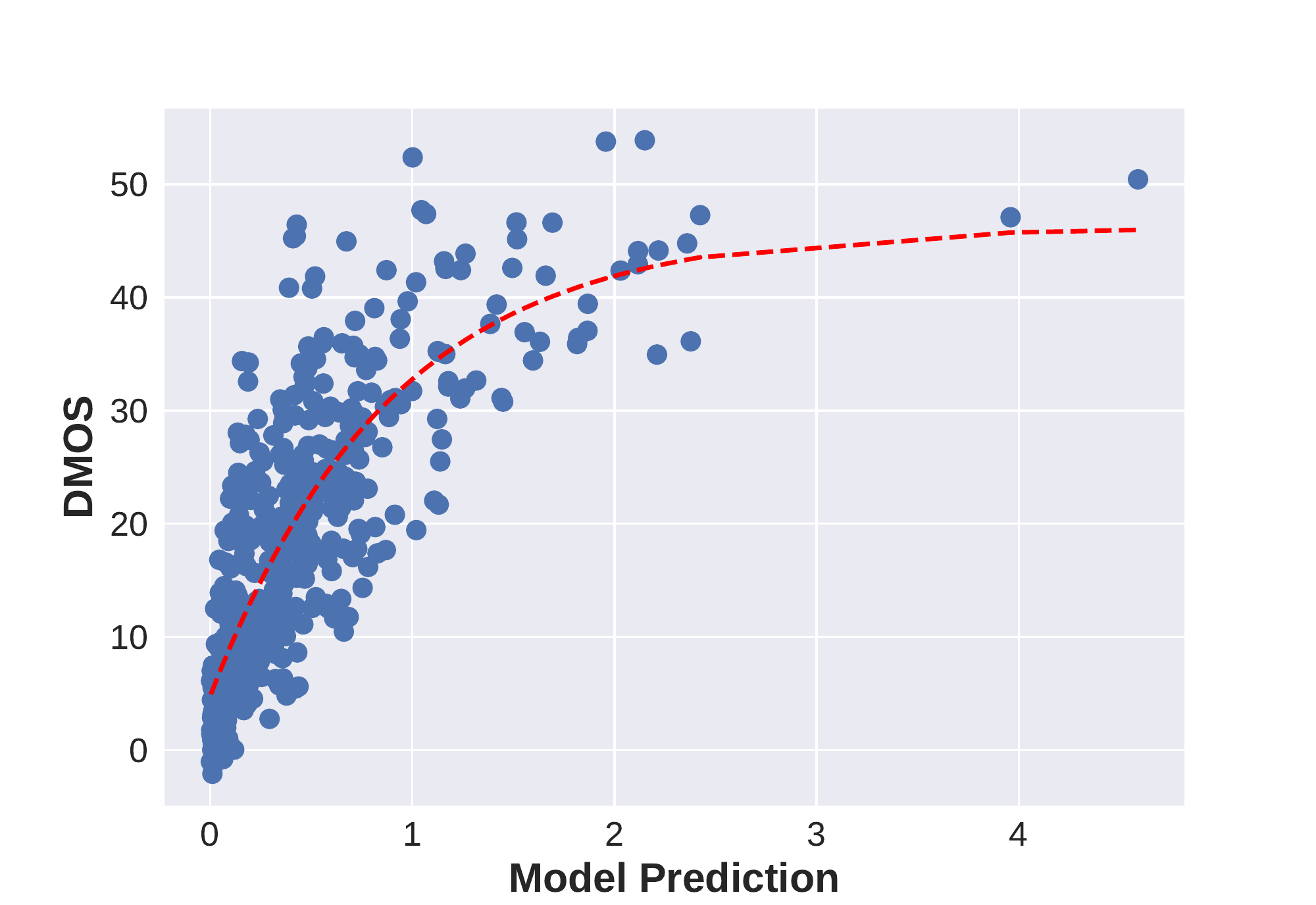}}
	\caption{Scatter plots of objective VQA scores vs DMOS across all videos in LIVE-YT-HFR database. The broken red line depicts the best fitting logistic function.}
	\label{fig:FR_scatter}
\end{figure*}

\subsection{Statistical Significance}
We analyzed the statistical significance of the subjective scores obtained from the human study, by performing a t-test between the Gaussian distributions centered at MOS values (and also employing the standard deviation of MOS) to infer the significance of individual frame rates at the 95\% confidence level. Since the condition being studied is a function of content, we performed our experiments separately on each content. In Table \ref{table:fps_significance} a value of `1' signifies that the row-condition was statistically superior (better visual quality) to the column-condition, while a value of `0' denotes the row is worse than a column; a value of `-' indicates that row and column conditions were statistically equivalent. For example, in Table \ref{table:fps_significance}, on all 16 contents the 120 fps videos exhibited statistically better visual quality than the 24 fps and 30 fps videos.

In Table \ref{table:fps_significance} we assess whether the MOS values were statistically distinguishable across frame-rates via the t-test. From the Table, we may observe that lower frame rates exhibited high degrees of statistical separability, but this margin of difference reduced towards high frame rates, especially beyond 60 fps. This reinforces our previous findings regarding the influence of frame rate on MOS.

\section{Evaluation of Objective Quality Predictors}
\label{sec:objective_QA}
As a way of demonstrating the value of the new LIVE-YT-HFR Database, we evaluated a variety of relevant objective VQA models on it. We employed four performance criteria: Spearman's rank order correlation coefficient (SROCC), Kendall's rank order correlation coefficient (KROCC), Pearson's linear correlation coefficient (PLCC) and the root mean squared error (RMSE) to the evaluate VQA models. Before computing PLCC and RMSE, the predicted scores were passed through a four-parameter logistic non-linearity as described in \cite{VQEG2000}
\begin{align}
	Q(x) = \beta_2 + \frac{\beta_1 - \beta_2}{1 + \exp\Bigg(-\Big(\frac{x - \beta_3}{|\beta_4|}\Big)\Bigg)}.
	\label{eqn:logistic_non}
\end{align}
Since we obtained MOS values from the human study, our database can be employed to create and/or test both FR and NR VQA models.

\subsection{FR-VQA Models}
To conduct FR model evaluations, we used the DMOS values obtained from equation \ref{eqn:DMOS}, considering the original lossless 120 fps videos as references. We began by testing four 4 FR-IQA methods: PSNR, SSIM \cite{wang2004image}, MS-SSIM \cite{wang2003multiscale} and FSIM \cite{zhang2011fsim}. These are image quality models, hence do not take into account any temporal information. These were computed on every frame, and the frame scores averaged across all frames to obtain the final video scores. We also studied five popular FR-VQA models: ST-RRED \cite{soundararajan2012video}, SpEED \cite{bampis2017speed}, FRQM\cite{zhang2017frame}, VMAF\footnote{We used the pretrained VMAF model available at \url{https://github.com/Netflix/vmaf}} \cite{VMAF2016}, and deepVQA \cite{kim2018deep}. Further, we also include a prototype model we recently devised, called the Generalized Spatio-Temporal Index (GSTI) \cite{pavan2020gsti}, which is designed to capture artifacts arising from frame rate variations, while also being responsive to other distortions. When evaluating deepVQA, we only used stage-1 of the pretrained model (trained on the LIVE-VQA \cite{seshadrinathan2010study} database) obtained from the code released by the authors. Among the above VQA models, only FRQM and GSTI allow for frame rate variations, while the rest require the reference and corresponding distorted sequences to have the same frame rate. When there were differing frame rates, we performed naive temporal upsampling by frame duplication to match the reference frame rate. Although we could have downsampled the reference, we avoided this method since it could potentially introduce artifacts (\eg judder) in the reference which is not desirable. We also do not consider any specialized temporal upsampling technique (\eg motion compensated temporal interpolation), as the performance can be very sensitive to the choice of interpolation method.

\begin{table}[t]
	\caption{Performance comparison of FR-VQA algorithms on the LIVE-YT-HFR database. The Distorted videos were temporally upsampled to match the reference frame rate. In each column first and second best models are boldfaced.}
	\label{Table:MOS_comparison}
	\centering
	\begin{tabular}{|c||c|c|c|c|}
		\hline
		~    & SROCC $\uparrow$ & KROCC $\uparrow$ & PLCC $\uparrow$ & RMSE $\downarrow$ \\ \hline \hline
		PSNR & 0.6950 & 0.5071 & 0.6685 & 9.023 \\ 
		SSIM \cite{wang2004image} & 0.4494 & 0.3102 & 0.4526 & 10.819 \\ 
		MS-SSIM \cite{wang2003multiscale} & 0.4898 & 0.3407 & 0.4673 & 10.726 \\ 
		FSIM \cite{zhang2011fsim} & 0.5251 & 0.3655 & 0.5008 & 10.502 \\ 
		ST-RRED \cite{soundararajan2012video} & 0.5531 & 0.3800 & 0.5107 & 10.431 \\ 
		SpEED \cite{bampis2017speed} & 0.4861 & 0.3409 & 0.4449 & 10.866 \\ 
		FRQM \cite{zhang2017frame} & 0.4216 & 0.2956 & 0.452 & 10.804 \\ 
		VMAF \cite{VMAF2016}& \textbf{0.7303} & \textbf{0.5358} & \textbf{0.7071} & \textbf{8.587} \\
		deepVQA \cite{kim2018deep} & 0.3463 & 0.2371 & 0.3329 & 11.441 \\
		GSTI \cite{pavan2020gsti} & \textbf{0.7909} & \textbf{0.5979} & \textbf{0.7910} & \textbf{7.422} \\
		\hline
	\end{tabular}
\end{table}

\begin{table*}[t]
	\caption{Performance comparison of various FR methods for individual frame rates in the HFR database. The Distorted videos were temporally upsampled to match the reference frame rate. In each column first and second best values are marked boldface.}
	\label{Table:FPS_comparison}
	\centering
	\scriptsize
	\scalebox{0.88}{
		\begin{tabular}{|c||c|c|c|c|c|c|c|c|c|c|c|c|c|c|}
			\hline
			& \multicolumn{2}{|c|}{24 fps} & \multicolumn{2}{|c|}{30 fps} & \multicolumn{2}{|c|}{60 fps} & \multicolumn{2}{|c|}{82 fps} & \multicolumn{2}{|c|}{98 fps} & \multicolumn{2}{|c|}{120 fps} & \multicolumn{2}{|c|}{Overall} \\
			\cline{2-15}
			~ & SROCC$\uparrow$ & PLCC$\uparrow$ & SROCC$\uparrow$ & PLCC$\uparrow$ & SROCC$\uparrow$ & PLCC$\uparrow$ & SROCC$\uparrow$ & PLCC$\uparrow$ & SROCC$\uparrow$ & PLCC$\uparrow$ & SROCC$\uparrow$ & PLCC$\uparrow$ & SROCC$\uparrow$ & PLCC$\uparrow$\\ \hline \hline
			PSNR & \textbf{0.4101} & \textbf{0.3647} & \textbf{0.4414} & \textbf{0.4179} & \textbf{0.6202} & 0.5719 & \textbf{0.6878} & 0.6431 & 0.7171 & 0.6489 & 0.6019 & 0.5937 & 0.6950 & 0.6685 \\ 
			SSIM \cite{wang2004image} & 0.1277 & 0.0949 & 0.1108 & 0.0816 & 0.2123 & 0.1845 & 0.2079 & 0.2430 & 0.3876 & 0.3964 & \textbf{0.7485} & 0.6726 & 0.4494 & 0.4526 \\ 
			MS-SSIM \cite{wang2003multiscale} & 0.2221 & 0.1500 & 0.1929 & 0.1112 & 0.2516 & 0.1900 & 0.2906 & 0.2549 & 0.4237 & 0.4007 & 0.6165 & 0.5843 & 0.4898 & 0.4673\\ 
			FSIM \cite{zhang2011fsim} & 0.2337 & 0.1882 & 0.2487 & 0.1786 & 0.3450 & 0.2776 & 0.2984 & 0.2839 & 0.5089 & 0.4735 & 0.7053 & 0.6368 & 0.5251 & 0.5008\\ 
			ST-RRED \cite{soundararajan2012video} & 0.1541 & 0.0369 & 0.1188 & 0.0307 & 0.5062 & 0.4457 & 0.3394 & 0.3271 & 0.4962 & 0.4556 & 0.6745 & 0.5906 & 0.5531 & 0.5107\\ 
			SpEED \cite{bampis2017speed} & 0.2591 & 0.1237 & 0.2278 & 0.0896 & 0.1824 & 0.1110 & 0.2955 & 0.2425 & 0.4118 & 0.3295 & 0.6827 & 0.6097 & 0.4861 & 0.4449\\ 
			FRQM \cite{zhang2017frame} & 0.1556 & 0.2089 & 0.0983 & 0.0854 & 0.0947 & 0.0309 & 0.0137 & 0.0035 & 0.0317 & 0.0100 & - & - & 0.4216 & 0.4520\\ 
			VMAF \cite{VMAF2016} & 0.1743 & 0.2669 & 0.2855 & 0.3740 & 0.5408 & \textbf{0.6015} & 0.6820 & \textbf{0.7390} & \textbf{0.8214} & \textbf{0.8128} & \textbf{0.7943} & \textbf{0.7844} & \textbf{0.7303} & \textbf{0.7071}\\ 
			deepVQA \cite{kim2018deep} & 0.1144 & 0.0495 & 0.1353 & 0.1059 & 0.2527 & 0.1652 & 0.1803 & 0.1515 & 0.2816 & 0.2654 & 0.6865 & 0.6209 & 0.3463 & 0.3329\\
			GSTI \cite{pavan2020gsti} & \textbf{0.4538} & \textbf{0.5935} & \textbf{0.4758} & \textbf{0.6689} & \textbf{0.6552} & \textbf{0.7566} & \textbf{0.7633} & \textbf{0.8183} & \textbf{0.7844} & \textbf{0.7775} & 0.7390 & \textbf{0.7003} & \textbf{0.7909} & \textbf{0.7910}\\
			\hline
		\end{tabular}
	}
\end{table*}

\begin{table*}
	\caption{Results of F-test between residuals of model predictions and DMOS values across various FR methods. Each cell contains 7 entries: 6 frame rates - 24, 30, 60, 82, 98, 120 fps and all videos, in that order. A value of `1' indicates that the row is statistically superior (better visual quality) than the column, while a value of `0' indicates that the column is statistically superior than the row. A value of `-' indicates statistical equivalence between row and column.}
	\label{table:FR_significance}
	\centering
	\begin{tabular}{|c||c|c|c|c|c|c|c|c|c|c|}
		\hline
		~ & PSNR & SSIM & MS-SSIM & FSIM & ST-RRED & SpEED & FRQM & VMAF & deepVQA & GSTI \\ \hline \hline
		PSNR & -\hspace{1pt}-\hspace{1pt}-\hspace{1pt}-\hspace{1pt}-\hspace{1pt}-\hspace{1pt}-\hspace{1pt} & -\hspace{1pt}-\hspace{1pt}1\hspace{1pt}1\hspace{1pt}1\hspace{1pt}-\hspace{1pt}1\hspace{1pt} & -\hspace{1pt}-\hspace{1pt}1\hspace{1pt}1\hspace{1pt}-\hspace{1pt}-\hspace{1pt}1\hspace{1pt} & -\hspace{1pt}-\hspace{1pt}-\hspace{1pt}1\hspace{1pt}-\hspace{1pt}-\hspace{1pt}1\hspace{1pt} & -\hspace{1pt}-\hspace{1pt}-\hspace{1pt}1\hspace{1pt}-\hspace{1pt}-\hspace{1pt}1\hspace{1pt} & -\hspace{1pt}-\hspace{1pt}1\hspace{1pt}1\hspace{1pt}1\hspace{1pt}-\hspace{1pt}1\hspace{1pt} & -\hspace{1pt}-\hspace{1pt}1\hspace{1pt}1\hspace{1pt}1\hspace{1pt}-\hspace{1pt}1\hspace{1pt} & -\hspace{1pt}-\hspace{1pt}-\hspace{1pt}-\hspace{1pt}0\hspace{1pt}-\hspace{1pt}-\hspace{1pt} & -\hspace{1pt}-\hspace{1pt}1\hspace{1pt}1\hspace{1pt}1\hspace{1pt}-\hspace{1pt}1\hspace{1pt} & 0\hspace{1pt}0\hspace{1pt}0\hspace{1pt}0\hspace{1pt}0\hspace{1pt}-\hspace{1pt}0\hspace{1pt} \\ 
		SSIM & -\hspace{1pt}-\hspace{1pt}0\hspace{1pt}0\hspace{1pt}0\hspace{1pt}-\hspace{1pt}0\hspace{1pt} & -\hspace{1pt}-\hspace{1pt}-\hspace{1pt}-\hspace{1pt}-\hspace{1pt}-\hspace{1pt}-\hspace{1pt} & -\hspace{1pt}-\hspace{1pt}-\hspace{1pt}-\hspace{1pt}-\hspace{1pt}-\hspace{1pt}-\hspace{1pt} & -\hspace{1pt}-\hspace{1pt}-\hspace{1pt}-\hspace{1pt}-\hspace{1pt}-\hspace{1pt}-\hspace{1pt} & -\hspace{1pt}-\hspace{1pt}-\hspace{1pt}-\hspace{1pt}-\hspace{1pt}-\hspace{1pt}-\hspace{1pt} & -\hspace{1pt}-\hspace{1pt}-\hspace{1pt}-\hspace{1pt}-\hspace{1pt}-\hspace{1pt}-\hspace{1pt} & -\hspace{1pt}-\hspace{1pt}-\hspace{1pt}-\hspace{1pt}-\hspace{1pt}-\hspace{1pt}-\hspace{1pt} & -\hspace{1pt}-\hspace{1pt}0\hspace{1pt}0\hspace{1pt}0\hspace{1pt}-\hspace{1pt}0\hspace{1pt} & -\hspace{1pt}-\hspace{1pt}-\hspace{1pt}-\hspace{1pt}-\hspace{1pt}-\hspace{1pt}-\hspace{1pt} & 0\hspace{1pt}0\hspace{1pt}0\hspace{1pt}0\hspace{1pt}0\hspace{1pt}-\hspace{1pt}0\hspace{1pt} \\ 
		MS-SSIM & -\hspace{1pt}-\hspace{1pt}0\hspace{1pt}0\hspace{1pt}-\hspace{1pt}-\hspace{1pt}0\hspace{1pt} & -\hspace{1pt}-\hspace{1pt}-\hspace{1pt}-\hspace{1pt}-\hspace{1pt}-\hspace{1pt}-\hspace{1pt} & -\hspace{1pt}-\hspace{1pt}-\hspace{1pt}-\hspace{1pt}-\hspace{1pt}-\hspace{1pt}-\hspace{1pt} & -\hspace{1pt}-\hspace{1pt}-\hspace{1pt}-\hspace{1pt}-\hspace{1pt}-\hspace{1pt}-\hspace{1pt} & -\hspace{1pt}-\hspace{1pt}-\hspace{1pt}-\hspace{1pt}-\hspace{1pt}-\hspace{1pt}-\hspace{1pt} & -\hspace{1pt}-\hspace{1pt}-\hspace{1pt}-\hspace{1pt}-\hspace{1pt}-\hspace{1pt}-\hspace{1pt} & -\hspace{1pt}-\hspace{1pt}-\hspace{1pt}-\hspace{1pt}-\hspace{1pt}-\hspace{1pt}-\hspace{1pt} & -\hspace{1pt}-\hspace{1pt}0\hspace{1pt}0\hspace{1pt}0\hspace{1pt}-\hspace{1pt}0\hspace{1pt} & -\hspace{1pt}-\hspace{1pt}-\hspace{1pt}-\hspace{1pt}-\hspace{1pt}-\hspace{1pt}-\hspace{1pt} & 0\hspace{1pt}0\hspace{1pt}0\hspace{1pt}0\hspace{1pt}0\hspace{1pt}-\hspace{1pt}0\hspace{1pt} \\ 
		FSIM & -\hspace{1pt}-\hspace{1pt}-\hspace{1pt}0\hspace{1pt}-\hspace{1pt}-\hspace{1pt}0\hspace{1pt} & -\hspace{1pt}-\hspace{1pt}-\hspace{1pt}-\hspace{1pt}-\hspace{1pt}-\hspace{1pt}-\hspace{1pt} & -\hspace{1pt}-\hspace{1pt}-\hspace{1pt}-\hspace{1pt}-\hspace{1pt}-\hspace{1pt}-\hspace{1pt} & -\hspace{1pt}-\hspace{1pt}-\hspace{1pt}-\hspace{1pt}-\hspace{1pt}-\hspace{1pt}-\hspace{1pt} & -\hspace{1pt}-\hspace{1pt}-\hspace{1pt}-\hspace{1pt}-\hspace{1pt}-\hspace{1pt}-\hspace{1pt} & -\hspace{1pt}-\hspace{1pt}-\hspace{1pt}-\hspace{1pt}-\hspace{1pt}-\hspace{1pt}-\hspace{1pt} & -\hspace{1pt}-\hspace{1pt}-\hspace{1pt}-\hspace{1pt}-\hspace{1pt}-\hspace{1pt}-\hspace{1pt} & -\hspace{1pt}-\hspace{1pt}0\hspace{1pt}0\hspace{1pt}0\hspace{1pt}-\hspace{1pt}0\hspace{1pt} & -\hspace{1pt}-\hspace{1pt}-\hspace{1pt}-\hspace{1pt}-\hspace{1pt}-\hspace{1pt}1\hspace{1pt} & 0\hspace{1pt}0\hspace{1pt}0\hspace{1pt}0\hspace{1pt}0\hspace{1pt}-\hspace{1pt}0\hspace{1pt} \\ 
		ST-RRED & -\hspace{1pt}-\hspace{1pt}-\hspace{1pt}0\hspace{1pt}-\hspace{1pt}-\hspace{1pt}0\hspace{1pt} & -\hspace{1pt}-\hspace{1pt}-\hspace{1pt}-\hspace{1pt}-\hspace{1pt}-\hspace{1pt}-\hspace{1pt} & -\hspace{1pt}-\hspace{1pt}-\hspace{1pt}-\hspace{1pt}-\hspace{1pt}-\hspace{1pt}-\hspace{1pt} & -\hspace{1pt}-\hspace{1pt}-\hspace{1pt}-\hspace{1pt}-\hspace{1pt}-\hspace{1pt}-\hspace{1pt} & -\hspace{1pt}-\hspace{1pt}-\hspace{1pt}-\hspace{1pt}-\hspace{1pt}-\hspace{1pt}-\hspace{1pt} & -\hspace{1pt}-\hspace{1pt}-\hspace{1pt}-\hspace{1pt}-\hspace{1pt}-\hspace{1pt}-\hspace{1pt} & -\hspace{1pt}-\hspace{1pt}-\hspace{1pt}-\hspace{1pt}-\hspace{1pt}-\hspace{1pt}-\hspace{1pt} & -\hspace{1pt}-\hspace{1pt}-\hspace{1pt}0\hspace{1pt}0\hspace{1pt}-\hspace{1pt}0\hspace{1pt} & -\hspace{1pt}-\hspace{1pt}-\hspace{1pt}-\hspace{1pt}-\hspace{1pt}-\hspace{1pt}1\hspace{1pt} & 0\hspace{1pt}0\hspace{1pt}0\hspace{1pt}0\hspace{1pt}0\hspace{1pt}-\hspace{1pt}0\hspace{1pt} \\ 
		SpEED & -\hspace{1pt}-\hspace{1pt}0\hspace{1pt}0\hspace{1pt}0\hspace{1pt}-\hspace{1pt}0\hspace{1pt} & -\hspace{1pt}-\hspace{1pt}-\hspace{1pt}-\hspace{1pt}-\hspace{1pt}-\hspace{1pt}-\hspace{1pt} & -\hspace{1pt}-\hspace{1pt}-\hspace{1pt}-\hspace{1pt}-\hspace{1pt}-\hspace{1pt}-\hspace{1pt} & -\hspace{1pt}-\hspace{1pt}-\hspace{1pt}-\hspace{1pt}-\hspace{1pt}-\hspace{1pt}-\hspace{1pt} & -\hspace{1pt}-\hspace{1pt}-\hspace{1pt}-\hspace{1pt}-\hspace{1pt}-\hspace{1pt}-\hspace{1pt} & -\hspace{1pt}-\hspace{1pt}-\hspace{1pt}-\hspace{1pt}-\hspace{1pt}-\hspace{1pt}-\hspace{1pt} & -\hspace{1pt}-\hspace{1pt}-\hspace{1pt}-\hspace{1pt}-\hspace{1pt}-\hspace{1pt}-\hspace{1pt} & -\hspace{1pt}-\hspace{1pt}0\hspace{1pt}0\hspace{1pt}0\hspace{1pt}-\hspace{1pt}0\hspace{1pt} & -\hspace{1pt}-\hspace{1pt}-\hspace{1pt}-\hspace{1pt}-\hspace{1pt}-\hspace{1pt}-\hspace{1pt} & 0\hspace{1pt}0\hspace{1pt}0\hspace{1pt}0\hspace{1pt}0\hspace{1pt}-\hspace{1pt}0\hspace{1pt} \\ 
		FRQM & -\hspace{1pt}-\hspace{1pt}0\hspace{1pt}0\hspace{1pt}0\hspace{1pt}-\hspace{1pt}0\hspace{1pt} & -\hspace{1pt}-\hspace{1pt}-\hspace{1pt}-\hspace{1pt}-\hspace{1pt}-\hspace{1pt}-\hspace{1pt} & -\hspace{1pt}-\hspace{1pt}-\hspace{1pt}-\hspace{1pt}-\hspace{1pt}-\hspace{1pt}-\hspace{1pt} & -\hspace{1pt}-\hspace{1pt}-\hspace{1pt}-\hspace{1pt}-\hspace{1pt}-\hspace{1pt}-\hspace{1pt} & -\hspace{1pt}-\hspace{1pt}-\hspace{1pt}-\hspace{1pt}-\hspace{1pt}-\hspace{1pt}-\hspace{1pt} & -\hspace{1pt}-\hspace{1pt}-\hspace{1pt}-\hspace{1pt}-\hspace{1pt}-\hspace{1pt}-\hspace{1pt} & -\hspace{1pt}-\hspace{1pt}-\hspace{1pt}-\hspace{1pt}-\hspace{1pt}-\hspace{1pt}-\hspace{1pt} & -\hspace{1pt}-\hspace{1pt}0\hspace{1pt}0\hspace{1pt}0\hspace{1pt}-\hspace{1pt}0\hspace{1pt} & -\hspace{1pt}-\hspace{1pt}-\hspace{1pt}-\hspace{1pt}-\hspace{1pt}-\hspace{1pt}-\hspace{1pt} & -\hspace{1pt}0\hspace{1pt}0\hspace{1pt}0\hspace{1pt}0\hspace{1pt}-\hspace{1pt}0\hspace{1pt} \\ 
		VMAF & -\hspace{1pt}-\hspace{1pt}-\hspace{1pt}-\hspace{1pt}1\hspace{1pt}-\hspace{1pt}-\hspace{1pt} & -\hspace{1pt}-\hspace{1pt}1\hspace{1pt}1\hspace{1pt}1\hspace{1pt}-\hspace{1pt}1\hspace{1pt} & -\hspace{1pt}-\hspace{1pt}1\hspace{1pt}1\hspace{1pt}1\hspace{1pt}-\hspace{1pt}1\hspace{1pt} & -\hspace{1pt}-\hspace{1pt}1\hspace{1pt}1\hspace{1pt}1\hspace{1pt}-\hspace{1pt}1\hspace{1pt} & -\hspace{1pt}-\hspace{1pt}-\hspace{1pt}1\hspace{1pt}1\hspace{1pt}-\hspace{1pt}1\hspace{1pt} & -\hspace{1pt}-\hspace{1pt}1\hspace{1pt}1\hspace{1pt}1\hspace{1pt}-\hspace{1pt}1\hspace{1pt} & -\hspace{1pt}-\hspace{1pt}1\hspace{1pt}1\hspace{1pt}1\hspace{1pt}-\hspace{1pt}1\hspace{1pt} & -\hspace{1pt}-\hspace{1pt}-\hspace{1pt}-\hspace{1pt}-\hspace{1pt}-\hspace{1pt}-\hspace{1pt} & -\hspace{1pt}-\hspace{1pt}1\hspace{1pt}1\hspace{1pt}1\hspace{1pt}-\hspace{1pt}1\hspace{1pt} & 0\hspace{1pt}0\hspace{1pt}0\hspace{1pt}-\hspace{1pt}-\hspace{1pt}-\hspace{1pt}0\hspace{1pt} \\ 
		deepVQA & -\hspace{1pt}-\hspace{1pt}0\hspace{1pt}0\hspace{1pt}0\hspace{1pt}-\hspace{1pt}0\hspace{1pt} & -\hspace{1pt}-\hspace{1pt}-\hspace{1pt}-\hspace{1pt}-\hspace{1pt}-\hspace{1pt}-\hspace{1pt} & -\hspace{1pt}-\hspace{1pt}-\hspace{1pt}-\hspace{1pt}-\hspace{1pt}-\hspace{1pt}-\hspace{1pt} & -\hspace{1pt}-\hspace{1pt}-\hspace{1pt}-\hspace{1pt}-\hspace{1pt}-\hspace{1pt}0\hspace{1pt} & -\hspace{1pt}-\hspace{1pt}-\hspace{1pt}-\hspace{1pt}-\hspace{1pt}-\hspace{1pt}0\hspace{1pt} & -\hspace{1pt}-\hspace{1pt}-\hspace{1pt}-\hspace{1pt}-\hspace{1pt}-\hspace{1pt}-\hspace{1pt} & -\hspace{1pt}-\hspace{1pt}-\hspace{1pt}-\hspace{1pt}-\hspace{1pt}-\hspace{1pt}-\hspace{1pt} & -\hspace{1pt}-\hspace{1pt}0\hspace{1pt}0\hspace{1pt}0\hspace{1pt}-\hspace{1pt}0\hspace{1pt} & -\hspace{1pt}-\hspace{1pt}-\hspace{1pt}-\hspace{1pt}-\hspace{1pt}-\hspace{1pt}-\hspace{1pt} & 0\hspace{1pt}0\hspace{1pt}0\hspace{1pt}0\hspace{1pt}0\hspace{1pt}-\hspace{1pt}0\hspace{1pt} \\ 
		GSTI & 1\hspace{1pt}1\hspace{1pt}1\hspace{1pt}1\hspace{1pt}1\hspace{1pt}-\hspace{1pt}1\hspace{1pt} & 1\hspace{1pt}1\hspace{1pt}1\hspace{1pt}1\hspace{1pt}1\hspace{1pt}-\hspace{1pt}1\hspace{1pt} & 1\hspace{1pt}1\hspace{1pt}1\hspace{1pt}1\hspace{1pt}1\hspace{1pt}-\hspace{1pt}1\hspace{1pt} & 1\hspace{1pt}1\hspace{1pt}1\hspace{1pt}1\hspace{1pt}1\hspace{1pt}-\hspace{1pt}1\hspace{1pt} & 1\hspace{1pt}1\hspace{1pt}1\hspace{1pt}1\hspace{1pt}1\hspace{1pt}-\hspace{1pt}1\hspace{1pt} & 1\hspace{1pt}1\hspace{1pt}1\hspace{1pt}1\hspace{1pt}1\hspace{1pt}-\hspace{1pt}1\hspace{1pt} & -\hspace{1pt}1\hspace{1pt}1\hspace{1pt}1\hspace{1pt}1\hspace{1pt}-\hspace{1pt}1\hspace{1pt} & 1\hspace{1pt}1\hspace{1pt}1\hspace{1pt}-\hspace{1pt}-\hspace{1pt}-\hspace{1pt}1\hspace{1pt} & 1\hspace{1pt}1\hspace{1pt}1\hspace{1pt}1\hspace{1pt}1\hspace{1pt}-\hspace{1pt}1\hspace{1pt} & -\hspace{1pt}-\hspace{1pt}-\hspace{1pt}-\hspace{1pt}-\hspace{1pt}-\hspace{1pt}-\hspace{1pt} \\
		\hline
	\end{tabular}
\end{table*}

The performances of the various FR methods is shown in Table \ref{Table:MOS_comparison}. In Fig. \ref{fig:FR_scatter}, scatter plots of the objective VQA scores against DMOS are shown for all of the FR-VQA models, along with the best fitting logistic function obtained from equation \ref{eqn:logistic_non}. GSTI was the best performing VQA model amongst the compared models across all performance criteria. The poor correlation values of the FR-IQA indices PSNR, SSIM, MS-SSIM and FSIM highlight the importance of the efficacy of crucial temporal information for VQA in HFR scenarios. The inferior performance of other existing VQA models is also indicative of the fundamental limitations encountered when reference and distorted sequences have differing frame rates.

In order to individually analyze performance against each frame rate, we subdivided the database into sets of videos having the same frame rates. The performance comparison is shown in Table \ref{Table:FPS_comparison}. To avoid clutter, we only included SROCC and PLCC scores in the evaluation. However, KROCC and RMSE were observed to follow the same trends as in Table \ref{Table:FPS_comparison}. It may be seen that VMAF and GSTI performed well across almost all frame rates. We also observed an interesting anomaly, whereby the PSNR achieved higher performance at lower frame rates as compared to some of the other models. This seemed surprising, given that PSNR has been shown to correlate relatively poorly with human judgments of quality \cite{wang2009mean}, even when the distortions are purely spatial. However, it achieved higher correlations at lower frame-rates, without access to temporal distortion. This is very likely because algorithms like SSIM estimate the spatial aspects of distortion very accurately, causing a ``spatial bias" when high spatial quality is combined with low temporal quality that is not assessed. In such instances, the spatial quality measurements unduly influence the overall video quality prediction. This is not an advantage of PSNR so much as it is a disadvantage of frame-based models, and provides one of the storngest arguments for modeling temporal quality.


From Table \ref{Table:FPS_comparison} it maybe observed that almost all VQA models have comparatively lower correlations for 24/30 fps sequences. We hypothesize that the presence of contents involving sports and/or high motion severely affects the perceived video quality, particularly when viewed at these frame rates. Existing VQA models generally do not account for these type of artifacts. VMAF, although trained on cinematic content, fails to capture judder/stutter artifacts, since these cinematic sequences are generally devoid of those distortions because of the careful involvement of the cinematographer, colorist, and editor. The FRQM index correlates very poorly when analyzed at fixed frame rates. This is because it only captures frame rate variations, hence is insensitive to other artifacts. Moreover, FRQM can only be calculated between videos having differing frame rates.

\subsection{Statistical Evaluation}
Next we addressed the question of whether the observed differences in performance in Table \ref{Table:MOS_comparison} are statistically significant. We employed an F-test on residuals between DMOS and the objective scores predicted by VQA models after applying logistic non-linearity \cite{seshadrinathan2010study}. The main underlying assumption is that residuals follow a Gaussian distribution with zero mean. An F-test was conducted on the ratios of variances of the residuals between each pair of objective models. Statistical equivalence is achieved if the variances of residuals from the two objective models are equal at the 95\% significance level. The results of the statistical significance tests are reported in Table \ref{table:FR_significance}. We followed similar convention as used in Table \ref{table:fps_significance} in determining statistical superiority. Each cell in Table \ref{table:FR_significance} consists of 7 entries: 6 frame rates - 24, 30, 60, 82, 98, 120 fps and all videos, in that order.

To summarize the results in Table \ref{table:FR_significance}, the performance of GSTI was statistically superior to the other FR-VQA models across all frame rates.

\begin{table*}[t]
	\caption{Performance comparison of various NR models for individual frame rates in the HFR database. The numbers denote median values for $500$ iterations of randomly chosen train and test sets (subjective MOS vs predicted MOS). The values inside the brackets denote standard deviation. Top two performing models in each column are highlighted.}
	\label{Table:FPS_comparison_NR}
	\centering
	\scriptsize
	\scalebox{0.775}{
		\begin{tabular}{|c||c|c|c|c|c|c|c|c|c|c|c|c|c|c|}
			\hline
			& \multicolumn{2}{|c|}{24 fps} & \multicolumn{2}{|c|}{30 fps} & \multicolumn{2}{|c|}{60 fps} & \multicolumn{2}{|c|}{82 fps} & \multicolumn{2}{|c|}{98 fps} & \multicolumn{2}{|c|}{120 fps} & \multicolumn{2}{|c|}{Overall} \\
			\cline{2-15}
			~ & SROCC$\uparrow$ & PLCC$\uparrow$ & SROCC$\uparrow$ & PLCC$\uparrow$ & SROCC$\uparrow$ & PLCC$\uparrow$ & SROCC$\uparrow$ & PLCC$\uparrow$ & SROCC$\uparrow$ & PLCC$\uparrow$ & SROCC$\uparrow$ & PLCC$\uparrow$ & SROCC$\uparrow$ & PLCC$\uparrow$\\ \hline \hline
			BRISQUE \cite{mittal2012no} & \textbf{0.53(0.27)} & \textbf{0.45(0.27)} & \textbf{0.56(0.26)} & \textbf{0.49(0.26)} & \textbf{0.56(0.25)} & \textbf{0.51(0.24)} & \textbf{0.52(0.24)} & \textbf{0.46(0.24)} & \textbf{0.55(0.24)} & \textbf{0.48(0.24)} & \textbf{0.51(0.26)} & \textbf{0.48(0.25)} & 0.38(0.2) & 0.38(0.2) \\ 
			NIQE \cite{mittal2013making} & 0.3(0.35) & 0.32(0.32) & 0.32(0.34) & 0.35(0.3) & 0.41(0.25) & 0.33(0.25) & 0.42(0.25) & 0.38(0.25) & 0.45(0.24) & 0.39(0.24) & 0.46(0.19) & 0.39(0.21) & 0.28(0.18) & 0.25(0.2) \\ 
			V-BLIINDS \cite{saad2014blind} & 0.39(0.33) & 0.29(0.29) & 0.34(0.33) & 0.26(0.28) & 0.36(0.26) & 0.28(0.24) & 0.43(0.27) & 0.36(0.23) & 0.43(0.25) & 0.35(0.22) & 0.46(0.24) & 0.41(0.22) & \textbf{0.43(0.21)} & \textbf{0.4(0.20)} \\ 
			TLVQM \cite{korhonen2019two} & 0.26(0.35) & 0.28(0.29) & 0.26(0.35) & 0.22(0.26) & 0.26(0.32) & 0.23(0.27) & 0.31(0.29) & 0.25(0.26) & 0.28(0.28) & 0.22(0.25) & 0.27(0.29) & 0.21(0.28) & 0.32(0.25) & 0.29(0.23)\\ 
			Li \etal \cite{li2019quality} & \textbf{0.49(0.25)} & \textbf{0.67(0.13)} & \textbf{0.47(0.28)} & \textbf{0.64(0.15)} & \textbf{0.46(0.25)} & \textbf{0.59(0.16)} & \textbf{0.54(0.22)} & \textbf{0.58(0.16)} & \textbf{0.56(0.21)} & \textbf{0.6(0.15)} & \textbf{0.49(0.20)} & \textbf{0.53(0.17)} & \textbf{0.42(0.19)} & \textbf{0.54(0.13)}\\
			\hline
		\end{tabular}
	}
\end{table*}

\subsection{NR-VQA Models}
\begin{table}[t]
	\caption{Median values of SROCC, KROCC, PLCC and RMSE with No Reference QA Algorithms for $500$ iterations of randomly chosen train and test sets (subjective MOS vs predicted MOS). The values inside the brackets denote standard deviation. Top two performing models are highlighted.}
	\label{table:NR_MOS}
	\centering
	\scriptsize
	\begin{tabular}{|c||c|c|c|c|}
		\hline 
		~    & SROCC $\uparrow$ & KROCC $\uparrow$ & PLCC $\uparrow$ & RMSE $\downarrow$ \\ \hline \hline
		BRISQUE \cite{mittal2012no} & 0.376(0.2) & 0.255(0.14) & 0.384(0.2) & 12.47(4.44) \\ 
		NIQE \cite{mittal2013making} & 0.278(0.18) & 0.2(0.12) & 0.255(0.2) & 12.71(1.33) \\ 
		V-BLIINDS \cite{saad2014blind} & \textbf{0.430(0.21)} & \textbf{0.312(0.15)} & \textbf{0.403(0.20)} & \textbf{16.84(5.36)} \\ 
		TLVQM \cite{korhonen2019two} & 0.320(0.25) & 0.241(0.17) & 0.289(0.23) & 17.61(6.24) \\ 
		Li \etal \cite{li2019quality} & \textbf{0.418(0.19)} & \textbf{0.315(0.14)} & \textbf{0.536(0.13)} & \textbf{11.9(2.34)} \\
		\hline
	\end{tabular}
\end{table}

Since we also obtained MOS values on every video, we were able to evaluate NR-VQA models on the new database. We compared the performance of several NR-VQA models, including BRISQUE \cite{mittal2012no}, NIQE \cite{mittal2013making}, V-BLIINDS \cite{saad2014blind} and TLVQM \cite{korhonen2019two} as reported in Table \ref{table:NR_MOS}. All of these models employ handcrafted features. The former three derive from Natural Scene Statistics (NSS) models, while the latter uses a combination of low and high complexity features. We also included the recently proposed model by \cite{li2019quality}, which employs a deep CNN along with a Gated Recurrent Unit (GRU) for blind video quality evaluation. To evaluate this model on our database, we employed a pretrained model (trained on the KonViD-1K \cite{hosu2017konstanz} database) released by the authors. We report the performance of BRISQUE, V-BLIINDS and TLVQM features, when trained on the LIVE-YT-HFR database using a Support Vector Regressor (SVR) with Radial Basis Function (RBF) kernel. For training purposes we divided the LIVE-YT-HFR database content-wise into two random subsets: 80\% for training and the remaining 20\% for testing - ensuring that there existed no overlap between the contents present in the train and test subsets. For fair analysis, we repeated this random train-test division 500 times, and report the median performance in Table \ref{table:NR_MOS}. Since BRISQUE is an image quality model, we calculated features on every frame, and averaged the features across frames to obtain video level features. When computing NIQE, scores were obtained on every frame, then averaged to obtain overall video scores. It may be observed that V-BLIINDS and \cite{li2019quality} were the top performing NR methods. There were substantial differences between the correlations obtained by FR and NR models, indicating the significance of reference information.

\begin{figure}[t]
	\centering
	\includegraphics[width=0.45\textwidth]{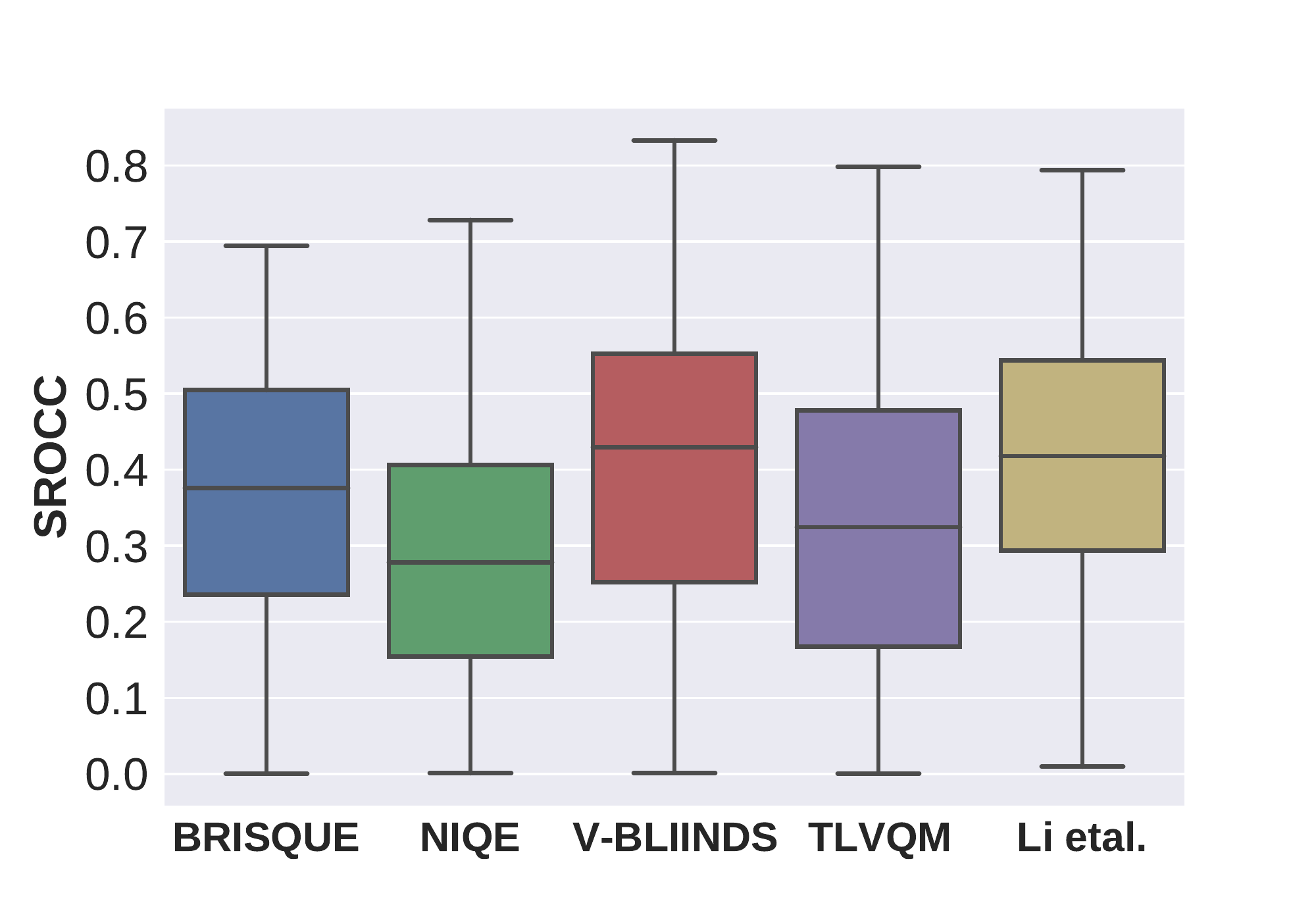}
	\caption{Boxplots of SROCC distributions for multiple NR-VQA algorithms}
	\label{fig:NR_boxplot}
\end{figure}

In Table \ref{Table:FPS_comparison_NR} the performances of the NR models for fixed frame rates is analyzed. It may be observed that \cite{li2019quality} achieved the highest correlation across all frame rates. Interestingly, BRISQUE, although an IQA model, achieved high correlations for individual frame rates, but when analyzed collectively across frame rates yielded poor correlation. Since sets of individual frame rates only differ by the amount of compression, BRISQUE might effectively differentiate them, but its overall efficacy was reduced by its inability to capture frame rate quality variations. In Fig. \ref{fig:NR_boxplot}, boxplots depicting the spreads of SROCC values for each NR algorithm are shown, illustrating the reduced spread of scores of the method in \cite{li2019quality}, as also reported in Table \ref{Table:FPS_comparison_NR}.

\section{Discussion and Conclusion}
\label{sec:conclusion} 
We constructed a large HFR database comprising of 480 videos spanning six different frame rates and five compression levels, obtained from 16 diverse contents involving both HD and UHD spatial resolutions. We used these to conduct a human study involving 85 volunteer subjects. The LIVE-YT-HFR Database is unique with respect to the number of frame rates, and the joint presence of compression artifacts and frame rate variations. We also presented a comprehensive evaluation of existing FR and NR-VQA models and benchmarked their performance on the new database.

Important and obvious conclusions of our analysis are that frame rate has considerable influence on human subjective judgments of video quality, and that humans prefer higher frame rates over lower ones. Further, this preference of higher frame rates is not ubiquitous, but depends on the content being viewed. Videos involving significant camera motion almost always received higher quality scores at high frame rates, as compared to low frame rates. Moreover, the quality gain associated with frame rate increases diminishes somewhat above 60 fps. This might be expected, since videos at lower frame rates suffer from judder/strobing artifacts, while quality variations at higher frame rates, \eg 98 and 120 fps, are more subtle, becoming noticeable only when there is high motion.

The results of objective VQA model testing were particularly encouraging. The majority of the IQA methods faltered, underscoring the importance of capturing temporal information. The tested FR-VQA models mainly suffered from two shortcomings: 1) Almost all FR-VQA algorithms require the same frame rate for reference and distorted videos, thus a temporal upsampling step is needed, which can influence the outcome. 2) When analyzed separately on fixed frame rates, model performance varied across frame rates. The tested NR-VQA models also failed to capture temporal artifacts arising from frame rate changes, since the features they use do not explicitly address these type of distortions.

We believe this new HFR database will benefit the research community towards advancing and understanding the complex relationships associated with frame rate and perceptual video quality. We also believe that these relationships are not limited to HFR content, and much may be learned regarding temporal information in generic VQA models.

\section{Acknowledgment}
The authors would like to thank all the volunteers who took part in the human study.

\bibliographystyle{IEEEtran}
\bibliography{template_4}

\begin{IEEEbiography}[{\includegraphics[width=1in,height=1.25in,clip,keepaspectratio]{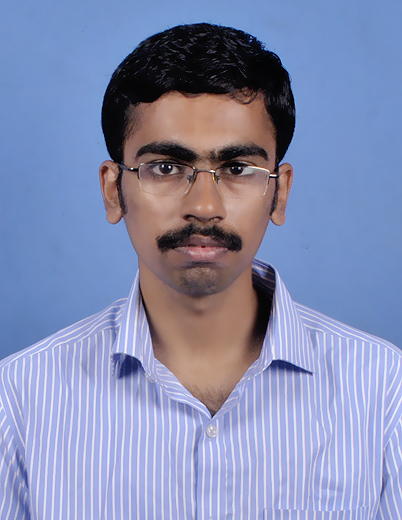}}]{Pavan C. Madhusudana}
received the B.Tech. degree in Electronics and Communication Engineering from The National Institute of Technology Karnataka (NITK), Surathkal, India, in 2016, and the M.Tech. (Research) degree in Electrical and Communication Engineering from the Indian Institute of Science (IISc), Bangalore, India in 2018. He is currently pursuing the Ph.D. degree in Electrical and Computer engineering with The University of Texas at Austin, USA. His research interests include image and video signal processing, computer vision, and machine learning.
\end{IEEEbiography}

\begin{IEEEbiography}[{\includegraphics[width=1in,height=1.25in,clip,keepaspectratio]{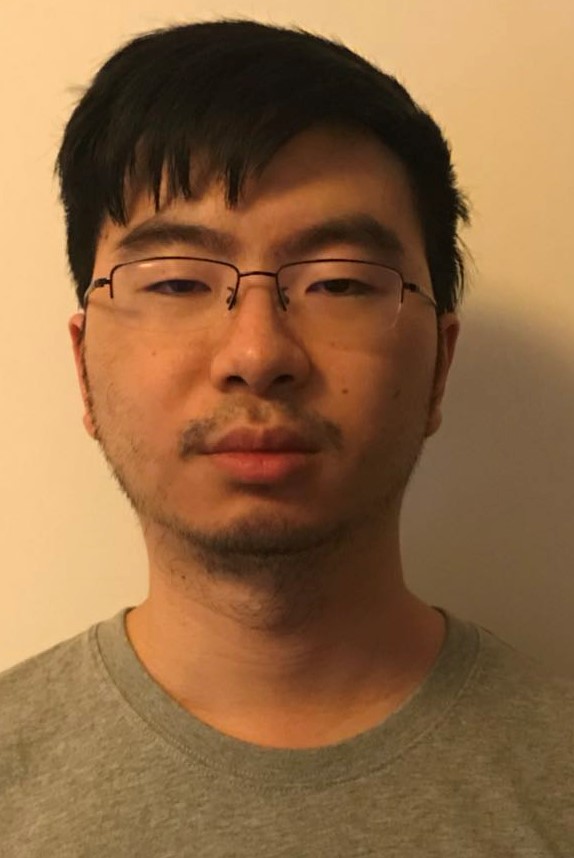}}]{Xiangxu Yu}
received the B.Eng in Electronic and Information Engineering from The Hong Kong Polytechnic University, Hongkong, China, and the M.S. degree in Electrical and Computer Engineering from The University of Texas at Austin, Austin, in 2015 and 2018, respectively. He is currently pursuing the Ph.D. degree with the Laboratory for Image and Video Engineering, The University of Texas at Austin. His research interests focus on image and video processing, and machine learning.
\end{IEEEbiography}

\begin{IEEEbiography}
[{\includegraphics[width=1in,height=1.25in,clip,keepaspectratio]{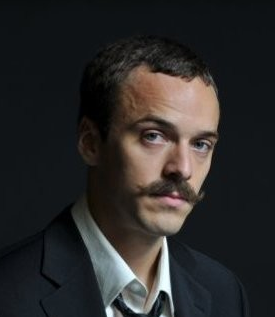}}]{Neil Birkbeck} obtained his Ph.D from the University of Alberta in 2011 working on topics in computer vision, graphics and robotics, with a specific focus on image-based modeling and rendering. He went on to become a Research Scientist at Siemens corporate research working on automatic detection and segmentation of anatomical structures in full body medical images. He is now a software engineer in the Media Algorithms team at YouTube/Google, with research interests in perceptual video processing, video coding, and video quality assessment.
\end{IEEEbiography}

\begin{IEEEbiography}
[{\includegraphics[width=1in,height=1.25in,clip,keepaspectratio]{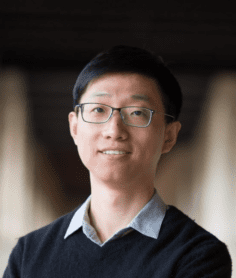}}]{Yilin Wang}
received B.S. and M.S. degrees in Computer Science from Nanjing University, China, in 2005 and 2008 respectively, PhD degree in Computer Science from the University of North Carolina at Chapel Hill in 2014, working on topics in computer vision and image processing. After graduation, he joined the Media Algorithm team in Youtube/Google. His research fields include video processing infrastructure, video quality assessment, and video compression.
\end{IEEEbiography}

\begin{IEEEbiography}
[{\includegraphics[width=1in,height=1.25in,clip,keepaspectratio]{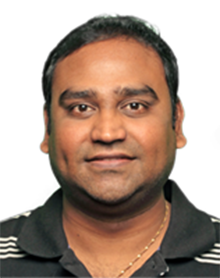}}]{Balu Adsumilli} manages and leads the Media Algorithms group at YouTube/Google. He did his masters in University of Wisconsin Madison in 2002, and his PhD at University of California Santa Barbara in 2005, on watermark-based error resilience in video communications. From 2005 to 2011, he was Sr. Research Scientist at Citrix Online, and from 2011-2016, he was Sr. Manager Advanced Technology at GoPro, at both places developing algorithms for images/video quality enhancement, compression, capture, and streaming. He is an active member of IEEE (and MMSP TC), ACM, SPIE, and VES, and has co-authored more than 120 papers and patents. His fields of research include image/video processing, machine vision, video compression, spherical capture, VR/AR, visual effects, and related areas.
\end{IEEEbiography}

\begin{IEEEbiography}[{\includegraphics[width=1in,height=1.25in,clip,keepaspectratio]{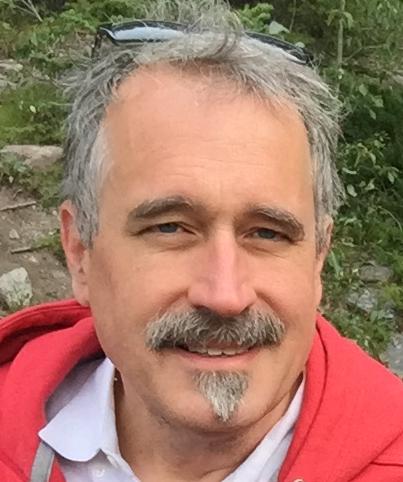}}]{Alan C. Bovik} (F ’95) is the Cockrell Family Regents Endowed Chair Professor at The University of Texas at Austin. His research interests include image processing, digital photography, digital television, digital streaming video, social media, and visual perception. For his work in these areas he has been the recipient of the 2019 Progress Medal from The Royal Photographic Society, the 2019 IEEE Fourier Award, the 2017 Edwin H. Land Medal from The Optical Society, a 2015 Primetime Emmy Award for Outstanding Achievement in Engineering Development from the Television Academy, a 2020 Technology and Engineering Emmy Award from the National Academy for Television Arts and Sciences, and the Norbert Wiener Society Award and the Karl Friedrich Gauss Education Award from the IEEE Signal Processing Society. He has also received about 10 ‘best journal paper’ awards, including the 2016 IEEE Signal Processing Society Sustained Impact Award. His books include The Essential Guides to Image and Video Processing. He co-founded and was longest-serving Editor-in-Chief of the IEEE Transactions on Image Processing, and also created/Chaired the IEEE International Conference on Image Processing which was first held in Austin, Texas, 1994.
\end{IEEEbiography}

\EOD

\end{document}